\documentclass{jfm}
\usepackage{graphicx}
\usepackage{epstopdf, epsfig}
\usepackage{amsmath}
\usepackage{array}
\usepackage{multirow}
\usepackage{psfrag}
\usepackage[font=footnotesize,format=plain,labelsep=period,justification=raggedright]{caption}
\usepackage{subcaption}

\usepackage[utf8]{inputenc}
\usepackage{combelow}
\usepackage{amssymb}
\usepackage{braket}
\usepackage{bm}
\usepackage{color}
\usepackage{tikz}
\usepackage{xr}
\definecolor{final}{rgb}{0,0.5,0.75}

\usetikzlibrary{decorations.pathreplacing}
\usetikzlibrary{shapes,arrows,chains}

\tikzstyle{block} = [rectangle, draw, text centered, rounded corners]
\tikzstyle{line} = [draw, -latex']

\def\hati{{\hat{\imath}}}
\def\hatj{{\hat{\jmath}}}
\def\hatk{{\hat{k}}}
\def\hatl{{\hat{\ell}}}
\def\hvarphi{{\hat{\varphi}}}
\def\htheta{{\hat{\theta}}}

\makeatletter
\renewcommand*\env@matrix[1][\arraystretch]{%
  \edef\arraystretch{#1}%
  \hskip -\arraycolsep
  \let\@ifnextchar\new@ifnextchar
  \array{*\c@MaxMatrixCols c}}
\makeatother

\shorttitle{Axisymmetric flows on the torus}
\shortauthor{S. Busuioc, H. Kusumaatmaja, V. E. Ambrus}

\title{Axisymmetric flows on the torus geometry}

\author{
Sergiu Busuioc\aff{1}, \and
H. Kusumaatmaja\aff{2}, \corresp{\email{halim.kusumaatmaja@durham.ac.uk}} \and
Victor E. Ambru\cb{s}\aff{3,4} \corresp{\email{victor.ambrus@e-uvt.ro}}
}

\affiliation{
\aff{1}School of Engineering, University of Edinburgh, Edinburgh, EH9 3FB, UK
\aff{2}Department of Physics, Durham University, Durham, DH1 3LE, UK
\aff{3}Department of Physics, West University of Timi\cb{s}oara, Timi\cb{s}oara, 300223, Romania
\aff{4}Department of Mathematics and Statistics, Old Dominion University, Norfolk, VA 23529, USA
}

\begin{document}

\maketitle

\begin{abstract}
We present a series of analytically solvable axisymmetric flows on the torus geometry.
For the single-component flows, we describe the propagation of sound waves for perfect fluids, 
as well as the viscous damping of shear and longitudinal waves for isothermal and thermal fluids. 
Unlike the case of planar geometry, the non-uniform curvature on a torus necessitates a distinct 
spectrum of eigenfrequencies and their corresponding basis functions. This has several 
interesting consequences, including breaking the degeneracy between even and odd modes,
a lack of periodicity even in the flows of perfect fluids and the loss of Galilean invariance
for flows with velocity components in the poloidal direction. 
For the multi-component flows, we study the equilibrium configurations and relaxation dynamics 
of axisymmetric fluid stripes, described using the Cahn-Hilliard equation. We find a second-order
phase transition in the equilibrium location of the stripe as a function of its area $\Delta A$.
This phase transition leads 
to a complex dependence of the Laplace pressure on $\Delta A$. We also derive the
underdamped oscillatory dynamics as the stripes approach equilibrium. 
Furthermore, relaxing the assumption of axial symmetry, we derive the conditions under 
which the stripes become unstable.
In all cases, the analytical results are confirmed numerically using a finite-difference 
Navier-Stokes solver.
\end{abstract}

\section{Introduction}\label{sec:intro}

In recent years there has been a growing interest in studying and understanding hydrodynamic flows on curved surfaces, supported by increasing evidence for their relevance in a wide range of problems in nature and engineering. Examples include phenomena in materials science, such as the motion of electrons in graphene \citep{Giordanelli18}, interface rheology in foams \citep{Cox04} and the dynamics of confined active matter \citep{Keber14,Henkes18,Janssen17,Pearce19}; in biophysics, such as flows on curved biomembranes \citep{Henle10,arroyo09,Izzi18,Fonda2018} or fluid deformable surfaces \citep{torres-sanchez19,voigt19}; in fusion technology, such as plasma motion under toroidal confinement \citep{Boozer05}; and in geophysics, such as zonal flows on planets and the Sun \citep{sasaki15}. 

In this work, we consider a series of axisymmetric flows on the torus geometry (i.e. flows which are homogeneous with respect to the azimuthal torus coordinate) for which analytic solutions can be derived. The torus is chosen as it represents one of the simplest geometries with non-uniform curvature. On the one hand, these flows allow us to identify novel flow phenomena arising from the presence of non-uniform curvature, which are absent on planar geometries. Importantly, our analytical calculations allow us to identify the key ingredients for observing these phenomena. On the other hand, this work can provide several non-trivial benchmark problems suitable for developing computational methods for flows on curved surfaces. To date, a number of numerical approaches have been developed to solve the fluid equations of motion on curved manifolds, including using finite-element \citep{Dziuk07,Dziuk13}, level set \citep{Bertalmio01}, phase-field  \citep{Ratz06}, closest point \citep{Macdonald10} and lattice Boltzmann \citep{Ambrus19} methods. 
Recently, interest has been shown also for fluid systems on evolving curved manifolds both for incompressible \citep{koba17,Nitschke19} and compressible \citep{koba18} fluids. 
However, despite the availability of these various methods, to date there is still a lack of systematic comparisons to assess and compare their accuracy and robustness. Here, we directly compare all the analytical derivations against numerical simulations obtained using a finite-difference Navier-Stokes solver.

In total we discuss five problems with increasing complexity. First, we start with the propagation of sound waves for a perfect fluid on a torus. Then, we consider viscous damping. We study shear wave damping, where the fluid velocity is in the azimuthal direction of the torus, as well as the damping of longitudinal waves, where the fluid velocity is in the poloidal direction. These three problems have been regularly studied for the planar geometry, and they are popular benchmark case studies for Navier-Stokes solvers \citep{sofonea03,rembiasz17,sofonea18pre,busuioc20camwa}. Here, for their torus equivalent, we analyse the flows by deriving their distinct discrete spectrum of eigenfrequencies and corresponding basis functions. We carry out these studies for isothermal and thermal single-component fluids, as well as for multicomponent fluids described by the Cahn-Hilliard equation. Interestingly, we find that the degeneracy between odd and even modes is broken, which can be observed both in the oscillation frequencies and decay rates of those modes. Due to the non-uniform curvature, we will also show that Galilean invariance and flow periodicity, as commonly observed in the planar geometry, can be lost.

Next, we focus on an axisymmetric fluid stripe embedded on a torus. Focussing on the static configurations, the spatial symmetry is broken in the poloidal direction and we find a second-order phase transition in the location of the minimum energy configurations depending on the area of the fluid stripes. We further derive the equivalent of a Laplace pressure on a torus geometry, where additional terms are present due to the underlying curved metric. As a consequence of the phase transition, the Laplace pressure of a fluid stripe in equilibrium has a complex dependence on its area. For completeness, we also discuss other configurations, available when the axisymmetry restriction is lifted, which may have lower energy compared to the stripe configuration under certain conditions. Furthermore, we derive
the regime of stability of the stripe configurations under small azimuthal 
perturbations.
We then study the relaxation dynamics of the fluid stripes.  When the Cahn-Hilliard equation is coupled with hydrodynamics, we find an underdamped oscillatory motion for the stripe dynamics. We derive the oscillation frequency and the exponential decay rate. The case in the absence of hydrodynamics, where the stripes simply relax exponentially to their equilibrium position, is discussed in Sec.~\ref{SM:sec:noh} of the supplementary material.

The paper is structured as follows. Sec.~\ref{sec:hydro} describes the hydrodynamic equations for flows 
on general curved surfaces, which are then specialised to the case of axisymmetric flows on the torus geometry. The five axisymmetric flow problems are introduced and presented in Secs.~\ref{sec:inv}-\ref{sec:CH}. Taken together, our series of axisymmetric flows cover single- and multi-component flows, static and dynamic aspects, instabilities under small perturbations, perfect and viscous fluids, isothermal and thermal cases and motion in the azimuthal and poloidal directions of the torus. A summary of the work and concluding remarks are finally presented in Sec.~\ref{sec:conc}.
The paper also includes two appendices. Appendix~\ref{app:conv} presents a convergence 
order analysis of the solver employed in this paper with respect to the first three benchmark tests,
discussed in Sections~\ref{sec:inv}, \ref{sec:shear} and \ref{sec:damp}. Appendix~\ref{app:modes} discusses 
the perturbative procedure that we use to obtain the mode solutions necessary for the 
spatial part of the linearised hydrodynamic equations, which are employed in the main text.

The supplementary material (SM) \citep{SM:JFM} contains three sections. 
Section~\ref{SM:numsch} provides details on the implementation of our numerical scheme.
Section~\ref{SM:CH} contains mathematical complements for the analysis of the Cahn-Hilliard model
on the torus geometry. Finally, Section~\ref{SM:EF} applies the procedure described in Appendix~\ref{app:modes} 
to derive expansions of the mode functions and related quantities up to ninth order with respect to the 
torus aspect ratio, $0 < a = r / R < 1$. 
These expansions are available for download 
under as gnuplot files ($funcs-inv.gpl$ and $funcs-shear.gpl$) and Mathematica notebooks ($funcs_inv.nb$ and $funcs_shear.nb$) in the supplementary material. In addition, two animations of the development of the instability of fluid stripes due to azimuthal perturbations, discussed in Sec.~\ref{sec:laplace:inst}, also provided in the supplementary material.

\section{Hydrodynamics on curved surfaces} \label{sec:hydro}

Over the past decades, there have been several attempts to formulate the 
hydrodynamic equations on curved surfaces \citep{serrin59,marsden94,taylor11}. 
In this paper, we take the strategy of first writing the fluid equations with 
respect to curvilinear coordinates in covariant form. Employing the orthonormal 
vielbein vector field $\{\bm{e}_{\hat{\alpha}}, \alpha = 1, 2, 3\}$, we then take the 
first two vectors, $\bm{e}_{\hati}$ ($i = 1, 2$) to be tangent to the manifold 
and enforce that no dynamics occurs along the third vector, $\bm{e}_{\hat{3}}$.
This approach allows the fundamental conservation equations for mass, momentum and energy for fluids on a curved surface to be written in covariant form as follows:
\begin{subequations}\label{eq:hydro}
\begin{align}
 \frac{D\rho}{Dt} + \rho \nabla_\hati u^\hati =& 0, \label{eq:hydro_cont}\\
 \rho \frac{Du^\hati}{Dt} + \nabla_{\hatj} \mathsf{T}^{\hati\hatj} =&  
 \rho f^\hati,\label{eq:hydro_cauchy}\\
 \rho \frac{De}{Dt} + \mathsf{T}^{\hati\hatj} \nabla_\hati u_\hatj =& 
 - \nabla_\hati q^\hati,\label{eq:hydro_en}
\end{align}
\end{subequations}
where $1 \le i,j \le 2$ cover the tensor components along the directions 
which are tangent to the surface. 
In the above, $\rho$ is the fluid mass density,
$\bm{u} = u^\hati \bm{e}_\hati$ is the fluid velocity,
$\nabla_\hati$ is the covariant derivative, 
$D/Dt = \partial_t + u^\hatj \nabla_\hatj$ is the material (convective) derivative, 
$\mathsf{T}^{\hati\hatj}$ is the pressure tensor,
$f^\hati$ is the external force per unit mass (which 
we neglect for the remainder of this paper),
$e = c_v T$ is the internal energy per unit mass,
$c_v$ is the specific heat capacity, 
$T$ is the fluid temperature and 
$q^\hati$ is the heat flux.
The set of relations \eqref{eq:hydro_cont}--\eqref{eq:hydro_en} are compatible 
with those derived from kinetic theory in curvilinear coordinates \citep{busuioc19}
or on curved manifolds \citep{Ambrus19}. Furthermore, the computation of the 
divergence of the stress tensor in a covariant way ensures the compatibility 
with the approaches currently taken in the literature 
\citep{arroyo09,taylor11,nitschke17,gross18}.

The hydrodynamic equations, Eq.~\eqref{eq:hydro}, are not closed 
unless the pressure tensor $\mathsf{T}^{\hati\hatj}$ and heat flux 
$q^{\hati}$ are known. The specific models employed in this paper for 
these quantities are discussed below in Subsec. \ref{sec:hydro:P} 
and \ref{sec:hydro:q}, respectively. After briefly introducing the relevant
differential operators in Subsec. \ref{eq:hydro:dtorus},
we explicitly write the equations of motion for axisymmetric flows on the torus geometry
in Subsec. \ref{eq:hydro:torus}.

\subsection{Models for the pressure tensor}\label{sec:hydro:P}

We restrict our analysis to the case of Newtonian fluids, for which the pressure tensor can be decomposed as
\begin{equation}
 \mathsf{T}^{\hati\hatj} = P_{\rm b} \delta^{\hati\hatj}+ 
 \mathsf{P}_{\kappa}^{\hati\hatj} - 
 \tau^{\hati\hatj}. \label{eq:Tab_dec}
\end{equation}
The dissipative part $\tau^{\hati\hatj} = \tau^{\hati\hatj}_{\rm dyn} 
+ \tau^{\hati\hatj}_{\rm bulk}$ of the 
pressure tensor for a two-dimensional Newtonian 
fluid reads
\begin{equation}
 \tau^{\hati\hatj}_{\rm dyn} = \eta\left(\nabla^\hati u^\hatj + 
 \nabla^\hatj u^\hati - 
 \delta^{\hati\hatj} \nabla_\hatk u^\hatk\right), \qquad 
 \tau^{\hati\hatj}_{\rm bulk} = \eta_v \delta^{\hati\hatj} 
 \nabla_\hatk u^\hatk,
 \label{eq:tau_def}
\end{equation}
where $\eta$ and $\eta_v$ are the dynamic and bulk (volumetric) viscosity coefficients, respectively. 
For the applications considered in this work, the dependence of 
the transport coefficients on the flow properties is not important.
Hence, we adopt the usual 
model in which the kinematic viscosities $\nu$ and $\nu_v$ are 
constant, such that $\eta$ and $\eta_v$ are computed using 
\begin{equation}
 \eta = \nu \rho,\qquad 
 \eta_v = \nu_v \rho.\label{eq:eta}
\end{equation}

For the first two terms in Eq.~\eqref{eq:Tab_dec}, $\mathsf{P}_{\rm b}$ is the isotropic bulk pressure and $P_{\kappa}^{\hati\hatj}$ is responsible for the surface tension, which is relevant in the case of multicomponent systems. For ideal single-component fluids, the bulk pressure is the ideal gas pressure and the surface tension part vanishes
\begin{equation}
 P_{\rm b} = P_{\rm i} = \frac{\rho k_B T}{m}, \qquad 
 \mathsf{P}^{\hati\hatj}_{\kappa} = 0,
\end{equation}
where $m$ is the average particle mass. In this paper, we always use units such that $m=1$.

For multicomponent flows, we consider a binary mixture of fluids $\mathcal{A}$ and $\mathcal{B}$, characterised by an order parameter $\phi$, such that $\phi = 1$ corresponds to a bulk $\mathcal{A}$ fluid and $\phi = -1$ to a bulk $\mathcal{B}$ fluid.
The coexistence of these two bulk fluids can be realised by using a simple form for the Helmholtz free energy $\Psi$:
\begin{equation}
 \Psi = \int_V dV\, (\psi_{\rm b} + \psi_{\rm g}),
 \label{eq:Landau}
\end{equation}
where the bulk $\psi_{\rm b}$ and the gradient $\psi_{\rm g}$ free energy densities are \citep{Briant04,KrugerBook}:
\begin{equation}
 \psi_{\rm b} = \frac{\text{A}}{4} (1-\phi^2)^2, \qquad 
 \psi_{\rm g} = \frac{\kappa}{2} (\nabla \phi)^2.
 \label{eq:psi}
\end{equation}
Here, $\text{A}$ and $\kappa$ are free parameters, which are related to 
the interface width $\xi$ and surface tension $\sigma$ through
\begin{equation}
 \xi = \sqrt{\frac{\kappa}{\text{A}}}, \qquad 
 \sigma = \sqrt{\frac{8 \kappa \text{A}}{9}}.
 \label{eq:xi_gamma}
\end{equation}
For simplicity, we consider the case when 
$\text{A}$ and $\kappa$ have constant values throughout the fluid.
The chemical potential can be derived by taking the functional 
derivative of the free energy with respect to the order parameter, giving
\begin{equation}
 \mu = \frac{\delta \Psi}{\delta \phi} = 
 \mu_{\rm b} + \mu_{\rm g}, \qquad
 \mu_{\rm b} = -\text{A}\phi(1- \phi^2), \qquad 
 \mu_{\rm g} = - \kappa \Delta \phi.
\label{eq:mu}
\end{equation}

The additional contributions to the pressure tensor arising from this free energy model can be found by imposing
\begin{equation}
 \nabla_\hatj [P_{\rm CH} \delta^{\hati\hatj} + 
 \mathsf{P}_{{\rm CH}; \kappa}^{\hati\hatj}] = 
 \phi \nabla^\hati \mu,
\end{equation}
which leads to
\begin{align}
 & P_{\rm b} = P_{\rm i} +  P_{\rm CH} = \frac{\rho k_B T}{m} -\text{A} \left(\frac{\phi^2}{2} - 
 \frac{3\phi^4}{4}\right), \nonumber\\
& \mathsf{P}^{\hati\hatj}_{\kappa} = \mathsf{P}^{\hati\hatj}_{{\rm CH}; \kappa} = 
 \kappa \nabla^\hati \phi \nabla^\hatj \phi - 
 \kappa \delta^{\hati\hatj} \left[
 \phi \Delta \phi  +\frac{1}{2} (\nabla \phi)^2\right].
 \label{eq:pbin}
\end{align}

For multicomponent flows, in addition to the hydrodynamic equations in Eq.~\eqref{eq:hydro},
another equation of motion is needed to capture the evolution of the order parameter $\phi$.
Here it is governed by the Cahn-Hilliard equation
\begin{equation}
 \frac{D\phi}{Dt} + \phi \nabla_\hati u^\hati = 
 \nabla_\hati (M \nabla^\hati \mu),
 \label{eq:CH}
\end{equation}
where $M$ is the mobility parameter,
$D/Dt = \partial_t + u^\hati \nabla_\hati$ is the material derivative
and the fluid velocity $\bm{u}$ is a solution of the hydrodynamic 
equations Eq.~\eqref{eq:hydro}. For simplicity, we assume for simplicity that $M$ takes 
a constant value throughout the fluid.

\subsection{Model for the heat flux}\label{sec:hydro:q}

We consider fluids for which the heat 
flux is given via Fourier's law
\begin{equation}
 q^{\hati} = - \text{k} \nabla^\hati T.
\end{equation}
The heat conductivity $\text{k}$ is related to the 
dynamic viscosity through the Prandtl number ${\rm Pr}$:
\begin{equation}
 {\rm Pr} = c_p \frac{\eta}{\text{k}} = 
 c_v \frac{\gamma \eta}{\text{k}},
\end{equation}
where $c_p$ is the specific heat at constant pressure 
and $\gamma$ is the adiabatic index.
For definiteness, we assume that 
${\rm Pr}$ is a constant number in this work.

When considering isothermal flows, 
the temperature is assumed to remain constant and 
the heat flux vanishes
\begin{equation}
 T = T_{\rm Iso} = {\rm const}, \qquad 
 q^\hati_{\rm Iso} = 0.
\end{equation}
In this case, the energy equation is no longer taken 
into consideration.

\subsection{Differential operators on the torus geometry}
\label{eq:hydro:dtorus}
\begin{figure}
\begin{center}
\includegraphics[width=0.65\linewidth]{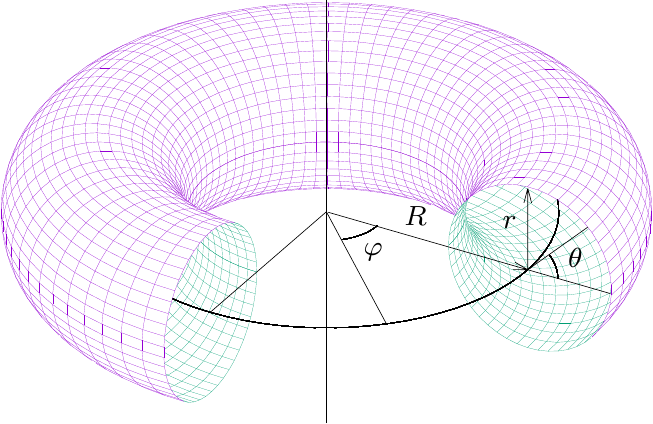}
\caption{Spatial discretisation of the torus geometry.\label{fig:tor_coord}}
\end{center}
\end{figure}

In this subsection we provide a brief introduction to the differential
geometry approach we have used to analyse the fluid flows.
For concreteness, we consider the parametrisation of a torus 
of outer radius $R$ and inner radius $r$ using 
the coordinates $q^i \in \{\varphi, \theta\}$
($i$ represents a coordinate index) as follows:
\begin{align}
 x =& (R + r \cos \theta) \cos \varphi,\nonumber\\
 y =& (R + r \cos \theta) \sin \varphi,\nonumber\\
 z =& r \sin \theta. \label{eq:torus}
\end{align}
Here, $\varphi$ and $\theta$ are the azimuthal and 
the poloidal angles, respectively, and 
the system is periodic with respect to both angles 
with period of $2\pi$. Figure~\ref{fig:tor_coord} depicts 
the coordinates and the equidistant spatial discretisation in $\varphi$ and $\theta$.

The line element on the torus can be written 
with respect to $\theta$ and $\varphi$ as follows:
\begin{equation}
 ds^2 = (R + r \cos\theta)^2 d\varphi^2 + r^2 d\theta^2.
\end{equation}
The metric tensor associated with the above line element 
has the following non-vanishing components:
\begin{equation}
 \mathsf{g}_{\varphi\varphi} = (R + r \cos\theta)^2, \qquad 
 \mathsf{g}_{\theta\theta} = r^2.
\end{equation}

Similar to the approach taken by other authors \citep{Nitschke12,reuther18}, 
it is convenient to introduce the vielbein vector frame 
$\{\bm{e}_{\hvarphi}, \bm{e}_{\htheta}\}$, 
where $\bm{e}_\hati = e_\hati^i \bm{\partial}_i$ 
is the notation for a vector tangent to the surface. 
The components $e_\hati^i$ satisfy
\begin{equation}
 \mathsf{g}_{ij} e^i_\hati e^j_\hatj = \delta_{\hati\hatj}.
 \label{eq:vielbein}
\end{equation}
The natural choice for the vielbein on the torus 
geometry is
\begin{equation}
 \bm{e}_\hvarphi = \frac{\bm{\partial}_\varphi}{R(1 +a \cos\theta)}, \qquad 
 \bm{e}_\htheta = \frac{\bm{\partial}_\theta}{r},
\end{equation}
where the following notation was introduced for future convenience:
\begin{equation}
 a = \frac{r}{R}.
\end{equation}
The corresponding vielbein co-frame, comprised 
of the one-forms $\bm{\omega}^\hati = \omega^{\hati}_i \bm{dq}^i$,
is given by
\begin{equation}
 \bm{\omega}^\hvarphi = R(1 + a\cos\theta) \bm{d\varphi}, \qquad 
 \bm{\omega}^\htheta = r \bm{d\theta},
\end{equation}
such that
\begin{equation}
 \omega^\hati_i e_{\hatj}^i = \delta^{\hati}{}_\hatj, \qquad 
 \delta_{\hati\hatj} \omega^\hati_i \omega^\hatj_j = g_{ij}.
\end{equation}

The algebraic rules to compute the terms appearing 
in Eq.~\eqref{eq:hydro} are described below.
The gradient $\nabla_\hati F = e_\hati^i \partial_i F$ 
of a scalar function $F$ has the following components:
\begin{equation}
 \nabla_\hvarphi F = \frac{\partial_\varphi F}{R(1 + a\cos\theta)},
 \qquad \nabla_\htheta F = \frac{1}{r} \partial_\theta F.
\end{equation}
For a vector field $A^\hati$, the covariant derivative is 
\begin{equation}
 \nabla_\hatj A^\hati = e_\hatj^j \partial_j A^{\hati} + 
 \Gamma^\hati{}_{\hatk\hatj} A^\hatk,
\end{equation}
and when the vector index is lowered, it becomes
\begin{equation}
 \nabla_\hatj A_\hati = e_\hatj^j \partial_j A_\hati - 
 \Gamma^\hatk{}_{\hati\hatj} A_\hatk.
\end{equation}
For the computation of the covariant derivatives, the connection coefficients 
$\Gamma^\hati{}_{\hatk\hatj} = \delta^{\hati\hatl} \Gamma_{\hatl\hatk\hatj}$ 
are defined as
\begin{equation}
 \Gamma_{\hatl\hatk\hatj} = \frac{1}{2}(c_{\hatl\hatk\hatj} + 
 c_{\hatl\hatj\hatk} - c_{\hatk\hatj\hatl}), 
\end{equation}
with the Cartan coefficients $c_{\hati\hatj}{}^{\hatk} = 
\delta^{\hatk\hatl} c_{\hati\hatj\hatl}$ to be computed 
from the commutator of the vectors of the vielbein field
\begin{equation}
 [\bm{e}_{\hati}, \bm{e}_{\hatj}] = c_{\hati\hatj}{}^{\hatk} \bm{e}_\hatk,
\end{equation}
where the components of the commutator are
$ ([\bm{e}_{\hati}, \bm{e}_{\hatj}])^d = 
 e_\hati^k \partial_k e_\hatj^d - 
 e_\hatj^k \partial_k e_\hati^d$.
We can also invert the above relation to get
\begin{equation}
 c_{\hati\hatj}{}^{\hatk} = \braket{[\bm{e}_{\hati}, \bm{e}_{\hatj}], 
 \bm{\omega}^{\hatk}} = 
 (e_\hati^d e_\hatj^k - e_\hatj^d e_\hati^k) \partial_k \omega_d^{\hatk}, 
\end{equation}
where $\braket{\bm{u}, \bm{w}} = u^\hati w_\hati$ is the 
inner product between a vector field $\bm{u} = u^\hati \bm{e}_\hati$ 
and a one-form $\bm{w} = w_\hati \bm{\omega}^\hati$.

Let us now apply these definitions for the case of a torus. 
The commutator of the vielbein vectors 
$\bm{e}_\htheta$ and $\bm{e}_\hvarphi$ is
\begin{equation}
 [\bm{e}_\htheta, \bm{e}_\hvarphi] = -[\bm{e}_\hvarphi, \bm{e}_\htheta] = 
 \frac{\sin\theta}{R(1 + a\cos\theta)} \bm{e}_\hvarphi.
\end{equation}
Substituting these relations into the definition of the Cartan coefficients,
we find that the only non-vanishing Cartan coefficients are
\begin{equation}
 c_{\htheta\hvarphi}{}^{\hvarphi} = -c_{\hvarphi\htheta}{}^{\hvarphi} = 
 \frac{\sin\theta}{R(1 + a\cos\theta)},
\end{equation} 
and the ensuing connection coefficients read
\begin{equation}
 \Gamma_{\htheta\hvarphi\hvarphi} = -\Gamma_{\hvarphi\htheta\hvarphi} = 
 \frac{\sin\theta}{R(1 + a\cos\theta)}.\label{eq:conn_torus}
\end{equation}

Another important operator is the divergence of a vector field, where
the following relation applies:
\begin{equation}
 \nabla_\hati A^\hati = \frac{1}{\sqrt{g}} 
 \partial_i (\sqrt{g} e_\hati^i A^{\hati}) = 
 \frac{\partial_\varphi A^{\hvarphi}}{R(1 + a \cos\theta)} + 
 \frac{\partial_\theta[A^{\htheta}(1 + a \cos\theta)]}
 {r(1 + a \cos\theta)}.
\end{equation}
For the special case where $A^\hati = \nabla^\hati F$ 
is the gradient of a scalar function, the following relation 
may be employed:
\begin{equation}
 \Delta F = \nabla_\hati \nabla^\hati F = 
 \frac{1}{\sqrt{g}} \partial_i (\sqrt{g} g^{ij} \partial_j F) = 
 \frac{\partial_\varphi^2 A}{R^2 (1 + a \cos\theta)^2} + 
 \frac{\partial_\theta[(1 + a \cos\theta) \partial_\theta F]}
 {r^2(1 + a \cos\theta)}.\label{eq:laplace}
\end{equation}
Finally, the action of the covariant derivative on a 
tensor with two indices can be computed using
\begin{equation}
 \nabla_\hati \mathsf{M}^{\hatj\hatk} = 
 e_\hati^i \partial_i \mathsf{M}^{\hatj\hatk} + 
 \Gamma^{\hatj}{}_{\hatl\hati} \mathsf{M}^{\hatl\hatk} + 
 \Gamma^{\hatk}{}_{\hatl\hati} \mathsf{M}^{\hatj\hatl}.
\end{equation}

\subsection{Equations of motion for axisymmetric flows on the torus geometry}
\label{eq:hydro:torus}

In this paper, we focus on axisymmetric flows, for which all 
fluid quantities are independent of the $\varphi$ angular coordinate. 
In this case, the continuity equation [Eq.~\eqref{eq:hydro_cont}] becomes
\begin{equation}
 \frac{\partial \rho}{\partial t} + 
 \frac{\partial_\theta [\rho u^\htheta(1 + a \cos\theta)]}
 {r(1 + a \cos\theta)} = 0.\label{eq:hydro_cont_azim}
\end{equation}

To derive the Cauchy equation [Eq.~\eqref{eq:hydro_cauchy}],
let us first consider the viscous contributions to the pressure tensor. 
Taking the covariant derivatives in 
Eq.~\eqref{eq:tau_def}, the following expressions are obtained for 
the components of $\tau_{\rm dyn}^{\hati\hatj}$:
\begin{align}
 \tau^{\htheta\htheta}_{\rm dyn} = -\tau_{\rm dyn}^{\hvarphi\hvarphi} =& 
 \frac{\eta}{r} (1 + a \cos\theta) \frac{\partial}{\partial\theta} \left(
 \frac{u^{\htheta}}{1 + a \cos\theta}\right), \nonumber\\
 \tau_{\rm dyn}^{\htheta\hvarphi} = \tau_{\rm dyn}^{\hvarphi\htheta} =& 
 \frac{\eta}{r} (1 + a \cos\theta) \frac{\partial}{\partial\theta} \left(
 \frac{u^{\hvarphi}}{1 + a \cos\theta}\right),
 \label{eq:tau_dyn_azim}
\end{align}
while the volumetric parts are
\begin{equation}
 \tau^{\htheta\htheta}_{\rm bulk} = 
 \tau^{\hvarphi\hvarphi}_{\rm bulk} = \eta_v \nabla_\hatk u^\hatk 
 = \frac{\eta_v}{r(1 + a \cos\theta)} \frac{\partial}{\partial \theta} 
 [u^\htheta (1 +a \cos\theta)],
 \label{eq:tau_bulk_azim}
\end{equation}
with $\tau^{\htheta\hvarphi}_{\rm bulk} = \tau^{\hvarphi\htheta}_{\rm bulk} = 0$.
The divergence of $\tau^{\hati\hatj}$ is then
\begin{align}
 \nabla_{\hat{j}} \tau^{\htheta\hatj} =& 
 \frac{\partial_\theta\{\eta (1 + a\cos\theta)^3 
 \partial_\theta[u^\htheta/(1 + a\cos\theta)]\}}
 {r^2(1 + a \cos\theta)^2} + 
 \frac{1}{r^2} \frac{\partial}{\partial \theta}
 \frac{\eta_v \partial_\theta [u^\htheta(1 + a \cos\theta)]} 
 {1 + a \cos\theta},\nonumber\\
 \nabla_{\hat{j}} \tau^{\hvarphi\hatj} =& 
 \frac{\partial_\theta\{\eta (1 + a\cos\theta)^3 
 \partial_\theta[u^\hvarphi/(1 + a\cos\theta)]\}}
 {r^2(1 + a \cos\theta)^2}.
 \label{eq:cauchy_dtau}
\end{align}
For the non-dissipative contributions to the pressure tensor, 
the divergence $\nabla_\hatj \mathsf{P}^{\hati\hatj}_{\kappa}$ of the term 
involving surface tension can be evaluated using
\begin{equation}
 \nabla_\hatj \mathsf{P}^{\hati\hatj}_{\kappa} = -\phi \kappa \nabla^{\hati} \Delta \phi.
\end{equation}
Thus, the $\hvarphi$ component of the Cauchy equation reads as
\begin{subequations}\label{eq:hydro_cauchy_azim}
\begin{equation}
 \rho \left\{\frac{\partial u^\hvarphi}{\partial t} + 
 u^\htheta \frac{\partial_\theta[u^\hvarphi(1 + a \cos\theta)]}{r(1 + a\cos\theta)} 
 \right\}
 = \frac{\partial_\theta\{\eta (1 + a\cos\theta)^3 
 \partial_\theta[u^\hvarphi/(1 + a\cos\theta)]\}}
 {r^2(1 + a \cos\theta)^2},
 \label{eq:hydro_cauchyph_azim}
\end{equation}
while the $\htheta$ component can be written as:
\begin{multline}
 \rho\left[\frac{\partial u^\htheta}{\partial t} + 
 \frac{u^\htheta}{r} \frac{\partial u^\htheta}{\partial \theta} + 
 \frac{(u^\hvarphi)^2 \sin\theta}{R(1 + a \cos\theta)}\right]
 + \frac{1}{r} \frac{\partial P_{\rm b}}{\partial \theta}
 = \frac{\phi \kappa}{r^3} \frac{\partial}{\partial \theta} 
 \left\{\frac{\partial_\theta[(1 + a \cos\theta)\partial_\theta \phi]}
 {1 + a \cos\theta}\right\} \\
 + \frac{\partial_\theta\{\eta (1 + a\cos\theta)^3 
 \partial_\theta[u^\htheta/(1 + a\cos\theta)]\}}
 {r^2(1 + a \cos\theta)^2} 
 + \frac{1}{r^2} \frac{\partial}{\partial \theta}\left\{
 \eta_v \frac{\partial_\theta [u^\htheta(1 + a \cos\theta)]} 
 {1 + a \cos\theta}\right\}.
 \label{eq:hydro_cauchyth_azim}
\end{multline}
\end{subequations}

To derive the energy equation [Eq.~\eqref{eq:hydro_en}], the following 
contraction is useful:
\begin{align}
 \tau^{\hati\hatj} \nabla_\hati u_\hatj =& \frac{1}{2 \eta} 
 \tau_{\rm dyn}^{\hati\hatj} \tau^{\rm dyn}_{\hati\hatj} 
 + \frac{1}{2\eta_v} \tau_{\rm bulk}^{\hati\hatj} \tau^{\rm bulk}_{\hati\hatj} 
 \nonumber\\
 =& \frac{1}{\eta} \left[(\tau^{\htheta\htheta}_{\rm dyn})^2 + 
 (\tau_{\rm dyn}^{\hvarphi\htheta})^2\right] + 
 \frac{1}{\eta_v} (\tau_{\rm bulk}^{\htheta\htheta})^2,
 \label{eq:hydro_en_taudu}
\end{align}
where the properties $\tau_{\rm dyn}^{\hvarphi\hvarphi} = 
-\tau_{\rm dyn}^{\htheta\htheta}$ 
and $\tau^{\htheta\htheta}_{\rm bulk} = 
 \tau^{\hvarphi\hvarphi}_{\rm bulk}$
have been used.
Thus, the energy equation can be written as
\begin{multline}
 \rho \left(\frac{\partial e}{\partial t} + 
 \frac{u^\htheta}{r} \frac{\partial e}{\partial \theta}\right) + 
 \frac{P_{\rm b}\partial_\theta[u^\htheta(1 + a \cos\theta)]}
 {r(1 + a \cos\theta)} \\
 = \frac{1}{r^2} \frac{\partial_\theta[(1 + a \cos\theta) 
 \text{k} \partial_\theta T]}
 {1 + a\cos\theta} 
 + \frac{1}{\eta} \left[(\tau^{\htheta\htheta}_{\rm dyn})^2 + 
 (\tau_{\rm dyn}^{\hvarphi\htheta})^2\right] + 
 \frac{1}{\eta_v} (\tau_{\rm bulk}^{\htheta\htheta})^2.
 \label{eq:hydro_en_azim}
\end{multline}

Finally, on the torus, the Cahn-Hilliard equation, Eq.~\eqref{eq:CH}, reduces to
\begin{equation}
 \frac{\partial \phi}{\partial t} + 
 \frac{\partial_\theta[\phi u^\htheta(1 + a \cos\theta)]}
 {r(1 + a \cos\theta)} = \frac{M}{r^2} 
 \frac{\partial_\theta[(1 + a\cos\theta)\partial_\theta \mu]}
 {1 + a \cos\theta},
 \label{eq:CH_azim}
\end{equation}
where the chemical potential is computed using
\begin{equation}
 \mu = -\text{A} \phi(1- \phi^2) - \frac{\kappa}{r^2} 
 \frac{\partial_\theta[(1 + a\cos\theta)\partial_\theta \phi]}
 {1 + a \cos\theta}.\label{eq:mu_azim}
\end{equation}

\section{Sound speed for perfect fluids}\label{sec:inv}

The first problem we study in this work is sound wave propagation 
for perfect fluids on the torus geometry. 
In fluids, sound waves provide the basic mechanism of information 
propagation. Many interesting phenomena involving the properties 
of sound wave propagation form the object of focus in acoustics. 
In addition, due to their fundamental importance, sound wave propagation 
should be considered as a first benchmark for any hydrodynamics solver.
For perfect fluids, we neglect dissipative effects, such that the dynamic 
viscosity $\eta$ and the heat conductivity $\text{k}$ can be taken to be zero. For 
simplicity, we will also set the surface tension parameter $\kappa$ and the 
mobility $M$ in the Cahn-Hilliard equation to zero.

Focussing on sound wave propagation along the poloidal ($\theta$) direction
of the torus, we will show that the sound waves exhibit a discrete spectrum of harmonics.
The eigenfrequencies corresponding to these harmonics can be related to
those of the standard Fourier harmonics for periodic domains, but, surprisingly
the eigenfrequencies corresponding to odd and even modes have different 
values, unlike for a planar geometry \citep{rieutord15,busuioc20camwa}. The eigenfunctions describing the 
spatial dependence also generalise from the usual harmonic sine and cosine 
basis to more complex odd and even functions. We
determine the eigenfunctions using a perturbative approach, starting 
with the harmonic functions at zeroth order.

This section is structured as follows. The general solution for the 
propagation of longitudinal waves is presented in Subsec.~\ref{sec:inv:sol}.
Then, two benchmark problems are proposed in Subsecs.~\ref{sec:inv:b1} and 
\ref{sec:inv:b2}.

\subsection{General solution}\label{sec:inv:sol}

Let us consider small perturbations around a stationary, background state at
density $\rho_0$, internal energy $e_0$ and order parameter $\phi_0$, having 
bulk pressure $P_0 \equiv P_{\rm b}(\rho_0, e_0, \phi_0)$:
\begin{equation}
 \rho = \rho_0(1 + \delta \rho), \qquad 
 e = e_0(1 + \delta e), \qquad 
 P_{\rm b} = P_0(1 + \delta P), \qquad 
 \phi = \phi_0 + \delta \phi.
\end{equation}
The perturbations in the pressure $\delta P$ can be expressed as
\begin{equation}
 \delta P = \frac{\rho_0 P_{\rho,0}}{P_0} \delta \rho 
 + \frac{e_0 P_{e,0}}{P_0} \delta e + 
 \frac{P_{\phi,0}}{P_0} \delta \phi,
 \label{eq:deltaP}
\end{equation}
where for brevity the following notation is introduced:
\begin{equation}
 P_\rho = \frac{\partial P_{\rm b}}{\partial \rho}, \qquad 
 P_e = \frac{\partial P_{\rm b}}{\partial e}, \qquad 
 P_\phi = \frac{\partial P_{\rm b}}{\partial \phi}.\label{eq:dP} 
\end{equation}
The subscripts $0$ in Eq.~\eqref{eq:deltaP} indicate that the 
derivatives of the pressure are computed for the background state.

Assuming that the velocity components $u^\htheta$ and 
$u^\hvarphi$ are small, and neglecting all second-order 
terms of the perturbations introduced, the continuity [Eq.~\eqref{eq:hydro_cont_azim}], Cauchy [Eq.~\eqref{eq:hydro_cauchy_azim}], energy [Eq.~\eqref{eq:hydro_en_azim}] and Cahn-Hilliard [Eq.~\eqref{eq:CH_azim}] equations reduce to
\begin{align}
 \frac{\partial \delta \rho}{\partial t} + 
 \frac{\partial_\theta \mathcal{U}}{r(1 + a \cos\theta)} =& 0, \nonumber\\
 \frac{\partial \mathcal{U}}{\partial t} + 
 \frac{P_0(1 + a \cos\theta)}{\rho_0 r} 
 \frac{\partial \delta P}{\partial \theta} =& 0, \nonumber\\
 \frac{\partial \delta e}{\partial t} + 
 \frac{P_0}{\rho_0 e_0} \frac{\partial_\theta \mathcal{U}}
 {r(1 + a\cos\theta)} =& 0,\nonumber\\
 \frac{\partial \delta \phi}{\partial t} + \frac{\phi_0 \partial_\theta \mathcal{U}}{r(1+ a \cos\theta)} =& 0,
 \label{eq:inv_aux}
\end{align}
while $\partial_t u^\hvarphi = 0$. 
Note that, in the above, we introduced the following notation
\begin{equation}
 \mathcal{U} = u^\htheta (1 + a \cos\theta).\
\end{equation}
Taking the time derivative of the second relation in Eq.~\eqref{eq:inv_aux}
and replacing $\delta P$ with Eq.~\eqref{eq:deltaP} gives
\begin{equation}
 \frac{\partial^2 \mathcal{U}}{\partial t^2} - 
 \frac{c_{s,0}^2}{r^2} (1 + a \cos\theta) \frac{\partial}{\partial \theta}
 \left(\frac{1}{1 + a \cos\theta}\frac{\partial \mathcal{U}}{\partial \theta}\right) = 0.
 \label{eq:inv_soundeq}
\end{equation}
Eq.~\eqref{eq:inv_soundeq} represents the generalisation 
of the sound wave equation for axisymmetric flows on the torus 
geometry. 
We can recognise $c_{s,0}$ as the sound speed
corresponding to the background fluid parameters. In general,
$c_s^2$ can be computed using
\begin{equation}
 c_s^2 = P_\rho + 
 \frac{P_{\rm b}}{\rho^2} P_e + 
 \frac{\phi}{\rho} P_\phi.
 \label{eq:inv_cs2}
\end{equation}
For the  ideal gas, $P_{\rm b} = \rho k_B T / m$ and 
$c_s = \sqrt{\gamma P_{\rm b} / \rho}$, where 
$\gamma = 1 + k_B / m c_v$ is the adiabatic index 
(e.g. $\gamma = 2$ for a monoatomic ideal gas with $2$ translational 
degrees of freedom). The isothermal regime 
can be recovered by setting $c_v \rightarrow \infty$ and $\gamma \rightarrow 1$. 
For the isothermal ideal fluid, we recover
$c_s = \sqrt{k_B T / m}$.

Eq.~\eqref{eq:inv_soundeq} can be solved using the method of 
separation of variables with the following ansatz
\begin{equation}
 \mathcal{U}(t, \theta) \rightarrow 
 \mathcal{U}_n(t, \theta) = U_n(t) \Psi_n(\theta).
 \label{eq:inv_separation}
\end{equation}
The index $n$ reflects the fact that there are more than one possible solutions,
corresponding to a discrete set of eigenvalues $\lambda_n$.
The temporal function corresponds to simple harmonic oscillations of the form
\begin{equation}
 \ddot{U}_n = - \lambda_n^2 \frac{c_{s,0}^2}{r^2}  U_n.
 \label{eq:inv_separation_Ueq}
\end{equation}

The angular functions satisfy the differential equation
\begin{equation}
 (1 + a \cos\theta) \frac{d}{d\theta}\left(
 \frac{1}{1 + a \cos\theta} \frac{d\Psi_n}{d\theta}\right)
 + \lambda_n^2 \Psi_n = 0.\label{eq:inv_modes_eq}
\end{equation}
The functions $\Psi_n$ are twice differentiable periodic solutions with a discrete set of eigenvalues
$\lambda_n$.
Eq.~\eqref{eq:inv_modes_eq} has even and odd solutions, which we denote by $f_n(\theta)$ and $g_n(\theta)$.
It can be shown that these functions are orthogonal with respect to 
the inner product, which is defined below for 
two functions $\psi(\theta)$ and $\chi(\theta)$:
\begin{equation}
 \braket{\psi, \chi} = \int_{0}^{2\pi} \frac{d\theta}{2\pi} 
 \frac{\psi(\theta) \chi(\theta)}{1 + a \cos\theta}.
 \label{eq:inv_scprod}
\end{equation}
We seek solutions of unit norm, such that
\begin{equation}
 \braket{f_n, f_{n'}} = \delta_{n,n'}, \qquad 
 \braket{g_n, g_{n'}} = \delta_{n,n'}, \qquad 
 \braket{f_n, g_{n'}} = 0.\label{eq:inv_ortho}
\end{equation}

The zeroth mode solution, corresponding to $n = 0$ and $\lambda_0 = 0$, is straightforward to identify.
The solution is a constant. Exploiting the condition of unit norm, we can use the following integral
\begin{equation}
 \frac{1}{2\pi} \int_{0}^{2\pi} \frac{d\theta}{1 + a \cos\theta} = 
 \frac{1}{\sqrt{1 - a^2}}
\end{equation}
to obtain that 
\begin{equation}
 f_0(\theta) = (1 - a^2)^{1/4}. \label{eq:tor_f0}
\end{equation}
There is no antisymmetric solution corresponding to $n = 0$ and $\lambda_0 = 0$. 

We will now discuss the subsequent values of $\lambda_{c;n}$ and $\lambda_{s;n}$, 
the eigenvalues of the even ($f_n$) and odd ($g_n$) solutions. 
 More specifically, the pairs $(f_n, \lambda_{c;n})$ 
and $(g_n, \lambda_{s;n})$ satisfy Eq.~\eqref{eq:inv_modes_eq}:
\begin{align}
 (1 + a \cos\theta) \frac{d}{d\theta}\left(
 \frac{1}{1 + a \cos\theta} \frac{df_n}{d\theta}\right)
 + \lambda_{c;n}^2 f_n =& 0, & f_n(\theta) =& f_n(2\pi - \theta), \nonumber\\
 (1 + a \cos\theta) \frac{d}{d\theta}\left(
 \frac{1}{1 + a \cos\theta} \frac{dg_n}{d\theta}\right)
 + \lambda_{s;n}^2 g_n =& 0, & g_n(\theta) =& -g_n(2\pi - \theta).\label{eq:lambdacs_def}
\end{align}

We index the solutions incrementally
such that $f_{n+1}$ has an eigenvalue $\lambda_{c;n+1} > \lambda_{c;n}$, and similarly
for the odd solutions. 

Eq.~\eqref{eq:inv_modes_eq} can be solved analytically in the limit 
case $a = 0$ (corresponding to an infinitely wide torus, $R \rightarrow \infty$).
In this case, when $n > 0$, Eq.~\eqref{eq:inv_modes_eq} yields
the usual (normalised) harmonic basis encountered on a system with 
periodic coordinate $\theta$:
\begin{equation}
 f_n = \sqrt{2}\cos n\theta, \qquad 
 g_n = \sqrt{2}\sin n\theta.\label{eq:inv_modes_a0}
\end{equation}
Here, $\lambda_{c;n} = \lambda_{s;n} = n$. For $n = 0$, Eq.~\eqref{eq:tor_f0} 
reduces to $f_0(\theta) = 1$.

Another limit where the analytical solution is available is when $a = 1$.
In this case, the eigenfrequency spectrum is derived in 
Eqs.~\eqref{SM:eq:a1:inv_lambdac} and \eqref{SM:eq:a1:inv_lambdas} 
and is reproduced below, for convenience
\begin{equation}
 \lambda_{c;n}^2 = n^2 - \frac{1}{4},\quad \lambda_{s;n}^2 = n(n+1).
 \label{eq:inv_lambda_a1}
\end{equation}
The derivation and explicit form of the eigenfunctions for $a=1$ are given in Sec. \ref{SM:sec:modes:a1:inv} of the supplementary material.

For intermediate values of $a$ (i.e. for $0 < a < 1$) 
and $\lambda_n^2 > 0$, there is no known analytic 
solution of Eq.~\eqref{eq:inv_modes_eq}. However, given that 
$a < 1$, it is reasonable to seek for the solutions in 
a perturbative manner. 
Starting from the $a = 0$ solution in 
Eq.~\eqref{eq:inv_modes_a0}, for a given value of $n$,
we expect that the perturbation procedure will bring in 
harmonics corresponding to $n \pm 1$, $n \pm 2$, and so forth.
The eigenvalues $\lambda_{c;n}$ and $\lambda_{s;n}$ 
travel along a continuous path from $\lambda_{c;n} = n$ to 
$ \lambda_{c;n} = \sqrt{n^2 - \frac{1}{4}}$, and 
from $\lambda_{s;n} = n$ to $\lambda_{s;n} = \sqrt{n(n+1)}$, respectively, as $a$ goes from $0$ to $1$.
The perturbative procedure is discussed in
Appendix~\ref{app:modes} and the results for $1 \le n \le 4$ 
are given up to $O(a^9)$ in Eq.~\eqref{SM:eq:inv_modes} of 
the supplementary material.

In general, the eigenvalues $\lambda_{c;n}$ and $\lambda_{s;n}$
for the even and odd modes of the same order $n$ are not equal.
As discussed in Appendix~\ref{app:modes}, the difference 
between $\lambda_{c;n}$ and $\lambda_{s;n}$ appears via terms 
of order $O(a^{2n})$.
table~\ref{tab:inv_lambda} shows the values of $\lambda_{c;n}$ 
and $\lambda_{s;n}$ obtained using high precision numerical integration 
for the cases $a = 0.4$ and $a = 0.8$. It can be seen that the 
difference between $\lambda_{c;n}$ and $\lambda_{s;n}$ decreases 
as $n$ is increased and $a$ is kept fixed, or as $n$ is kept fixed 
and $a$ is decreased. This is in contrast to the flat geometry, 
where the eigenvalues for the even and odd modes of the same order 
$n$ are always identical.

\begin{figure}
\begin{center}
\begin{tabular}{ccc}
 \includegraphics[width=0.32\linewidth]{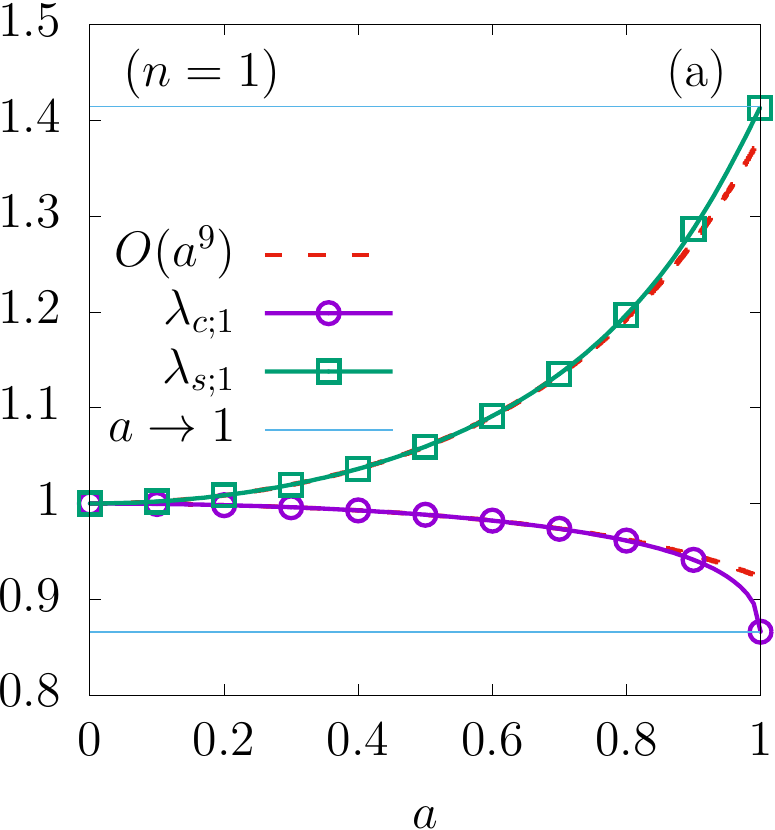} &
 \includegraphics[width=0.32\linewidth]{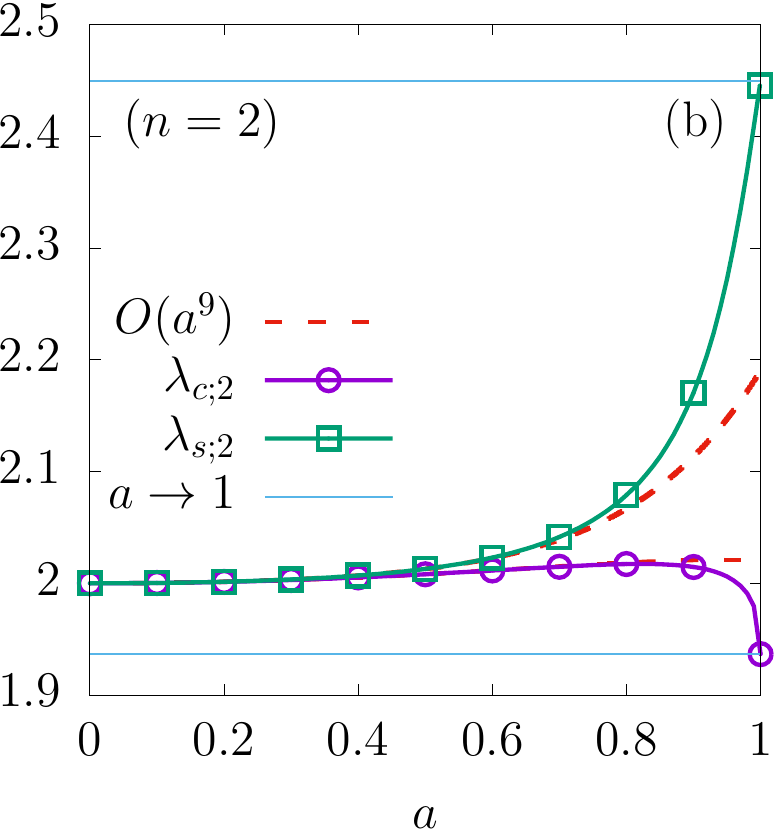} &
 \includegraphics[width=0.32\linewidth]{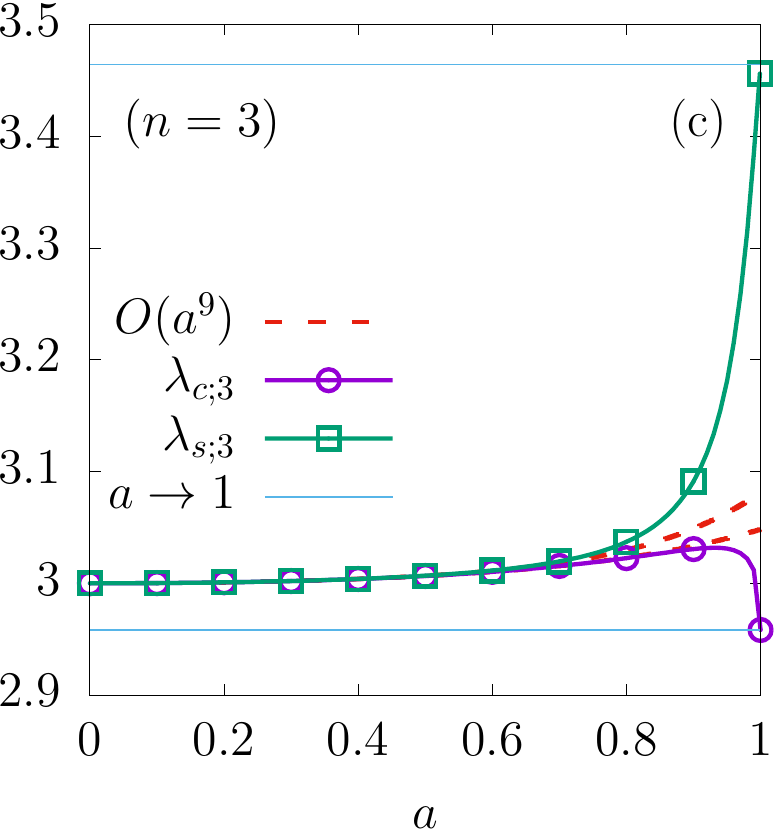} 
\end{tabular}
\end{center}
\caption{
The dependence of $\lambda_{c;n}$ and $\lambda_{s;n}$ 
on $a$ for $n = 1$ (a), $2$ (b) and $3$ (c), respectively. 
The solid lines with symbols represent the numerically evaluated values 
of the eigenfrequencies, 
while the dotted lines show the perturbative approximations with terms 
up to $O(a^9)$. The horizontal lines show the $a = 1$ limits given in 
Eq.~\eqref{eq:inv_lambda_a1}.
\label{fig:inv_lambda}}
\end{figure}

\begin{figure}
\begin{center}
\begin{tabular}{cc}
 \includegraphics[width=0.46\linewidth]{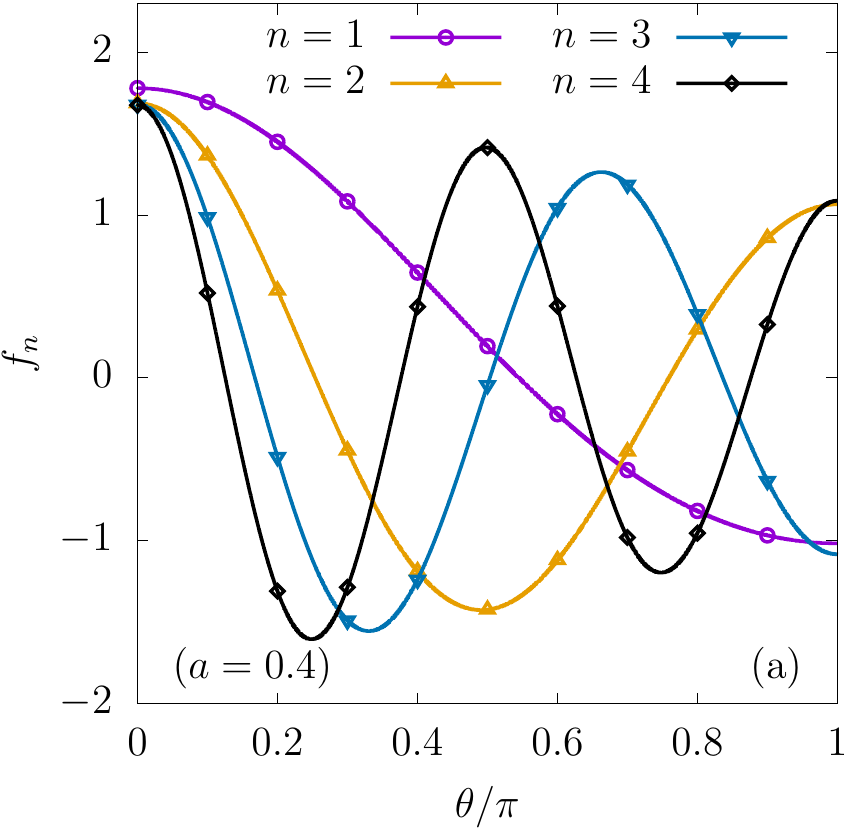} &
 \includegraphics[width=0.46\linewidth]{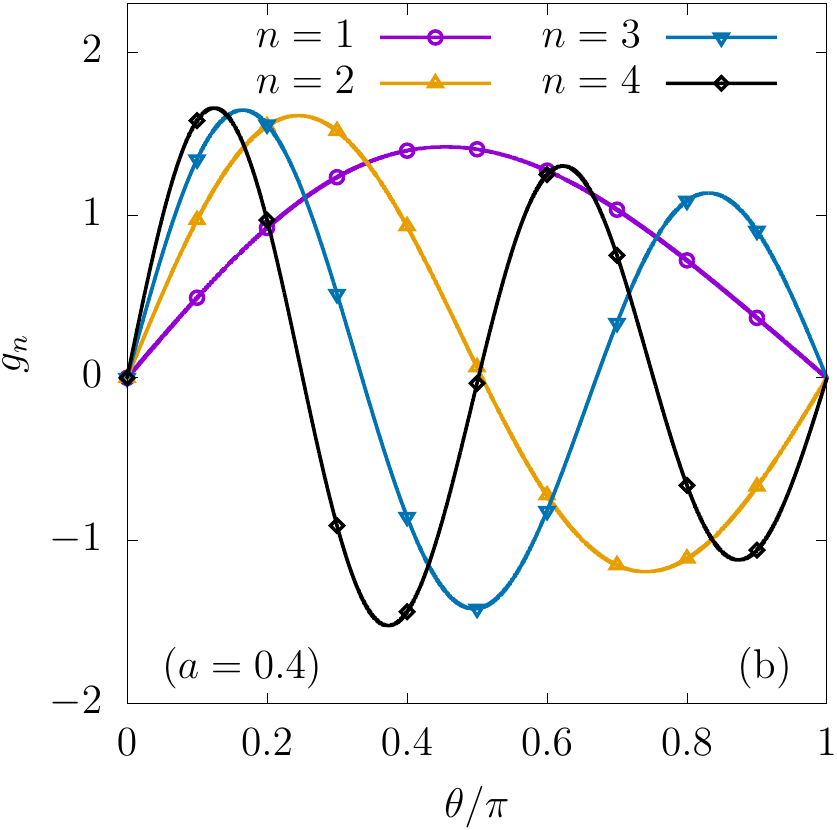} \\
 \includegraphics[width=0.46\linewidth]{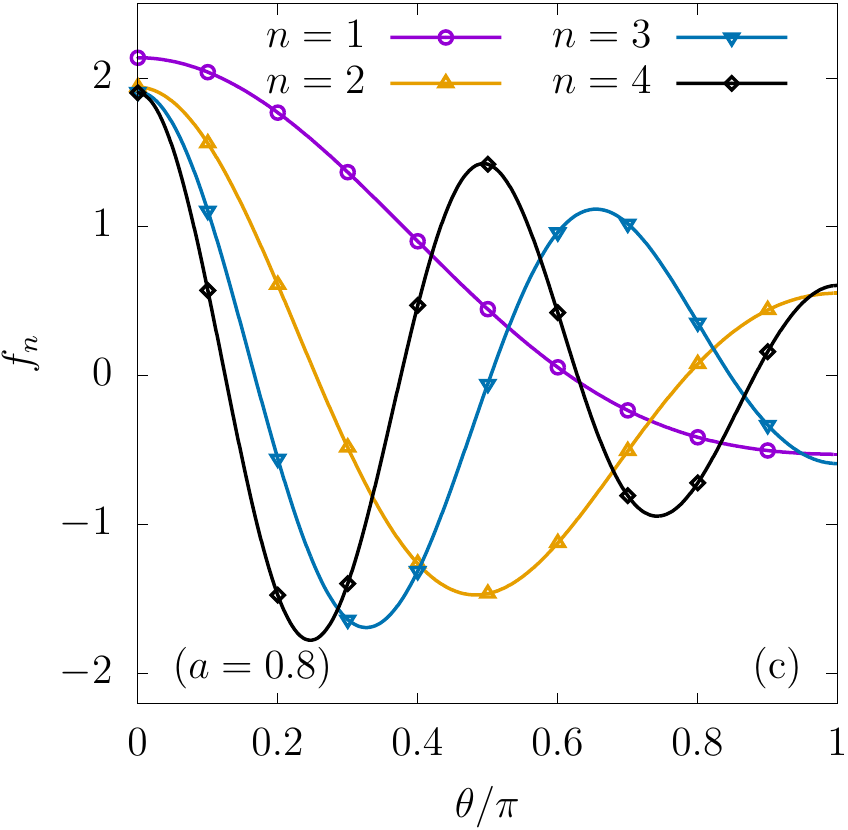} &
 \includegraphics[width=0.46\linewidth]{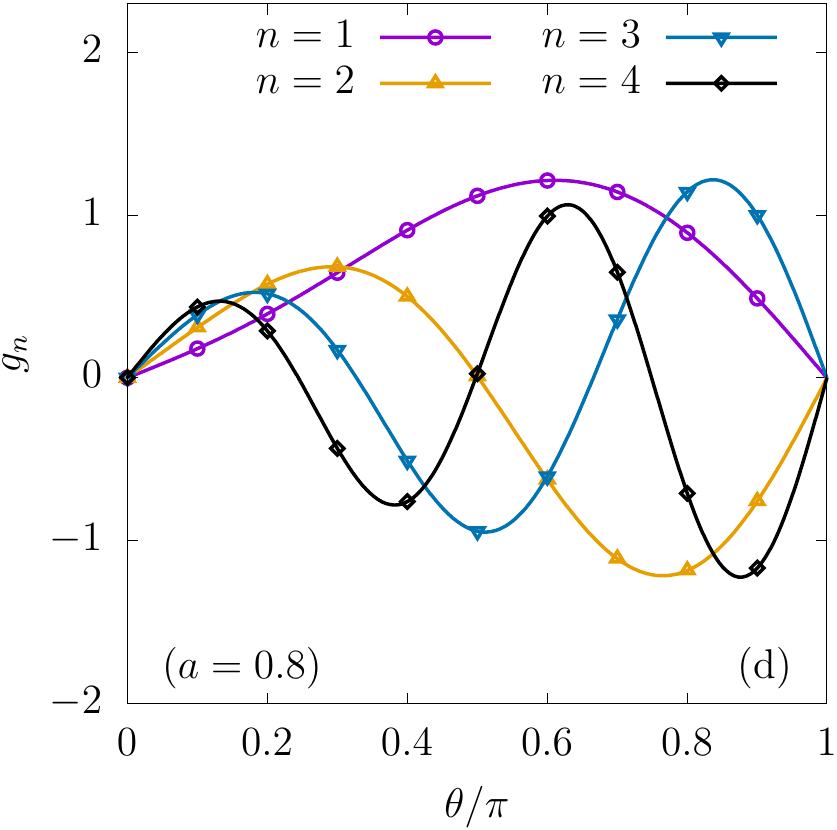}
\end{tabular}
\end{center}
\caption{The even and odd eigenfunctions $f_n$ (left) and $g_n$ 
(right) of Eq.~\eqref{eq:inv_modes_eq}, with $a = 0.4$ (top) and 
$a = 0.8$ (bottom), for $n = 1$, $2$, $3$ and $4$.
The eigenvalues are summarised in table~\ref{tab:inv_lambda}.
\label{fig:inv_funcs}}
\end{figure}

The dependence of $\lambda_{c;n}$ and $\lambda_{s;n}$ on $a$ is revealed in 
figures~\ref{fig:inv_lambda}(a)-\ref{fig:inv_lambda}(c) for $n = 1$, $2$ and $3$. 
It can be seen that, as $a \rightarrow 1$, $\lambda_{c;n}$ 
also has a strong variation with $a$. 
However, overall  the variation of $\lambda_{c;n}$ with $a$ is significantly milder than that of $\lambda_{s;n}$.
For comparison, the dotted lines corresponding to the perturbative approximations up to $O(a^9)$, and 
the limits $\lim_{a\rightarrow 1} \lambda_{c;n} = \sqrt{n^2 - \frac{1}{4}}$ 
and $\lim_{a\rightarrow 1} \lambda_{s;n} = \sqrt{n(n+1)}$ are also shown.

figures~\ref{fig:inv_funcs}(a) and \ref{fig:inv_funcs}(b) show the 
even and odd eigenfunctions $f_n$ and $g_n$ corresponding to $1 \le n \le 4$ 
over the half-domain $0 \le \theta \le \pi$ with $a = 0.4$.
Similarly, figures~\ref{fig:inv_funcs}(c) and \ref{fig:inv_funcs}(d) 
show $f_n$ and $g_n$ when $a = 0.8$. It can be seen that 
the amplitudes for the even harmonics $f_n$ become weaker towards $ \theta = \pi$ as $a$ 
is increased, while the amplitudes of the odd harmonics $g_n$ become weaker towards $\theta=0$.

\begin{table}
\begin{center}
\begin{tabular}{l|rr|rr}
 & \multicolumn{2}{c|}{$a = 0.4$} & \multicolumn{2}{c}{$a = 0.8$} \\
 $n$ & $\lambda_{c;n}$ & $\lambda_{s;n}$ & $\lambda_{c;n}$ & $\lambda_{s;n}$ \\\hline
 $1$ & $0.99283837$ & $1.03615819$ & $0.96123389$ & $1.19709137$ \\
 $2$ & $2.00528264$ & $2.00700233$ & $2.01720533$ & $2.07891859$ \\
 $3$ & $3.00388532$ & $3.00395489$ & $3.02259288$ & $3.03709989$ \\
 $4$ & $4.00289664$ & $4.00289952$ & $4.01992604$ & $4.02335307$ \\
 $5$ & $5.00230332$ & $5.00230344$ & $5.01664927$ & $5.01747046$ \\
 $6$ & $6.00191275$ & $6.00191276$ & $6.01401146$ & $6.01421028$ \\
 $7$ & $7.00163605$ & $7.00163605$ & $7.01201841$ & $7.01206689$ \\
 $8$ & $8.00142960$ & $8.00142960$ & $8.01050233$ & $8.01051420$ \\
 $9$ & $9.00126957$ & $9.00126957$ & $9.00932177$ & $9.00932469$ \\
$10$ &$10.00114185$ & $10.00114185$ & $10.00837943$ & $10.00838015$ \\\hline
\end{tabular}
\end{center}
\caption{Eigenvalues $\lambda_{c;n}$ and $\lambda_{s;n}$ corresponding 
to the even ($f_n$) and odd ($g_n$) solutions of Eq.~\eqref{eq:inv_modes_eq} 
with $a = 0.4$ (left) and $a = 0.8$ (right), 
for $0 < n \le 10$. The eigenvalue $\lambda_{c;0} = 0$, corresponding to 
\eqref{eq:tor_f0}, is not shown here.\label{tab:inv_lambda}}
\end{table}

Assuming that the functions $\{f_n, g_n\}$ form a complete set, 
the fluid velocity can in general be written as
\begin{equation}
 u^\htheta(t,\theta) = \frac{1}{1 + a \cos\theta} \sum_{n = 0}^\infty 
 \left[U_{c;n}(t) f_n(\theta) + U_{s;n}(t) g_n(\theta)\right].
 \label{eq:inv_utheta_series}
\end{equation}
Such an expansion is consistent when the inner product, 
Eq.~\eqref{eq:inv_scprod}, is dual to the following completeness 
relation:
\begin{equation}
 \sum_{n = 0}^\infty [f_n(\theta) f_n(\theta') + g_n(\theta) g_n(\theta')] = 
 2\pi (1 + a \cos\theta) \delta(\theta - \theta').
 \label{eq:inv_completeness}
\end{equation}
Solving Eq.~\eqref{eq:inv_separation_Ueq},
it can be seen that the even and odd solutions for the temporal function 
(for $n > 0$)
correspond to simple harmonic oscillations
\begin{equation}
 U_{c;n}(t) = U_{c;n;0} \cos\left(
 \omega_{c;n} t + \vartheta_{c;n}\right), \qquad 
 U_{s;n}(t) = U_{s;n;0} \sin\left(
 \omega_{s;n} t + \vartheta_{s;n}\right),
 \label{eq:inv_Usol}
\end{equation}
where $\omega_{c;n} = \lambda_{c;n} c_s / r$ and 
$\omega_{s;n} = \lambda_{s;n} c_s / r$.
The coefficients $U_{c;n;0}$ and $U_{s;n;0}$ and the 
phases $\vartheta_{c;n}$ and $\vartheta_{s;n}$
can be determined from the initial conditions
\begin{equation}
 u^\htheta(0,\theta) = u^\htheta_0(\theta), \qquad
 \dot{u}^\htheta(0, \theta) = -\frac{P_0}{\rho_0 r} 
 \partial_\theta \delta P_0(\theta),\label{eq:inv_init}
\end{equation}
where $u^\htheta_0(\theta)$ represents the initial velocity profile,
while $\delta P_0(\theta) = \delta P(0,\theta) = (P_{\mathrm{b}}(0,\theta) - P_0)/P_0$ 
represents the initial pressure fluctuations.
Projecting the above equations onto $f_n$ and $g_n$ yields
\begin{align}
 \begin{pmatrix}
  U_{c;n;0} \cos \vartheta_{c;n} \\
  U_{s;n;0} \sin \vartheta_{s;n} 
 \end{pmatrix} =& \int_{0}^{2\pi} \frac{d\theta}{2\pi} 
 \begin{pmatrix}
  f_n \\ g_n
 \end{pmatrix} u^\htheta_0(\theta), \nonumber\\
 \begin{pmatrix}
  U_{c;n;0} \sin \vartheta_{c;n} \\
  -U_{s;n;0} \cos \vartheta_{s;n} 
 \end{pmatrix} =& \frac{P_0}{\rho_0 c_s} \int_{0}^{2\pi} \frac{d\theta}{2\pi} 
 \begin{pmatrix}
  f_n / \lambda_{c;n} \\ g_n / \lambda_{s;n}
 \end{pmatrix} \frac{\partial \delta P_0}{\partial \theta}, \label{eq:project}
\end{align}
where the last equation applies only for $n > 0$. It is worth noting that the 
$n = 0$ term, corresponding to the incompressible flow profile
\begin{equation}
 \frac{U_{c;0} f_0(\theta)}{1 + a \cos\theta},\label{eq:tor_inv_vinc}
\end{equation}
is time-independent and its amplitude, $U_{c;0}$, is 
preserved at all times.
Thus, numerical methods developed for hydrodynamics 
on curved surfaces should ensure the preservation of the 
above profile.
In the Cartesian geometry, the incompressible flow profile along 
a single axis is a constant background velocity, which should 
be preserved due to the Galilean invariance of the theory.

For the rest of this work, we employ expansions 
of up to $a^9$ of the eigenfunctions, eigenvalues and all related quantities.
These expansions are given in Eq.~\eqref{SM:eq:inv_modes} 
of the supplementary material.
Although some expansions converge faster than the others, 
for consistency reasons, we choose to employ 
the same order of expansion for all quantities involved.

\subsection{First benchmark: Constant initial flow}\label{sec:inv:b1}

We now formulate a simple numerical experiment that can be 
used to benchmark the capabilities of numerical methods to 
capture sound wave propagation on curved geometries. 
The simplest configuration giving rise to sound wave propagation
corresponds to
\begin{equation}
 u^\htheta_0(\theta) = U_0, \qquad \delta P_0(\theta) = 0,
 \label{eq:inv_vinit}
\end{equation}
with $U_0$ a constant. Since the initial velocity profile 
is symmetric and the initial pressure is constant, 
$\vartheta_{c;n} = 0$ and $U_{s;n;0} = 0$. 
To calculate the coefficients of the even modes, we take advantage of the 
projections introduced in Eq. \eqref{eq:project}. 
The fluid velocity can then be written as
\begin{align}
 u^{\htheta}(t, \theta) =& \frac{1}{1 + a \cos\theta} 
 \left[U_0 \sqrt{1 - a^2} +
 \sum_{n = 1}^\infty U_{c;n}(t) f_n(\theta) \right],
 \nonumber\\
 U_{c;n}(t) =& U_0 I_{c;0;n} 
 \cos\left(\frac{c_{s,0} \lambda_{c;n}}{r}t\right),
 \label{eq:inv_sol}
\end{align}
where the eigenvalues $\lambda_{c;n}$ are given up to 
$9$th order with respect to $a$ in Eq.~\eqref{SM:eq:inv_modes}, 
and the integrals $I_{*;m; n}$ are defined as
\begin{equation}
 I_{c; m; n} = \int_0^{2\pi} \frac{d\theta}{2\pi} 
 \frac{f_n(\theta)}{(1 + a \cos\theta)^m}, \qquad
 I_{s; m; n} = \int_0^{2\pi} \frac{d\theta}{2\pi} 
 \frac{g_n(\theta) \sin\theta}{(1 + a \cos\theta)^m}.
 \label{eq:inv_In}
\end{equation}
In this section, we only need the case with $m = 0$,
for which $I_{c;0;0} = (1-a^2)^{1/4}$, 
while the first integrals ($1\leq n\leq4$) are given up to 
$9$th order with respect to $a$ in Eq.~\eqref{SM:eq:inv_modes}
of the supplementary material.
The integrals $I_{s;0;n}$ of the odd functions
will be employed later, in Subsec.~\ref{sec:inv:b2}.

In order to perform numerical simulations, we consider 
a non-dimensionalisation of physical quantities with respect to the 
background fluid parameters, such that 
$\rho_0 = T_0 = P_0 = 1$. Focussing on the torus with $a = r/ R = 0.4$,
we take the reference length scale such that $R = 2$. Setting 
the reference velocity naturally to $c_0 = \sqrt{P_0 / \rho_0}$,
we initialise the velocity by setting $u^\htheta_0(\theta) = U_0 = 10^{-5}$
in Eq.~\eqref{eq:inv_vinit}. Using the aforementioned reference velocity,
the non-dimensional sound speed is $c_{s,0} = 1$ for the isothermal case
and $c_{s,0} = \sqrt{2}$ for the thermal case when the adiabatic index 
is $\gamma = 2$. In addition, we also
consider an isothermal multicomponent fluid 
for which the sound speed is given by
\begin{equation}
 c_{s,0}^2 = \frac{k_B T_0}{m} - 
 \frac{\text{A} \phi_0^2}{\rho_0}(1 - 3 \phi_0^2).
\end{equation}
We choose $\text{A} = 1$; and consider values of $\phi_0 = 1$ and $\phi_0 = 0.8$,  
which are outside the spinodal region, $-\frac{1}{\sqrt{3}} < \phi_0 < \frac{1}{\sqrt{3}}$.
The resulting sound speeds are summarised in table~\ref{tab:inv}.

\begin{figure}
\begin{center}
\begin{tabular}{ccc}
 \includegraphics[width=0.3275\linewidth]{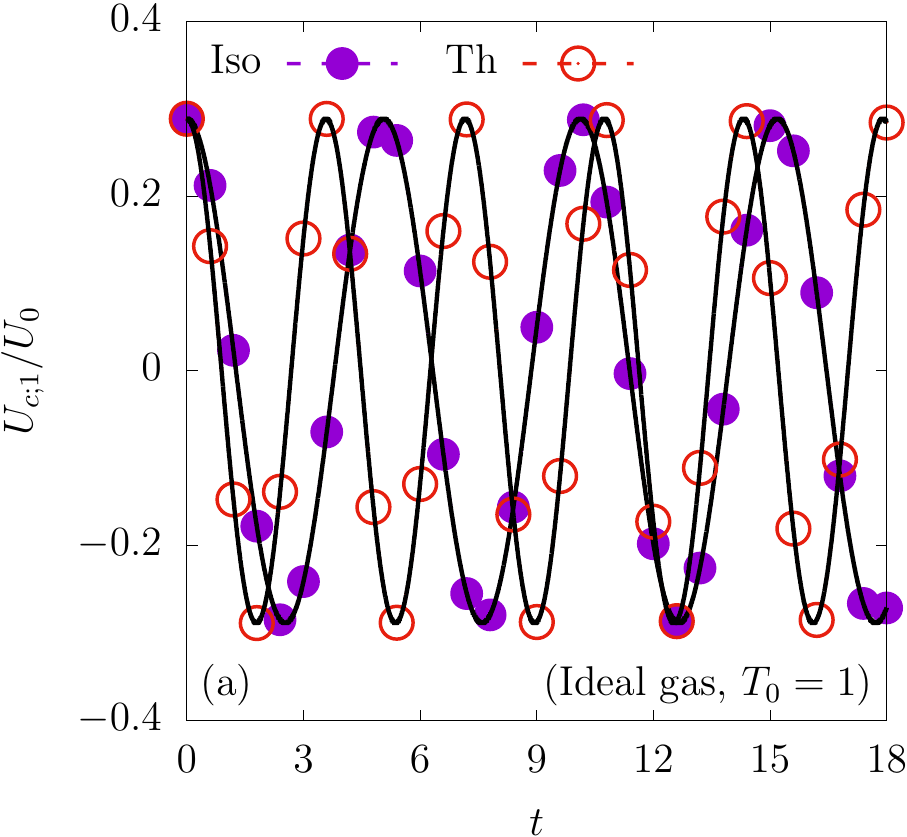} &
 \includegraphics[width=0.31\linewidth]{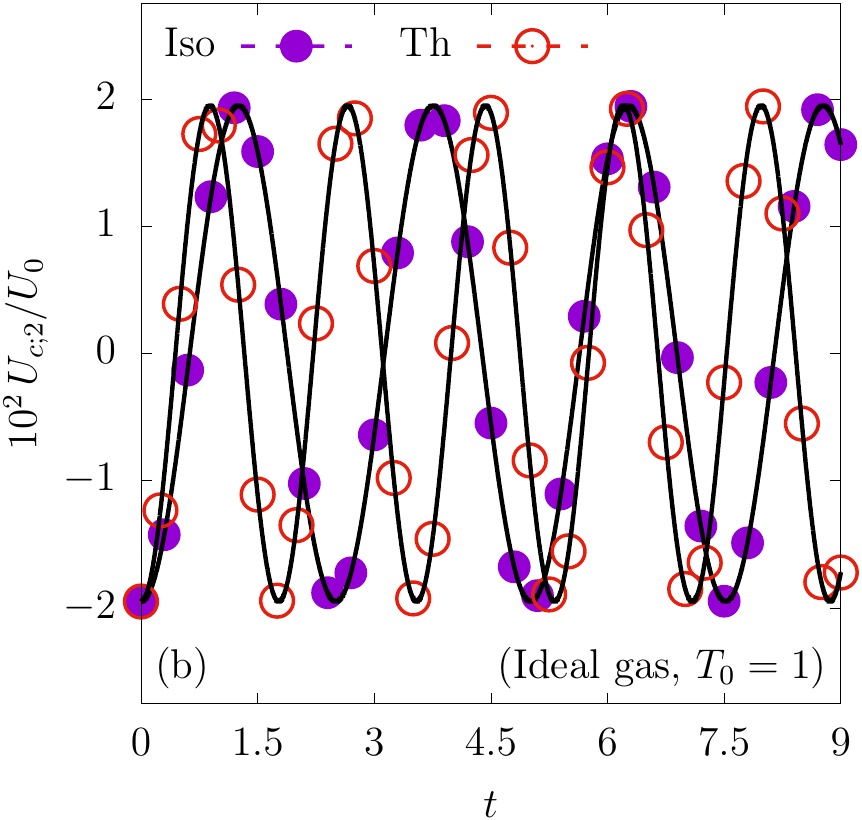} &
 \includegraphics[width=0.31\linewidth]{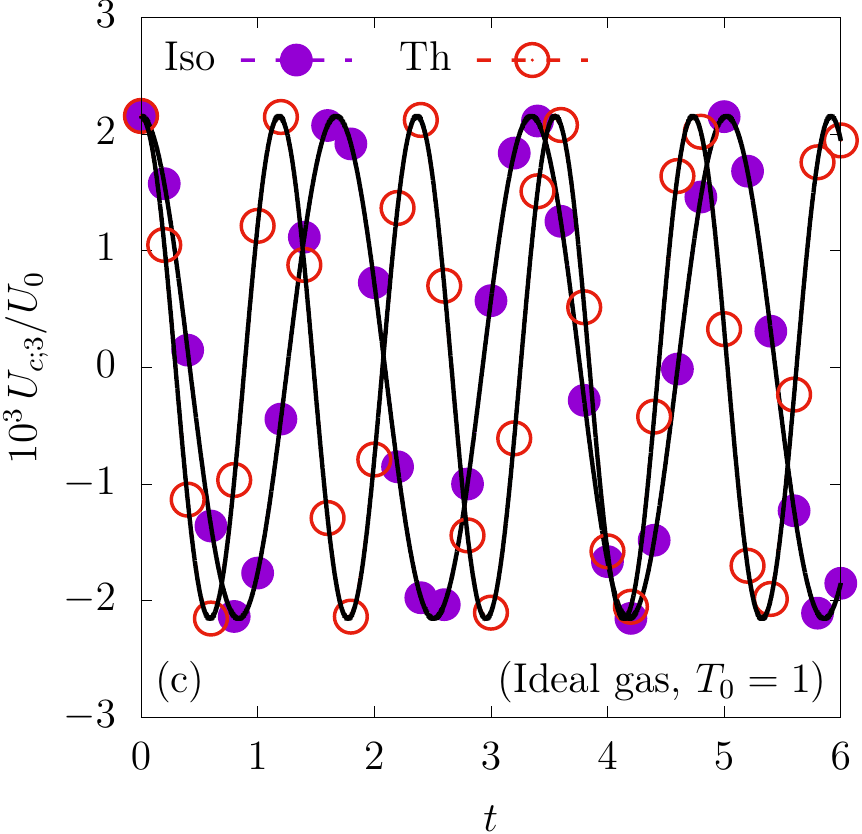} \\
 \includegraphics[width=0.3275\linewidth]{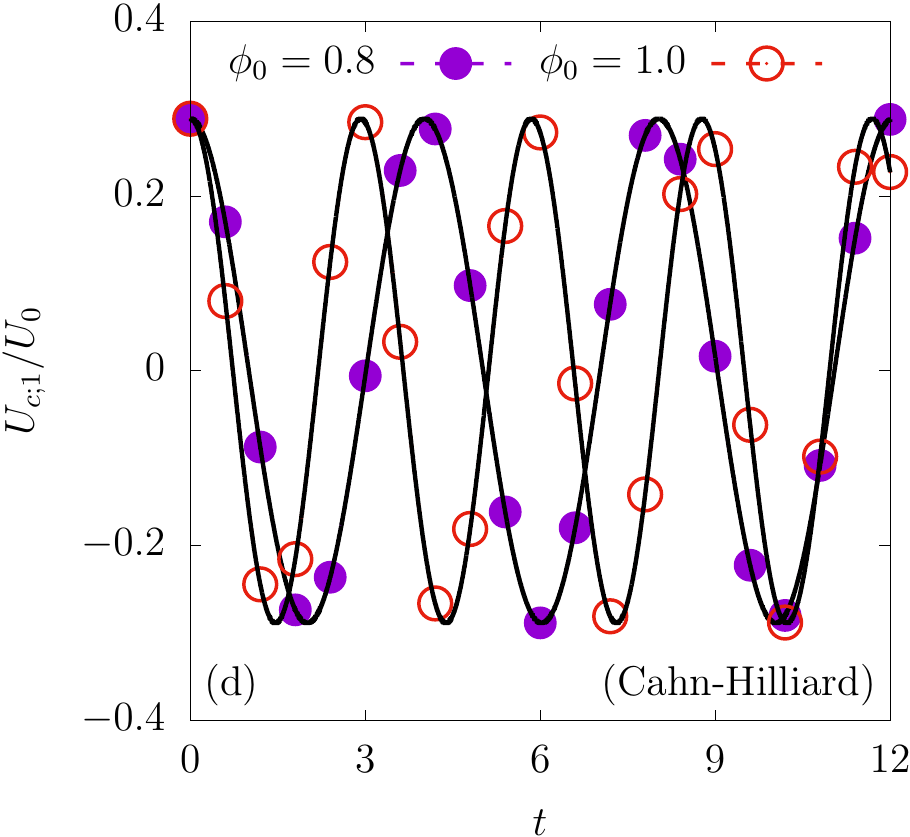} &
 \includegraphics[width=0.31\linewidth]{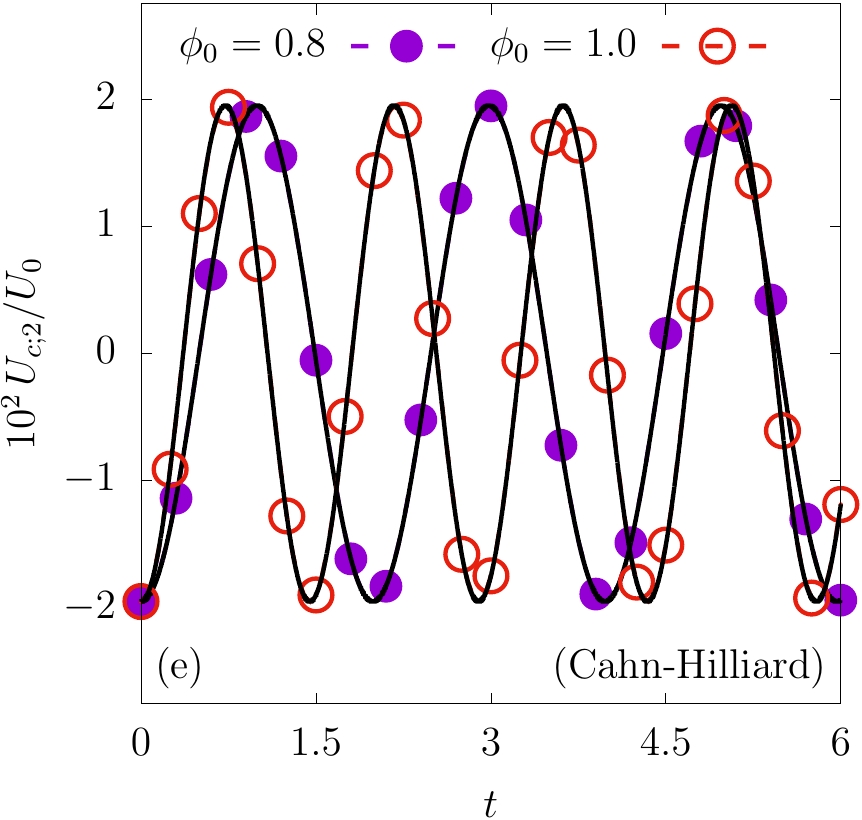} &
 \includegraphics[width=0.31\linewidth]{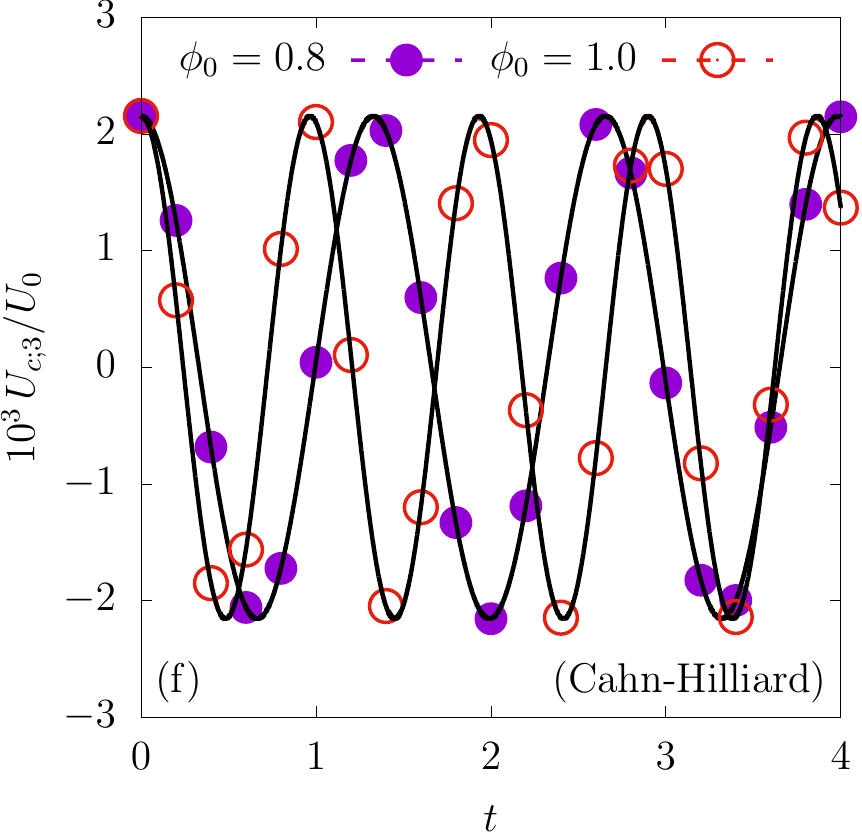}
\end{tabular}
\end{center}
\caption{Comparison between the numerical results (symbols) and 
analytical predictions (solid lines) for the evolution of $U_{c;n}(t)/U_0$,
as given in Eq.~\eqref{eq:inv_sol}.
The first row (a-c) is for isothermal (Iso) and thermal (Th) ideal fluids, while
the second row (d-f) is for Cahn-Hilliard multicomponent fluid.
The integrals $I_{c;0;n}$ given in Eq.~\eqref{SM:eq:inv_modes} have the 
values of $I_{c;0;1} \simeq 0.288$ (left); $I_{c;0;2} \simeq -0.0195$ (middle); 
and $I_{c;0;3} \simeq 0.00216$ (right).
\label{fig:inv}}
\end{figure}

For the four cases above with differing sound speeds, 
the system is evolved between $0 \le t \le 18$ on a grid with $N_\theta = 320$ equidistant 
nodes and a time step $\delta t = 5 \times 10^{-4}$. The velocity profile is 
projected onto the basis functions $f_1$, $f_2$ and 
$f_3$, as given in Eqs.~\eqref{SM:eq:inv_modes_1}, \eqref{SM:eq:inv_modes_2} 
and \eqref{SM:eq:inv_modes_3} of the supplementary material, 
respectively. The simulation results 
are shown using dashed lines and symbols in figure~\ref{fig:inv}. For comparison,
the corresponding analytical solutions in Eq.~\eqref{eq:inv_sol}
are shown in solid lines in figure~\ref{fig:inv}. 
The angular frequencies, $\omega_n = c_{s,0} \lambda_{c;n} / r$, for the 
first three harmonics are reported for convenience in table~\ref{tab:inv}.
The agreement between the analytical and numerical results is excellent. 
It is also worth noting that the angular frequencies on the torus differ from 
those for the flat geometry, and the deviations become more significant with increasing $a$.

\begin{table}
\begin{center}
\begin{tabular}{c|c|rrrr}
 Fluid type & Regime & $c_{s,0}$ & $\omega_1$ & $\omega_2$ & $\omega_3$ \\\hline
 Ideal gas & Iso & $1$ & $1.24104796$ & $2.50660330$ & $3.75485665$ \\
   & Th & $\sqrt{2}$ & $1.75510686$ & $3.54487238$ & $5.31016920$ \\\hline
 Cahn-Hilliard & $\phi_0 = 0.8$ & $1.26047610$ & $1.56431130$ & $3.15951355$ & $4.73290707$ \\
  multicomponent & $\phi_0 = 1.0$ & $1.73205081$ & $2.14955813$ & $4.34156426$ & $6.50360249$
\end{tabular}
\caption{Sound speed and angular frequencies 
$\omega_n = c_{s,0} \lambda_{c;n} / r$
for the first three harmonics of the oscillatory 
motion on the torus with $a = 0.4$, 
considered in figure~\ref{fig:inv}.\label{tab:inv}}
\end{center}
\end{table}

\subsection{Second benchmark test: Even and odd initial conditions} 
\label{sec:inv:b2}

\begin{table}
\begin{center}
\begin{tabular}{lrr}
 $n$ & $U^{\rm even}_{c;n;0} / U_0$ & $U^{\rm odd}_{s;n;0} / U_0$ \\\hline
 $1$ & $0.67162788$ & $0.55445076$ \\
 $2$ & $0.21755576$ & $-0.09426691$ \\
 $3$ & $-0.03806432$ & $-0.01672761$ \\
 $4$ & $0.01098141$ & $-0.00461361$ \\\hline
\end{tabular}
\end{center}
\caption{Values of the normalised amplitudes $U^{\rm even}_{c;n;0}/U_0$ and 
$U^{\rm odd}_{s;n;0}/U_0$ defined in Eq.~\eqref{eq:inv_b2_ints} 
for $a = 0.8$ and $1 \le n \le 4$.
\label{tab:inv_b2}}
\end{table}

The purpose of the second test is to highlight the difference in the 
period corresponding to the propagation of even and odd perturbations. 
As highlighted in figure~\ref{fig:inv_lambda}, the difference in the 
frequencies for the even and odd modes increases as $a$ is increased.
For this reason, in this example we consider $a = 0.8$. According to 
table~\ref{tab:inv_lambda}, the ratio $\lambda_{s;1} / \lambda_{c;1} 
\simeq 1.25$, therefore the $n = 1$ odd mode should exhibit $5$ periods 
for every $4$ periods of the $n = 1$ even mode. 

\begin{figure}
\begin{center}
\begin{tabular}{cc}
 \includegraphics[width=0.49\linewidth]{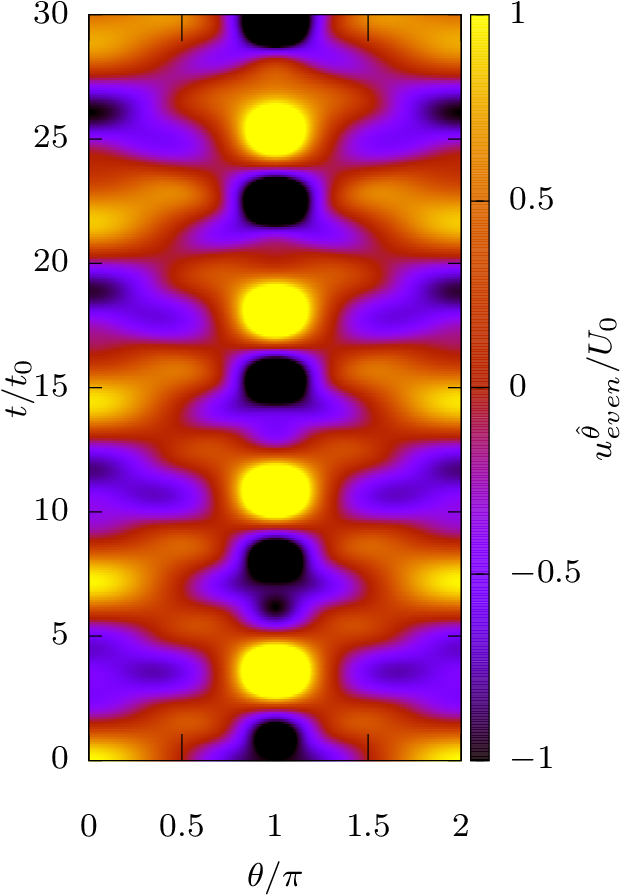} &
 \includegraphics[width=0.49\linewidth]{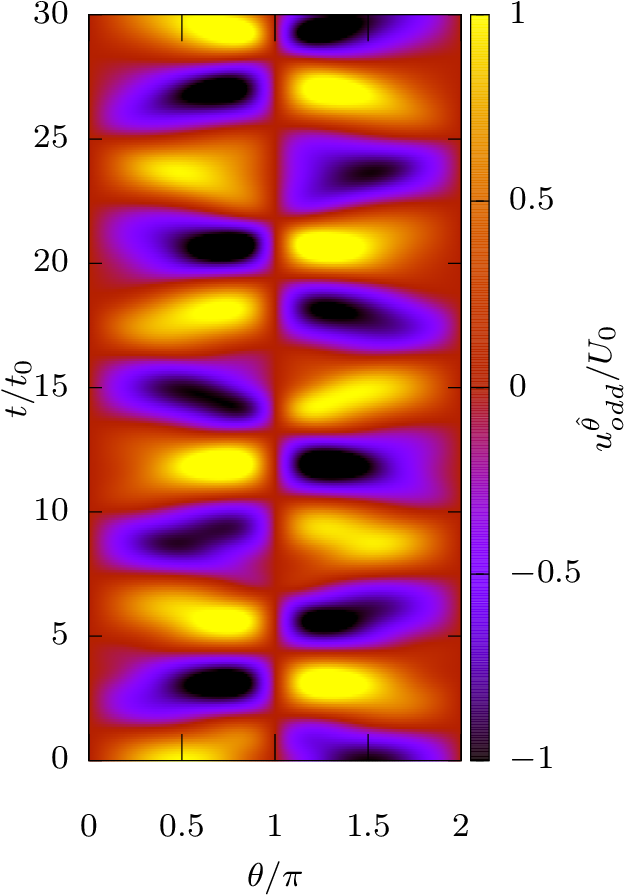} \\
 \hspace{-20pt}(a) &\hspace{-20pt} (b) 
\end{tabular}
\end{center}
\caption{Time evolution of $u^\htheta_{\rm even} / U_0$ (a) and 
$u^\htheta_{\rm odd} / U_0$ (b), defined in Eq.~\eqref{eq:inv_b2_utheta} 
on the torus with $a = 0.8$.
The horizontal axis represents the angular coordinate along the 
poloidal direction, normalised with respect to $\pi$.
The vertical axis shows the time coordinate $t$, normalised with respect to 
$t_0 = R / 2 c_0$, where
$c_0 = \sqrt{P_0 /\rho_0}$ is the reference speed. 
The colour map represents the value of $u^\htheta / U_0$ and is truncated 
to the interval $[-1,1]$.
\label{fig:inv_b2}}
\end{figure}

We consider two initial conditions, corresponding to 
even and odd initial velocity profiles
\begin{equation}
 u^\htheta_{0;\rm even}(\theta) = U_0 \cos\theta, \qquad
 u^\htheta_{0;\rm odd}(\theta) = U_0 \sin\theta,
 \label{eq:inv_b2_init}
\end{equation}
where $U_0$ is the (constant) initial amplitude.
As before, the initial pressure perturbation is assumed to vanish,
i.e. $\delta P_{0;\rm even}(\theta) = \delta P_{0;\rm odd}(\theta) = 0$.
According to Eq.~\eqref{eq:project}, this implies that the offset 
angles can be taken as $\vartheta_{c;n} = 0$ and 
$\vartheta_{s;n} = \pi / 2$. Furthermore, since 
$\int_0^{2\pi} d\theta\, \cos\theta = 0$, the 
coefficient $U_{c;n;0}^{\rm even}$ of the zeroth mode
(corresponding to $n = 0$) vanishes. 
This allows the velocity to be expanded in the two cases as follows:
\begin{equation}
 u^\htheta_{\rm even}(t, \theta) = \sum_{n = 1}^\infty 
 \frac{U_{c;n;0}^{\rm even} f_n(\theta)}{1 + a \cos\theta} 
 \cos(\omega_{c;n;0} t), \qquad
 u^\htheta_{\rm odd}(t, \theta) = \sum_{n = 1}^\infty 
 \frac{U_{s;n;0}^{\rm odd} g_n(\theta)}{1 + a \cos\theta} 
 \cos(\omega_{s;n;0} t),
 \label{eq:inv_b2_utheta}
\end{equation}
where $U_{s;n;0}^{\rm even} = U_{c;n;0}^{\rm odd} = 0$, while
\begin{align}
 U^{\rm even}_{c;n;0} =& U_0 \int_0^{2\pi} \frac{d\theta}{2\pi} f_n(\theta) \cos\theta =
 U_0 \frac{\lambda_{c;n}^2}{a(2-\lambda_{c;n}^2)} I_{c;0,n},\nonumber\\
 U^{\rm odd}_{s;n;0} =& U_0 \int_0^{2\pi} \frac{d\theta}{2\pi} g_n(\theta) \sin\theta 
 = U_0 I_{s;0;n},
 \label{eq:inv_b2_ints}
\end{align}
where the first relation follows from noting that 
$\cos\theta = a^{-1}(1 + a \cos\theta) - a^{-1}$, 
while the integral $I_{c;-1;n}$ can be expressed in terms of $I_{c;0;n}$ 
by multiplying the first line of Eq.~\eqref{eq:lambdacs_def}
with $(1 + a \cos\theta)/2\pi$ and integrating with respect to $\theta$:
\begin{equation}
 I_{c;-1;n} = \int_0^{2\pi} \frac{d\theta}{2\pi} (1 + a \cos\theta) f_n(\theta) = 
 \frac{2}{2-\lambda_{c;n}^2} I_{c;0;n}.
\end{equation}
As can be seen from table~\ref{tab:inv_b2}, at $a=0.8$, the coefficient of 
the $n = 1$ mode is dominant. For the even initial conditions, the amplitude of 
the $n = 2$ mode is almost a third of the amplitude of the $n = 1$ mode, thus 
it can be expected that a modulation due to this mode will show up in the solution.
This is less important for the odd initial conditions, since 
$U^{\rm odd}_{s;2;0}$ is almost $6$ times smaller in magnitude than 
$U^{\rm odd}_{s;1;0}$.

\begin{figure}
\begin{center}
\begin{tabular}{c}
(a)
\includegraphics[width=0.85\linewidth]{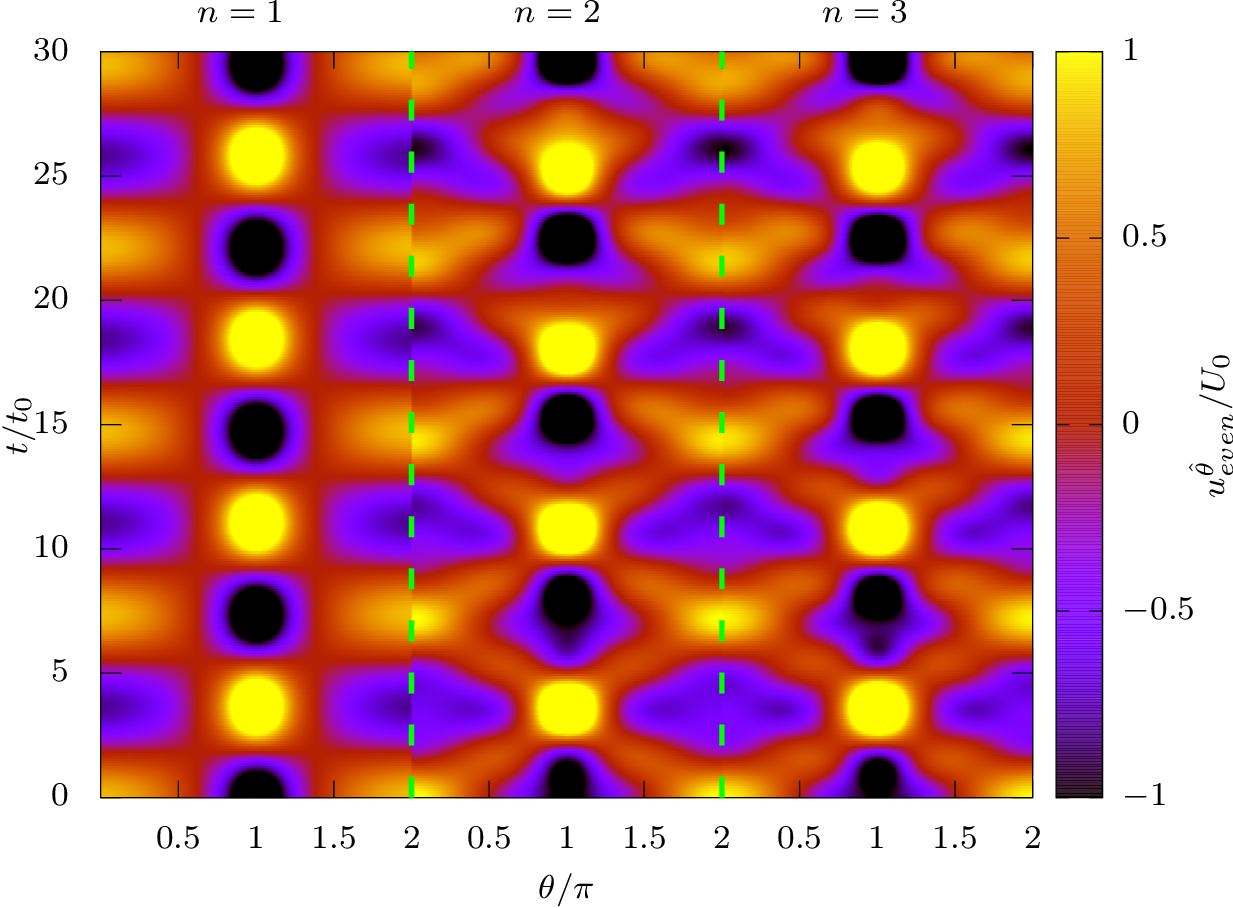} \\
(b)
\includegraphics[width=0.85\linewidth]{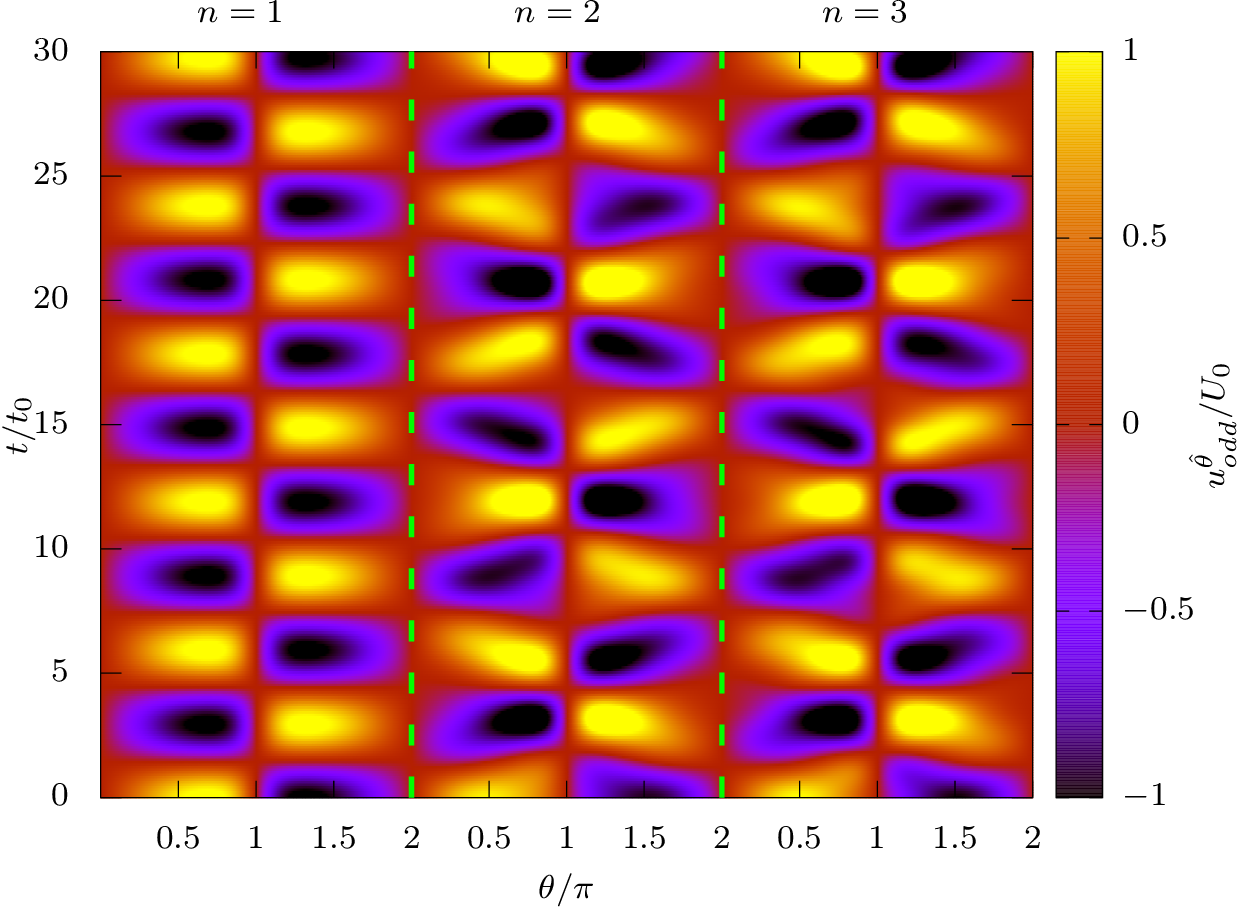}
\end{tabular}
\end{center}
\caption{Analytic solutions for $u^\htheta_{\rm even}(t,\theta) / U_0$ (a)
and $u^\htheta_{\rm odd}(t,\theta) / U_0$ (b),
reconstituted via Eq.~\eqref{eq:inv_b2_utheta} using the harmonics up 
to $n = 1$ (left), $2$ (middle) and $3$ (right).
\label{fig:inv_b2_an}}
\end{figure}
We now consider an ideal perfect thermal fluid with $\gamma = 2$ and employ 
the non-dimensionalisation according to which 
$\rho_0 = T_0 = P_0 = 1$, $R = 2$ ($r = 1.6$ such that $a = 0.8$),
and $c_0 = \sqrt{P_0 / \rho_0}$. The constant in Eq.~\eqref{eq:inv_b2_init}
is set to $U_0 = 10^{-5}$. In this case, the angular frequency for the first even mode 
is $\omega_{c;1} = c_{s;0} \lambda_{c;1} / r \simeq 0.85$
and the time required for $4$ periods for this mode is 
$8\pi / \omega_{c;1} \simeq 29.58$. 
The angular frequency for the first odd mode is 
$\omega_{s;1} = c_{s;0} \lambda_{s;1} / r \simeq 1.06$ and the time required for $5$ periods 
for this mode is $10\pi / \omega_{s;1} \simeq 29.69$. 
We thus perform simulations covering the time domain
$0 \le t \le 30$, using $N_{\theta} = 320$ nodes distributed 
equidistantly along the $\theta$ direction and a time step $\delta t = 10^{-3}$.
The velocity configuration is saved every $100$ time steps, yielding a total 
of $300$ snapshots, which are arranged in time lapses, as shown in 
figures~\ref{fig:inv_b2}(a) and \ref{fig:inv_b2}(b).
The ratio $u^\htheta/U_0$ is represented using a colour map, which is truncated 
to the values $[-1,1]$ for better visibility. It can be seen that 
the number of (quasi-)periods for the even and odd initial conditions 
are $4$ and $5$, as predicted based on the values of $\lambda_{c;1}$ 
and $\lambda_{s;1}$, respectively.

Finally, we discuss the emergence of the apparent periodicity breakdown
observed in figures~\ref{fig:inv_b2}(a) and \ref{fig:inv_b2}(b) for the even 
and odd initial conditions considered 
in this section. Figure~\ref{fig:inv_b2_an} shows the 
analytic solutions for $u^\htheta_{\rm even}$ (a) and $u^\htheta_{\rm odd}$ 
(b) derived in Eq.~\eqref{eq:inv_b2_utheta}, truncated at 
$n = 1$ (left), $2$ (middle) and $3$ (right). We note that 
the amplitude of the zeroth-order harmonic vanishes when the initial state 
is prepared according to Eq.~\eqref{eq:inv_b2_init}. The resulting 
configurations for different truncations are separated using dashed vertical green lines. It can 
be seen that the first-order harmonic exhibits the fundamental periodicity 
observed also in figure~\ref{fig:inv_b2}. Adding the second harmonic produces 
a visible disturbance since the amplitude ratios 
$U_{c;2;0} / U_{c;1;0} \simeq 0.324$ and $U_{s;2;0} / U_{s;1;0} \simeq -0.170$
are non-negligible.
Because the ratios $\omega_{c;2} / \omega_{c;1} \simeq 2.099$ 
and $\omega_{s;2} / \omega_{s;1} \simeq 1.737$ are irrational 
numbers, the resulting configurations become pseudo-periodic. 
This is different from the flat geometry case where
the ratios are integers, thereby conserving the periodicity of the solution.
The addition of the third-order harmonic has a significantly milder effect, since 
the ratios $U_{c;3;0} / U_{c;1;0} \simeq -0.057$ and 
$U_{s;3;0} / U_{s;1;0} \simeq -0.030$ are small. Therefore, the middle 
configuration presented in figure~\ref{fig:inv_b2_an} already provides a
reasonable approximation of the configurations observed in figure~\ref{fig:inv_b2}. 

\section{Viscous fluid: shear wave damping} \label{sec:shear}

In this section, we address the equivalent on the torus 
of a standard benchmark problem for viscous flow solvers. 
On the flat geometry, the shear wave setup typically consists of a system which is homogeneous 
in two directions, say the $y$ and $z$ axes. However, the fluid velocity in one of the directions, say the 
$y$ component, varies with respect to the $x$ axis.
Due to this dependence, layers which are adjacent with respect 
to the $x$ direction travel at different velocities along the $y$ direction.
Due to friction, the velocity difference between two 
such adjacent layers experiences a damping which is controlled 
by the kinematic viscosity of the fluid and is induced 
via the viscous part of the stress tensor. 
In the present case of the torus geometry, we 
consider that the poloidal component $u^\htheta$ of the fluid velocity vanishes,
while its azimuthal component $u^\hvarphi$ varies in magnitude as a function of 
the poloidal angle $\theta$.

This section is structured as follows. In Subsec.~\ref{sec:shear:sol}, 
the general solution for the shear wave damping problem on the torus 
is obtained. Subsections~\ref{sec:shear:b1} and \ref{sec:shear:b2} 
discuss two benchmark problems proposed in this context.

\subsection{General solution} \label{sec:shear:sol}

\begin{figure}
\begin{center}
\begin{tabular}{cc}
 \includegraphics[width=0.46\linewidth]{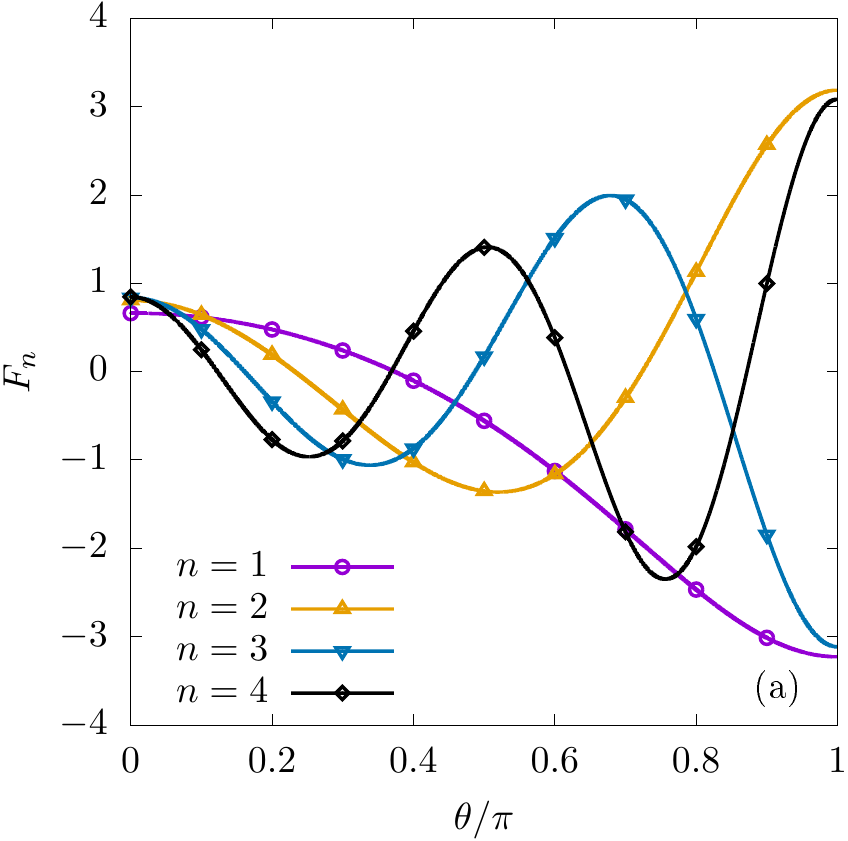} &
 \includegraphics[width=0.46\linewidth]{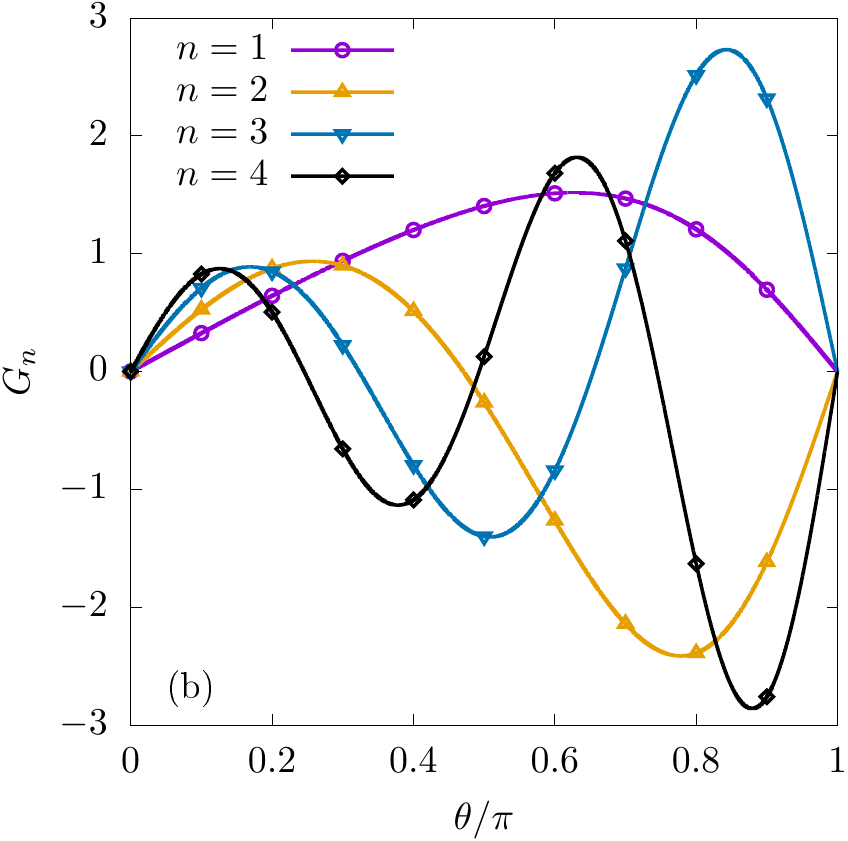}
\end{tabular}
\end{center}
\caption{The even and odd eigenfunctions $F_n$ (a) and 
$G_n$ (b) of Eq.~\eqref{eq:shear_FG_chi},
summarised in Eq.~\eqref{SM:eq:shear_modes}, with $a = 0.4$ for 
$n = 1$, $2$, $3$ and $4$.
The eigenvalues corresponding to $n = 1$, $2$, $3$ and $4$
are $\chi_{c;n} \simeq 1.185$, $2.055$, $3.035$ and $4.026$ for the even modes, 
and $\chi_{s;n} \simeq 1.060$, $2.054$, $3.035$ and $4.026$ for the odd modes.
\label{fig:shear_funcs}}
\end{figure}

For the torus geometry, we consider the axisymmetric flow of an ideal, 
single-component fluid with vanishing poloidal velocity ($u^\htheta = 0$). 
In this case, the linearised limit of the $\varphi$ component of the Cauchy 
equation [Eq.~\eqref{eq:hydro_cauchyph_azim}] reads
\begin{align}
 \partial_t u^\hvarphi = \frac{\nu}{r^2(1 + a \cos\theta)^2} 
 \frac{\partial}{\partial \theta} \left[ 
 (1 + a \cos\theta)^3 \frac{\partial}{\partial \theta} 
 \left( \frac{u^\hvarphi}{1 + a\cos\theta}\right)\right],
\end{align}
with $\rho \simeq \rho_0 = {\rm const}$ and $P \simeq P_0 = {\rm const}$. 
In the above, $\nu$ represents the kinematic 
viscosity, which we assume to be constant.
The above equation can be solved using separation of variables 
by letting 
\begin{equation}
 u^\hvarphi(t, \theta) \rightarrow 
 u_n^\hvarphi(t, \theta) = 
 V_n(t) \Lambda_n(\theta) (1 + a \cos\theta).
 \label{eq:tor_shear}
\end{equation}
Under this separation, the time-dependent 
amplitude satisfies the equation
\begin{equation}
 \partial_t V_n(t) = -\frac{\nu \chi_n^2}{r^2} V_n(t) \Rightarrow
 V_n(t) = V_{n,0} e^{-\nu \chi_n^2 t / r^2},
 \label{eq:shear_Vn_sol}
\end{equation}
where $\chi_n^2$ is a constant.
The spatial 
component in Eq.~\eqref{eq:tor_shear}, $\Lambda_n(\theta)$,
satisfies
\begin{equation}
 \frac{1}{(1 + a \cos\theta)^3} 
 \frac{\partial}{\partial \theta} \left[ 
 (1 + a \cos\theta)^3 \frac{\partial \Lambda_n}{\partial \theta} 
 \right] + \chi_n^2 \Lambda_n = 0.\label{eq:shear_modes_eq}
\end{equation}
Similar to the problem discussed in the previous section,
the above equation admits even and odd solutions, which we denote 
via $F_n(\theta)$ and $G_n(\theta)$, respectively. 
The index $n$ labels the discrete eigenmodes of Eq.~\eqref{eq:shear_modes_eq}.
We label the eigenvalues $\chi_{c;n}$ and $\chi_{s;n}$ for the 
even and odd modes, such that
\begin{align}
 \frac{1}{(1 + a \cos\theta)^3} 
 \frac{\partial}{\partial \theta} \left[ 
 (1 + a \cos\theta)^3 \frac{\partial F_n}{\partial \theta} 
 \right] + \chi_{c;n}^2 F_n =& 0,\nonumber\\
 \frac{1}{(1 + a \cos\theta)^3} 
 \frac{\partial}{\partial \theta} \left[ 
 (1 + a \cos\theta)^3 \frac{\partial G_n}{\partial \theta} 
 \right] + \chi_{s;n}^2 G_n =& 0. \label{eq:shear_FG_chi}
\end{align}

It can be shown that the modes corresponding to different indices $n$ and $n'$ are orthogonal.
We choose the overall normalisation constants by imposing unit norm
with respect to the inner product, $\braket{F_n, F_{n'}} = \braket{G_n, G_{n'}} = \delta_{n,n'}$. 
For two arbitrary functions $\Psi$ and $\Phi$, the inner product is defined as
\begin{equation}
 \braket{\Psi, \Phi} = \frac{1}{2\pi} \int_0^{2\pi} d\theta\, (1 + a \cos\theta)^3 
 \Psi(\theta) \Phi(\theta).\label{eq:visc_scprod}
\end{equation}

\begin{figure}
\begin{center}
\begin{tabular}{ccc}
 \includegraphics[width=0.32\linewidth]{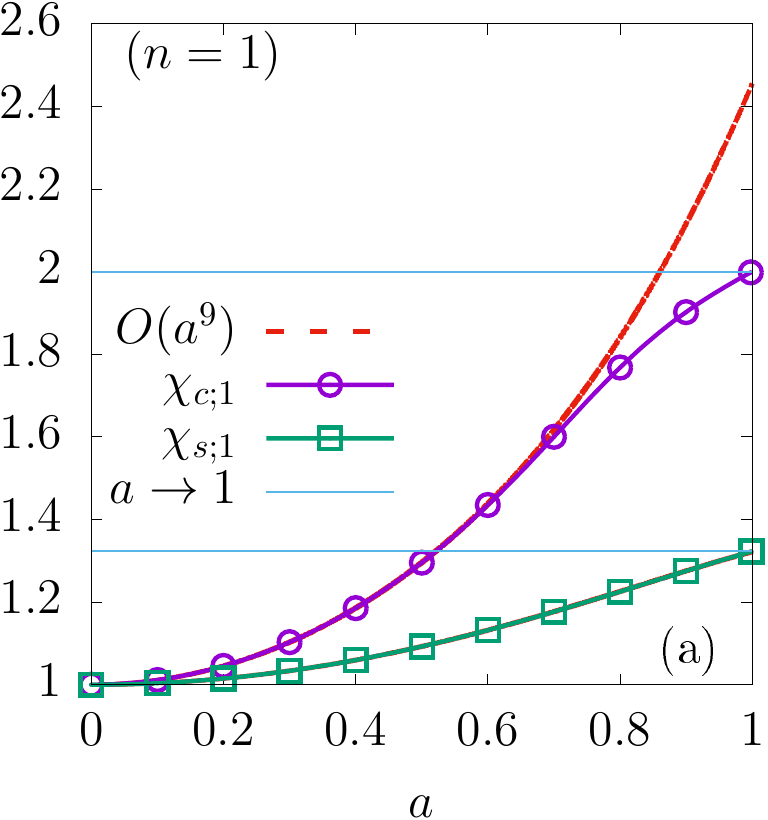} &
 \includegraphics[width=0.32\linewidth]{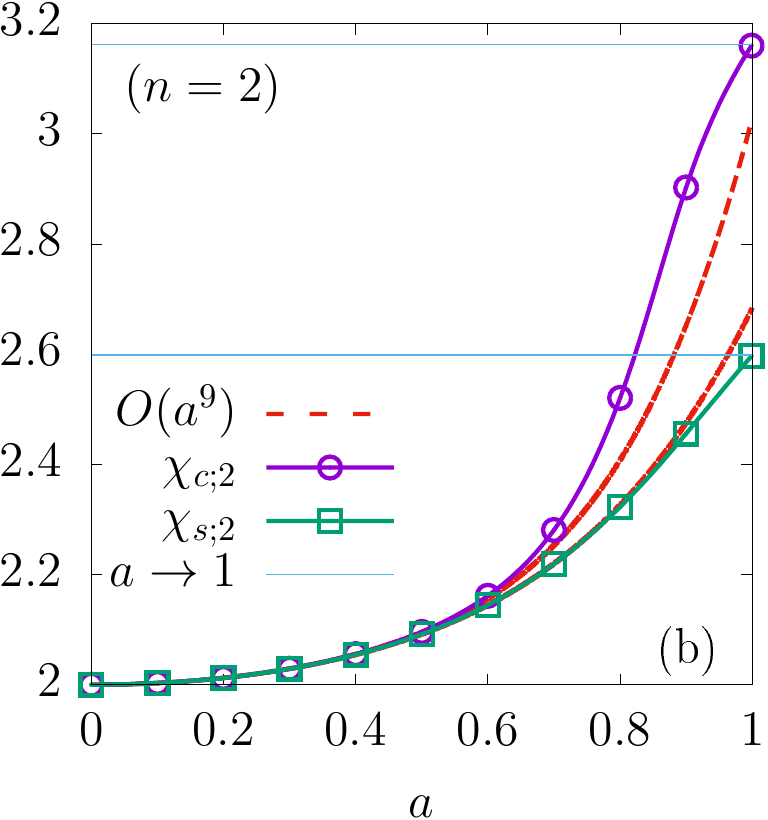} &
 \includegraphics[width=0.32\linewidth]{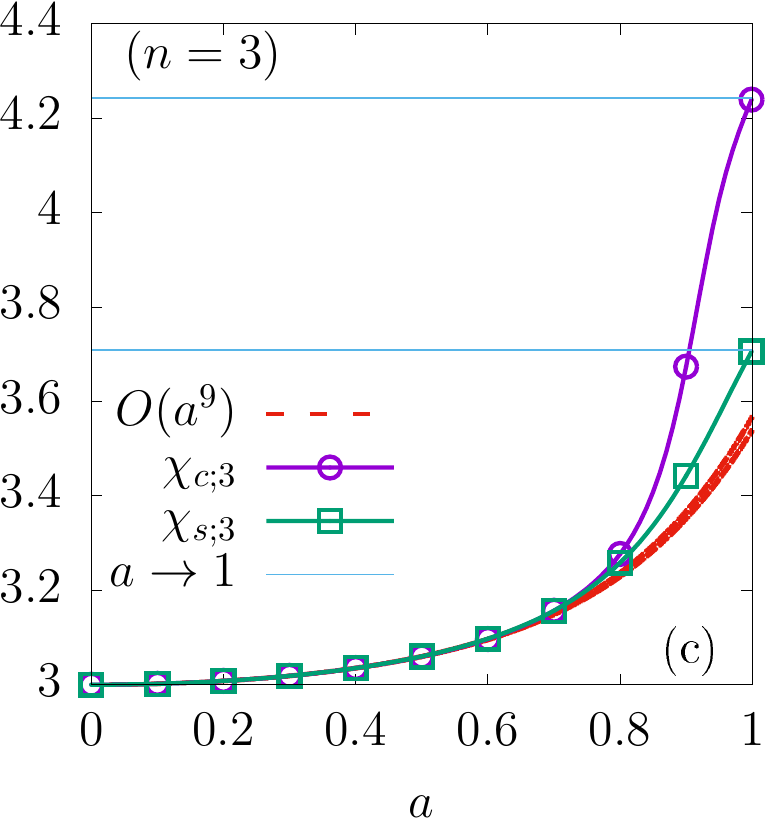} 
\end{tabular}
\end{center}
\caption{The dependence of $\chi_{c;n}$ and $\chi_{s;n}$ 
on $a$ for (a) $n = 1$, (b) $n = 2$ and (c) $n = 3$. 
The dotted lines show the perturbative approximations in 
Eq.~\eqref{SM:eq:shear_modes} with terms 
up to $O(a^9)$. 
\label{fig:shear_chi}}
\end{figure}

The solution of Eq.~\eqref{eq:shear_modes_eq} corresponding to $n = 0$ and 
$\chi = 0$ is even, being given by
\begin{equation}
 F_0 = \left[1 + \frac{3a^2}{2} \right]^{-1/2}.
 \label{eq:visc_F0}
\end{equation}
When $a = 0$, the eigenvalues are $\chi_{c;n}^2 = \chi_{s;n}^2 = n^2$, while 
the eigenmodes are given through
\begin{equation}
 F_n(\theta) = \sqrt{2} \cos(n\theta), \qquad 
 G_n(\theta) = \sqrt{2} \sin(n\theta),
\end{equation}
as was the case in Subsec.~\ref{sec:inv}. When $a = 1$, the eigenvalues are 
derived in Eq.~\eqref{SM:eq:a1:visc_chi} and are reproduced below, 
for convenience
\begin{equation}
 \chi_{c;n} = \sqrt{n(n + 3)}, \qquad 
 \chi_{s;n} = \sqrt{(n + \tfrac{5}{2})(n - \tfrac{1}{2})}.
 \label{eq:visc_chi_a1}
\end{equation}
The eigenfunctions and the detailed procedure used to obtain them 
are given in Sec.~\ref{SM:sec:modes:a1:visc} of the supplementary material.

When $0 < a < 1$, the eigenmodes can be obtained as power series 
with respect to $a$, as detailed in Appendix~\ref{app:modes}.
The eigenfunctions $F_n$ and $G_n$ are depicted 
graphically in figure~\ref{fig:shear_funcs} for $a = 0.4$ and 
$1 \le n \le 4$.
The eigenvalues $\chi_n^2$ can be obtained following the 
same perturbative procedure as described in the previous section. 
As in the inviscid case, 
the difference between the eigenvalues corresponding to the 
$n$'th odd and even modes appear at $O(a^{2n})$, as further discussed 
in Appendix~\ref{app:modes}. The dependence of $\chi_{*;n}$ 
($* \in \{c,s\}$, $1 \le n \le 3$) on $a$ is shown in 
figure~\ref{fig:shear_chi}, obtained using high precision
numerical integration. It can be seen that all eigenvalues exhibit 
a monotonic increase with respect to $a$ and the eigenvalues $\chi_{c;n}$ 
corresponding to the even modes become significantly larger than those 
corresponding to the odd modes as $a \rightarrow 1$,
as indicated in Eq.~\eqref{eq:visc_chi_a1}.
The dotted lines correspond to the perturbative approximations up to 
$O(a^9)$. This behaviour is 
contrary to that of the eigenvalues seen in the inviscid case, shown 
in figure~\ref{fig:inv_lambda}. In the inviscid case, the eigenvalues corresponding 
to the odd modes, $\lambda_{s;n}$, are generally larger than those corresponding 
to the even modes. Moreover, $\lambda_{c;n}$ has a non-monotonic behaviour,
increasing with $a$ at small $a$ (for $n > 1$) and decreasing as 
$a \rightarrow 1$.

\begin{figure}
\begin{center}
\begin{tabular}{cc}
 \includegraphics[width=0.48\linewidth]{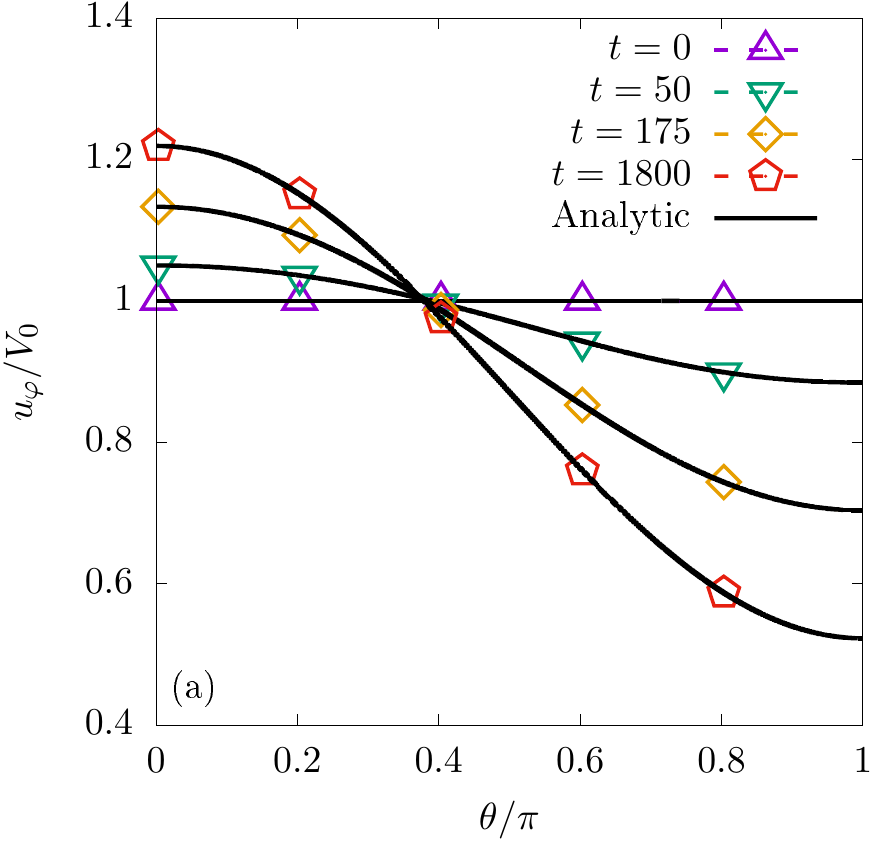} &
 \includegraphics[width=0.48\linewidth]{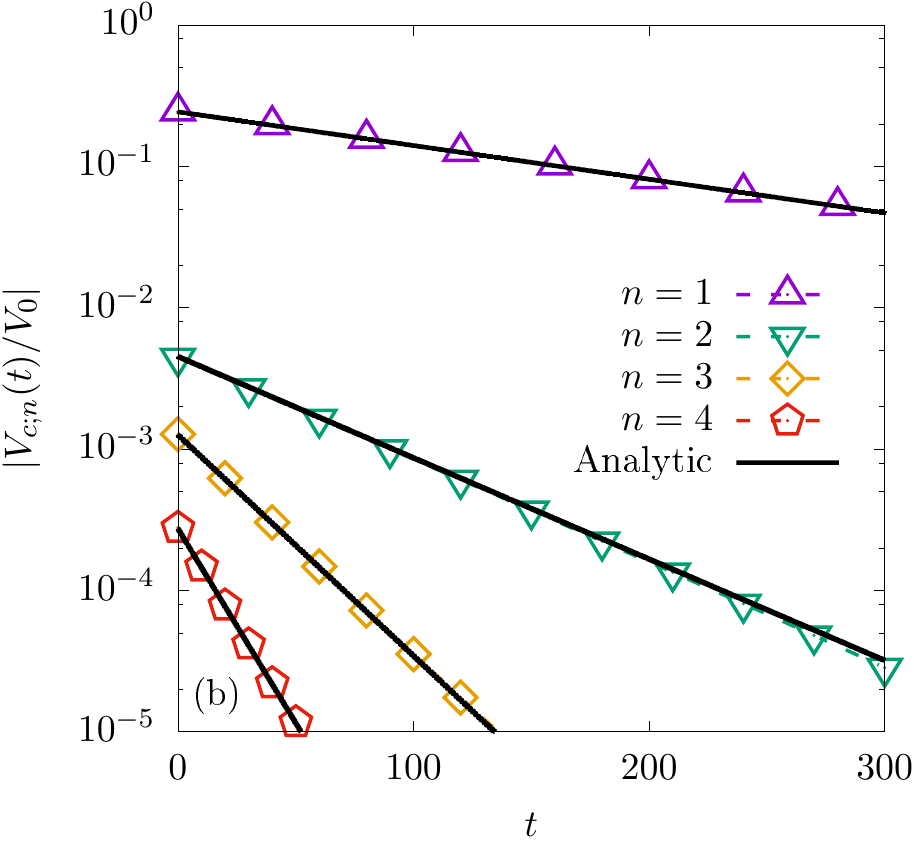}
\end{tabular}
\end{center}
\caption{(a) Time evolution of the ratio $u^\hvarphi/V_0$ 
of the azimuthal velocity $u^\hvarphi$ initialised 
according to Eq.~\eqref{eq:shear_init}, where 
$V_0$ is the initial amplitude. 
(b) Time evolution of the amplitudes 
$V_{c;n}(t)$ ($1 \le n \le 4$).
The numerical results are shown with dotted lines and points,
while the analytic prediction is summing only the terms with 
$0 \le n \le 4$ in Eq.~\eqref{eq:shear_sol}. The torus 
radii ratio is $a = 0.4$.
\label{fig:shear_bench}}
\end{figure}

Combining the solutions for the time and angular dependences,
the general solution can be written as
\begin{gather}
 u^\hvarphi(t, \theta) = (1 + a \cos\theta) 
 \sum_{n = 0}^\infty \left[ 
 V_{c;n}(t) F_n(\theta) + V_{s;n}(t) G_n(\theta)\right], \nonumber\\
 V_{c;n}(t) = V_{c;n;0} e^{-\nu \chi_{c;n}^2 t/r^2}, \qquad 
 V_{s;n}(t) = V_{s;n;0} e^{-\nu \chi_{s;n}^2 t/r^2}.
 \label{eq:shear_sol}
\end{gather}
The amplitudes $V_{c;n;0}$ and $V_{s;n;0}$ can be computed 
by integrating over the velocity profile at initial 
time, $u^\hvarphi_0(\theta) \equiv u^\hvarphi(0, \theta)$:
\begin{equation}
 \begin{pmatrix}
  V_{c;n;0} \\
  V_{s;n;0} 
 \end{pmatrix} = \int_0^{2\pi} \frac{d\theta}{2\pi} 
 (1 + a \cos\theta)^2 u^\hvarphi_0(\theta)
 \begin{pmatrix}
  F_n(\theta) \\ G_n(\theta)
 \end{pmatrix}.
\end{equation}

\subsection{First benchmark: Constant initial flow}\label{sec:shear:b1}

To verify the analytical theory developed in this section and to allow
comparisons against our numerical solutions, we consider a specific example
where the fluid on the torus has an initially constant velocity profile
\begin{equation}
 u^{\hvarphi}_0(\theta) = V_0.
 \label{eq:shear_init}
\end{equation}
In this case, it can be seen that the odd coefficients 
$V_{s;n;0}$ vanish, while the even coefficients can be computed as follows:
\begin{eqnarray}
V_{c;n;0} &=& V_0 \int_0^{2\pi} \frac{d\theta}{2\pi}
 (1 + a \cos\theta)^2 F_n(\theta)\nonumber\\
 &=& V_0 \left[ \frac{6}{(2 + \chi_{c;n}^2)^2} - \frac{1 - a^2}{2 + \chi_{c;n}^2}\right] 
 \mathcal{I}_{c;n}.\label{eq:shear_intndef_aux}
\end{eqnarray}
The second line in Eq.~\eqref{eq:shear_intndef_aux} is obtained by multiplying 
the first line in Eq.~\eqref{eq:shear_FG_chi} with $(1 + a \cos\theta)^2/2\pi$ and 
integrating with respect to $\theta$. For convenience, we also introduced
\begin{equation}
 \mathcal{I}_{c;n} = \int_0^{2\pi} \frac{d\theta}{2\pi} F_n(\theta), \qquad 
 \mathcal{I}_{s;n} = \int_0^{2\pi} \frac{d\theta}{2\pi} \sin\theta \, 
 G_n(\theta).
 \label{eq:shear_intndef}
\end{equation}
The result for $n = 0$ is exact: $\mathcal{I}_{c;0} = (1 + 3a^2 / 2)^{-1/2}$ and 
$V_{c;0;0} = V_0 (1+a^2/2)/\sqrt{1 + 3a^2/2}$.
For $1\leq n \leq 4$, the power series approximations of the $\mathcal{I}_{c;n}$ 
integrals can be found in Eq.~\eqref{SM:eq:shear_modes} of the supplementary material.

figure~\ref{fig:shear_bench}(a) shows the numerical solution
(dotted lines and points) and the analytic results 
obtained above (solid lines) for the fluid velocity in the azimuthal direction at four 
different values of the time coordinate. The agreement is excellent. We used an ideal, isothermal 
fluid with initial constant density $\rho_0 = 1$ and 
constant temperature $T_0 = 1$, on a grid with $N_\theta = 320$ 
equidistant points and a time step of $\delta t = 5 \times 10^{-3}$.
The reference speed is taken as $c_0 = \sqrt{P_0 / \rho_0}$, where 
$P_0 = \rho_0 k_B T_0 / m$ is the reference pressure and $m$ is the particle mass.
The kinematic viscosity is taken to be $\nu = 2.5 \times 10^{-3}$ with 
respect to the reference value $\nu_0 = c_0 L_0$, where $L_0 = R / 2$ is 
the reference length. With this convention, the non-dimensional torus parameters 
are $R = 2$ and $r = 0.8$, while the initial velocity amplitude in 
Eq.~\eqref{eq:shear_init} is $V_0 = 10^{-5}$.
Since the damping in Eq.~\eqref{eq:shear_Vn_sol} depends only on the 
fluid viscosity, the same results can be obtained when considering the 
thermal or the Cahn-Hilliard non-ideal fluids.

The amplitudes of the harmonics are extracted from the numerical solution 
by means of the orthogonality relation, Eq.~\eqref{eq:visc_scprod},
using the expansions of $F_n(\theta)$ given in 
Eq.~\eqref{SM:eq:shear_modes} of the supplementary material.
The analytic solution is that in Eq.~\eqref{eq:shear_sol}, with $V_{s;n;0} = 0$ 
and $V_{c;n;0}$ given in Eq.~\eqref{eq:shear_intndef_aux}. 
The eigenvalues $\chi_{c;n}$ controlling the damping of the amplitude $V_{c;n}(t)$,
as well as the integrals $\mathcal{I}_{c;n}$ ($1 \le n \le 4$) required to compute 
the initial amplitudes $V_{c;n;0}$ via Eq.~\eqref{eq:shear_intndef_aux}, 
are constructed using the mode expansions also found in 
Eq.~\eqref{SM:eq:shear_modes} of the supplementary material.

\subsection{Second benchmark test: Even and odd harmonics}
\label{sec:shear:b2}

\begin{figure}
\begin{center}
 \includegraphics[width=0.48\linewidth]{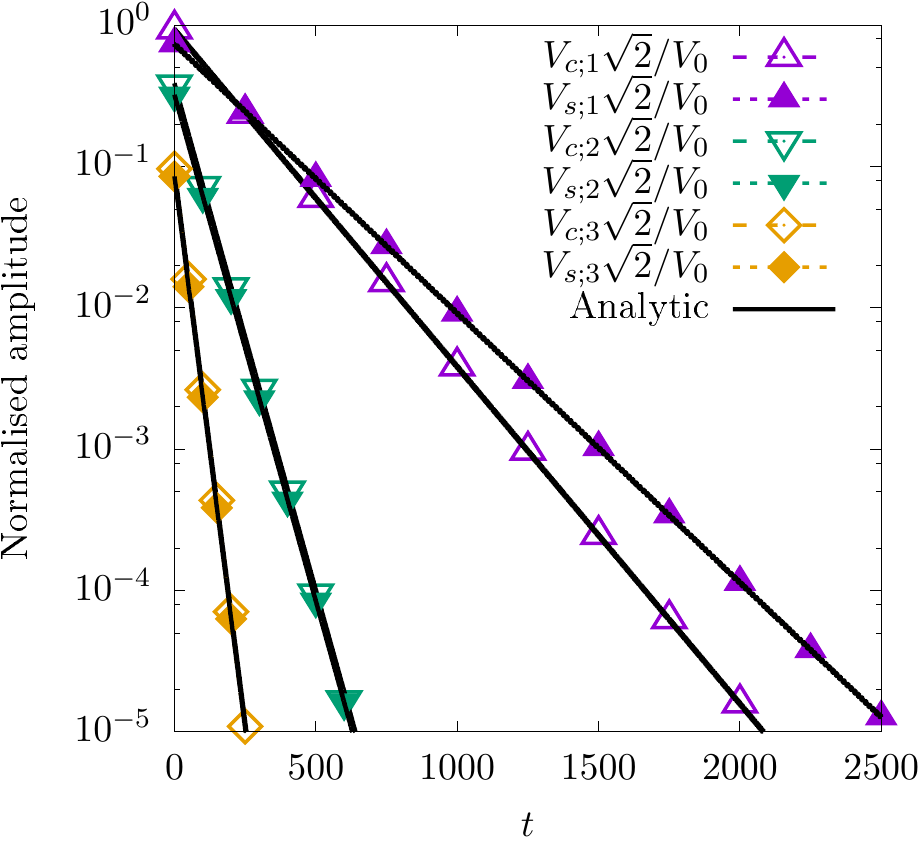}
\end{center}
\caption{Time evolution of the amplitudes 
$V_{c;n}(t)$ (dashed lines and empty symbols) and 
$V_{s;n}(t)$ (dotted lines and filled symbols) for
$n = 1$ (upper purple triangles), $2$ (lower green triangles)
and $3$ (orange rhombi) on the torus with $a = 0.4$.
The analytic prediction, Eq.~\eqref{eq:shear_sol_b2}, is shown 
with solid lines.
\label{fig:shear_harm2}}
\end{figure}

In this second benchmark test, we aim to highlight the difference between 
the rates of decay for the even and odd harmonics corresponding to the same 
order $n$. To this end, we consider initial conditions which are 
neither even nor odd, defined as a combination of harmonic functions
\begin{equation}
 u^\hvarphi_0(\theta) = \frac{V_0}{(1+a \cos\theta)^2 \sqrt{2}}  
 (\cos\theta + \sin\theta), \label{eq:shear_init_b2}
\end{equation}
where the overall $(1 + a \cos\theta)^{-2}$ was added to inhibit the 
development of the $n = 0$ harmonic. The initial amplitudes for the modes
$V_{c;n}(t)$ and $V_{s;n}(t)$ are
\begin{equation}
 V_{c;n;0} = -\frac{V_0 \chi_{c;n}^2}{a\sqrt{2} (2+\chi_{c;n}^2)} \mathcal{I}_{c;n}, \qquad
 V_{s;n;0} = \frac{V_0}{\sqrt{2}} \mathcal{I}_{s;n},
\end{equation}
where the notation $\mathcal{I}_{*;n}$ ($* \in \{c,s\}$)
was introduced in Eq.~\eqref{eq:shear_intndef}. 
The amplitudes $V_{c;n}(t)$ and $V_{s;n}(t)$ undergo exponential 
damping with their respective damping coefficients, 
$\nu \chi_{c;n}^2 / r^2$ and $\nu \chi_{s;n}^2 / r^2$, respectively.
The general solution can be written as
\begin{multline}
 u^{\hat{\varphi}}(t,\theta) = \frac{V_0}{\sqrt{2}}(1+a \cos\theta)
 \sum_{n = 1}^\infty \left[-\frac{\chi_{c;n}^2}{a(2+\chi_{c;n}^2)} \mathcal{I}_{c;n}
 e^{-\nu \chi_{c;n}^2 t / r^2} F_n(\theta) \right.\\
 \left.+ \mathcal{I}_{s;n} e^{-\nu \chi_{s;n}^2 t / r^2} G_n(\theta)\right].
 \label{eq:shear_sol_b2}
\end{multline}

Figure~\ref{fig:shear_harm2} shows the time dependence of the amplitudes 
$V_{c;n}(t)$ (dashed lines and empty symbols) and 
$V_{s;n}(t)$ (dotted lines and filled symbols) for
$n = 1$ (purple upper triangles), $2$ (green lower triangles) and $3$ 
(orange rhombi). As expected from figure~\ref{fig:shear_chi}, $V_{c;1}(t)$
decays at a faster rate than $V_{s;1}(t)$. However, at $a = 0.4$, the 
eigenvalues $\chi_{c;n}$ and $\chi_{s;n}$ have roughly the same values 
when $n \ge 2$. Therefore, the decay rates of $V_{c;2}(t)$ and $V_{c;3}(t)$
are very similar to those of $V_{s;2}(t)$ and $V_{s;3}(t)$, respectively.
In this benchmark test, the fluid and simulation parameters are the same 
as those employed in Subsec.~\ref{sec:shear:b1}.

\section{Viscous fluid: sound wave damping} \label{sec:damp}

In the previous sections, we considered the propagation of sound waves in 
the perfect fluid and the equivalent of shear wave damping
in a viscous fluid. This section presents an analysis of the damping of 
longitudinal waves propagating along the poloidal direction through a 
viscous fluid. For simplicity, we assume that the fluid velocity along the
azimuthal direction vanishes.

This section is structured as follows. The general solution for the 
damping of longitudinal waves propagating along the poloidal direction 
is presented in Subsec.~\ref{sec:damp:sol}.
Then, a benchmark test is proposed in Subsec.~\ref{sec:damp:bench}.

\subsection{General solution}\label{sec:damp:sol}

The starting point of the analysis in this section is the Cauchy equation 
in the poloidal direction, [Eq.~\eqref{eq:hydro_cauchyth_azim}], which can 
be linearised as follows:
\begin{multline}
 \frac{\partial u^\htheta}{\partial t} + 
 \frac{P_0}{\rho_0 r}\frac{\partial \delta P}{\partial \theta}
 = \frac{\kappa \phi_0}{\rho_0 r^3}
 \frac{\partial}{\partial \theta} 
 \left\{\frac{\partial_\theta[(1 + a \cos\theta)\partial_\theta \delta\phi]}
 {1 + a \cos\theta}\right\} \\
 + \frac{\nu}{r^2(1+a \cos\theta)^2}
 \frac{\partial}{\partial\theta} \left[(1 + a \cos\theta)^3 
 \frac{\partial}{\partial\theta}\left( \frac{u^{\htheta}}{1 + a \cos\theta}\right)
 \right] + \frac{\nu_{v}}{r^2} \frac{\partial}{\partial \theta}
 \left\{\frac{\partial_\theta[u^\htheta(1 + a \cos\theta)]}
 {1 + a \cos\theta}\right\}.
 \label{eq:damp_aux0}
\end{multline}
The left hand side of the above equation is similar to that encountered 
in the inviscid case, in Eq.~\eqref{eq:inv_aux}. On the right hand side,
one can see that the differential operator with respect to $\theta$ acting 
on $(1+ a\cos\theta)\partial_\theta \delta \phi$ and in the term proportional to $\nu_v$ 
is the one encountered in the inviscid case, defined in Eq.~\eqref{eq:inv_modes_eq}.
In the term proportional to $\nu$, one can recognise the operator encountered in the damping of the shear wave problem, presented in Eq.~\eqref{eq:shear_modes_eq}.
In principle, the normal modes analysis must be made with respect to the complete set of eigenfunctions and eigenvalues of only one operator. The set of eigenfunctions 
$\{f_n, g_n\}$ of the inviscid operator differs in general from the set $\{F_n, G_n\}$ corresponding 
to the viscous operator (they coincide only in the limit when $a \rightarrow 0$). 
Since the dominant phenomenon in the present setup is the 
wave propagation, it is natural to work with the basis given by the inviscid operator
and to treat the viscous operator as a perturbative effect. To this end,
we take advantage of the identity
\begin{multline}
 \frac{1}{(1+a \cos\theta)^2}
 \frac{\partial}{\partial\theta} \left[(1 + a \cos\theta)^3 
 \frac{\partial}{\partial\theta}\left( \frac{u^{\htheta}}{1 + a \cos\theta}\right)
 \right] \\
 = \frac{\partial}{\partial \theta}
 \left\{\frac{\partial_\theta[u^\htheta(1 + a \cos\theta)]}
 {1 + a \cos\theta}\right\} + \frac{2a \cos\theta}{1 + a \cos\theta} u^\htheta,
\end{multline}
which allows Eq.~\eqref{eq:damp_aux0} to be written as
\begin{multline}
 \frac{\partial u^\htheta}{\partial t} + 
 \frac{P_0}{\rho_0 r}\partial_\theta \delta P
 = \frac{\kappa \phi_0}{\rho_0 r^3}
 \frac{\partial}{\partial \theta} 
 \left\{\frac{\partial_\theta[(1 + a \cos\theta)\partial_\theta \delta\phi]}
 {1 + a \cos\theta}\right\} \\
 + \frac{\nu +\nu_{v}}{r^2} \partial_\theta 
 \left\{\frac{\partial_\theta[u^\htheta(1 + a \cos\theta)]}
 {1 + a \cos\theta}\right\} 
 + \frac{\nu}{r^2} \frac{2a \cos\theta}{1 + a \cos\theta} u^\htheta.
 \label{eq:damp_aux1}
\end{multline}

In principle, as was the case for the inviscid fluid, the sound wave equation 
can be obtained by taking the time derivative of Eq.~\eqref{eq:damp_aux1}. 
However, this approach is not insightful. Instead, starting from Eq.~\eqref{eq:deltaP}, the time derivative of the pressure deviation 
can be replaced using the continuity, energy 
and Cahn-Hilliard equations, reproduced below in the linearised limit
\begin{align}
 \frac{\partial \delta \rho}{\partial t} + 
 \frac{\partial_\theta \mathcal{U}}
 {r(1 + a \cos\theta)} =& 0,\nonumber\\
 \frac{\partial \delta e}{\partial t} + 
 \frac{P_0}{\rho_0 e_0} \frac{\partial_\theta\mathcal{U}}{r(1 + a \cos\theta)} 
 =& 
 \frac{\text{k}_0}{\rho_0 c_v} \frac{\partial_\theta[(1 + a \cos\theta) \partial_\theta \delta e]}
 {r^2(1 + a\cos\theta)}, \nonumber\\
 \frac{\partial \delta \phi}{\partial t} + 
 \phi_0 \frac{\partial_\theta \mathcal{U}}
 {r(1 + a \cos\theta)} =& \frac{M}{r^2} 
 \frac{\partial_\theta[(1 + a\cos\theta)\partial_\theta \delta \mu]}
 {1 + a \cos\theta}.
 \label{eq:damp_aux2}
\end{align}
We remind the readers that we consider small perturbations around a stationary, background state, which we denote by the subscript 0. We also introduced the notation $\mathcal{U} = u^\htheta (1 + a \cos\theta)$ and the deviation of the chemical potential from the background state $\delta \mu = \mu(\phi) - \mu(\phi_0)$ is given by
\begin{equation}
 \delta \mu = -\text{A} (1 - 3\phi_0^2) \delta \phi - \frac{\kappa}{r^2} 
 \frac{\partial_\theta[(1 + a \cos\theta) \partial_\theta \delta \phi]}
 {1 + a \cos\theta}.
\end{equation}
To solve the partial differential equations in Eq.~\eqref{eq:damp_aux2}, 
we seek normal solutions defined 
with respect to the complete set of modes 
$\{f_n, g_n\}$ introduced in Sec.~\ref{sec:inv}.
We introduce the following expansions:
\begin{align}
 \begin{pmatrix}
  u^\htheta & \partial_\theta \delta \rho & \partial_\theta \delta e\\
  \partial_\theta \delta \phi & \partial_\theta \delta \mu &
  \partial_\theta \delta P
 \end{pmatrix} = \sum_{n = 0}^\infty\frac{ f_n(\theta)}{1 + a \cos\theta}
 \begin{pmatrix}
  U_{c;n} & R_{c;n} & E_{c;n}\\
 \Phi_{c;n} & M_{c;n} & P_{c;n}
 \end{pmatrix},
 \label{eq:damp_dec}
\end{align}
where for simplicity we assume that the flow parameters are even with respect 
to $\theta$, such that the coefficients of the odd eigenfunctions
$g_n(\theta)$ vanish. The
amplitudes 
$\mathcal{A}_{c;n}(t)$ ($\mathcal{A} \in \{U, R, E, \Phi, M, P\}$)
have the following time dependence:
\begin{equation}
 \mathcal{A}_{c;n}(t) = \mathcal{A}_{c;n;0} e^{-\alpha_{c;n} t}.
\end{equation}
The real part of $\alpha_{c;n}$ controls the damping 
of the corresponding mode, while its imaginary part 
is responsible for its propagation.
The extension to the case of odd or general flow configurations 
is straightforward, but will not be discussed here for brevity.

In order to find the normal frequencies
$\alpha_{c;n}$, we multiply Eq.~\eqref{eq:damp_aux1} by 
$f_n(\theta)$ and integrate it with 
respect to $\theta$ between $0$ and $2\pi$. 
We obtain
\begin{equation}
 -\alpha_{c;n} U_{c;n;0} + \frac{P_0}{\rho_0 r} P_{c;n;0} =
 -\frac{\kappa \phi_0 \lambda_{c;n}^2}{\rho_0 r^3} \Phi_{c;n;0} -
 \frac{\nu+\nu_v}{r^2} \lambda_{c;n}^2 U_{c;n;0} - 
 \frac{2\nu}{r^2} \sum_{\ell = 0}^\infty \mathsf{M}_{n,\ell} U_{c;\ell;0},
 \label{eq:damp_aux3}
\end{equation}
where $\lambda_{c;n}^2$ is defined in Eq.~\eqref{eq:lambdacs_def}.
The infinite matrix $\bm{\mathsf{M}}$ mixes the 
normal modes due to the last term in Eq.~\eqref{eq:damp_aux1}.
Its components can be obtained as
\begin{align}
 \mathsf{M}_{n,\ell} =& -\int \frac{d\theta}{2\pi} 
 \frac{a \cos\theta }{(1 + a\cos\theta)^2} f_n(\theta) f_\ell(\theta)\nonumber\\
 =& \int \frac{d\theta}{2\pi} 
 \left[\frac{1}{(1 + a \cos\theta)^2} - \frac{1}{1 + a\cos\theta}\right]
 f_n(\theta) f_\ell(\theta).
 \label{eq:damp_M_def}
\end{align}
In the case $n = \ell = 0$, we find an analytic result
\begin{equation}
 \mathsf{M}_{0,0} = \frac{a^2}{1 - a^2}.\label{eq:damp_M00}
\end{equation}
When $\ell = 0$ and $n > 0$, the second term in the square brackets in Eq.~\eqref{eq:damp_M_def} 
does not contribute due to the orthogonality relation given in Eq.~\eqref{eq:inv_scprod}.
Comparing the first term with the definition of $I_{m;n}$ in Eq.~\eqref{eq:inv_In} 
for $m = 2$ and noting that $f_0(\theta) = (1-a^2)^{1/4}$ is a constant,
$\mathsf{M}_{n,0}$ can be written as:
\begin{equation}
 \mathsf{M}_{n,0} = (1 - a^2)^{1/4} I_{c;2;n} - \delta_{n,0}.\label{eq:damp_Mn0_aux}
\end{equation}
The integral $I_{c;2;n}$ ($n > 0$) can be obtained in terms of 
$I_{c;0;n}$ by integrating Eq.~\eqref{eq:lambdacs_def} with 
respect to $\theta$ and using integration by parts
\begin{align}
 I_{c;0;n} =& -\frac{1}{\lambda_{c;n}^2} \int_0^{2\pi} \frac{d\theta}{2\pi} 
 (1 + a \cos\theta) \frac{d}{d\theta} \left(\frac{df_n / d\theta}{1 + a \cos\theta}\right)\nonumber\\
 =& \frac{1}{\lambda_{c;n}^2} \int_0^{2\pi} \frac{d\theta}{2\pi} f_n(\theta) 
 \left[\frac{1}{1 + a\cos\theta} - \frac{1 - a^2}{(1 + a\cos\theta)^2}\right].
\end{align}
The first term in the square brackets on the last line of the above equation 
vanishes for $n > 0$. Setting $m =2$ in Eq.~\eqref{eq:inv_In}, it can be seen
that the second term can be expressed in terms of $I_{c;2;n}$, such that the 
following relation can be established:
\begin{equation}
 I_{c;2;n} = -\frac{\lambda_{c;n}^2}{1 - a^2} I_{c;0;n}. 
 \label{eq:inv_I2n}
\end{equation}
Putting together Eqs.~\eqref{eq:damp_M00}, \eqref{eq:damp_Mn0_aux} and 
\eqref{eq:inv_I2n} allows $\mathsf{M}_{n,0}$ to be expressed in 
the following form:
\begin{equation}
 \mathsf{M}_{n,0} = \frac{\delta_{n,0} a^2}{1 - a^2} 
 -\frac{\lambda_{c;n}^2}{(1 - a^2)^{3/4}} I_{0;n},
 \label{eq:damp_Mn0}
\end{equation}
which is also valid at $n= 0$ since 
the second term does not contribute due to the fact that 
$\lambda_{c;0} = 0$. Later in this section, the diagonal elements
$\mathsf{M}_{n,n}$ with $1\leq n\leq3$, will be necessary for the computation of the 
acoustic damping coefficient. Their analytic approximations up to $O(a^9)$ are 
given in Eq.~\eqref{SM:eq:damp_Mdiag} of the supplementary material. 

The next step is to find expressions for 
the quantities $P_{c;n;0}$ and $\Phi_{c;n;0}$ in 
Eq.~\eqref{eq:damp_aux3}. To this end, we insert
the decompositions in Eq.~\eqref{eq:damp_dec}
into Eq.~\eqref{eq:damp_aux2} and find
\begin{align}
 R_{c;n;0} =& -\frac{\lambda_{c;n}^2}{\alpha_{c;n} r} 
 U_{c;n;0},\nonumber\\
 E_{c;n;0} =& -\frac{P_0}{\rho_0 e_0} 
 \frac{\lambda_{c;n}^2}{\alpha_{c;n} r}
 \frac{U_{c;n;0}}{\widetilde{E}_{c;n;0}},\nonumber\\
 M_{c;n;0} =& \left[\frac{\lambda_{c;n}^2 \kappa}{r^2} - \text{A}(1 - 3\phi_0^2)\right] 
 \Phi_{c;n;0},\nonumber\\
 \Phi_{c;n;0} =& -\phi_0 \frac{\lambda_{c;n}^2}{\alpha_{c;n} r}
 \frac{U_{c;n;0}}{\widetilde{\Phi}_{c;n;0}},
 \label{eq:damp_ampl}
\end{align}
where we introduced the following dimensionless quantities:
\begin{equation}
 \widetilde{E}_{c;n;0} = 1 - \frac{\gamma \nu \lambda_{c;n}^2}
 {{\rm Pr}\, r^2 \alpha_{c;n}}, \qquad 
 \widetilde{\Phi}_{c;n;0} = 1 + \frac{M \lambda_{c;n}^2}{r^2 \alpha_{c;n}} 
 \left[\text{A}(1 - 3\phi_0^2) - \frac{\kappa \lambda_{c;n}^2}{r^2}\right].
\end{equation}
The pressure amplitude $P_{c;n;0}$ can be 
obtained by combining the above results in conjunction with Eq.~\eqref{eq:deltaP} via:
\begin{align}
 P_{c;n;0} =& \frac{\rho_0 P_{\rho,0}}{P_0} R_{c;n;0} + 
 \frac{e_0 P_{e,0}}{P_0} E_{c;n;0} + \frac{P_{\phi,0}}{P_0} \Phi_{c;n;0}\nonumber\\
 =& - \frac{\lambda_{c;n}^2 U_{c;n;0}}{\alpha_{c;n} r}
 \widetilde{P}_{c;n;0}.
 \label{eq:damp_Pcn0_aux}
\end{align}
The dimensionless quantity $\widetilde{P}_{c;n;0}$ was introduced for notational brevity,
being given by
\begin{equation}
 \widetilde{P}_{c;n;0} = \frac{\rho_0 P_{\rho,0}}{P_0} + 
 \frac{P_{e,0}}{\rho_0 \widetilde{E}_{c;n;0}} + 
 \frac{\phi_0 P_{\phi,0}}{P_0 \widetilde{\Phi}_{c;n;0}}.
\end{equation}

Using the expression for $P_{c;n;0}$ given in Eq.~\eqref{eq:damp_Pcn0_aux},
Eq.~\eqref{eq:damp_aux3} can be rearranged as a matrix equation
\begin{equation}
 \bm{\mathsf{A}} \bm{\mathsf{U}} = 0,\label{eq:damp_matrix_eq}
\end{equation}
where the column vector $\bm{\mathsf{U}}$ has elements 
$\mathsf{U}_n = U_{c;n;0}$, while the (infinite-dimensional)
matrix $\bm{\mathsf{A}}$ has the following components:
\begin{equation}
 \mathsf{A}_{n,m} = -\frac{\delta_{n,m}}{\alpha_{c;n}} 
 \left[\alpha_{c;n}^2 + \frac{\lambda_{c;n}^2 P_0}{r^2 \rho_0} 
 \widetilde{P}_{c;n;0} +
 \frac{\kappa \lambda_{c;n}^4 \phi_0^2}{\rho_0 r^4 \widetilde{\Phi}_{c;n;0}}
 - \frac{\alpha_{c;n}\lambda_{c;n}^2}{r^2} (\nu + \nu_{v})\right] 
 + \frac{2\nu}{r^2} \mathsf{M}_{n,m}.
 \label{eq:damp_matrix_el}
\end{equation}

Eq.~\eqref{eq:damp_matrix_eq} has non-trivial solutions 
when the determinant of the matrix $\bm{\mathsf{A}}$ vanishes. 
This condition selects a discrete set of values for 
the coefficients $\alpha_{c;n}$. 
In order to find these values, we make the assumption 
that the dissipative terms are small on their respective 
dimensional scale, i.e.:
$\nu, \nu_{v} \ll c_{s,0} / r$, 
$\kappa \ll r^2$, $M \ll r c_{s,0}$. To this end, we introduce the 
small parameter $\varepsilon$, which allows us to write:
\begin{equation}
 \nu = \varepsilon \overline{\nu}, \qquad 
 \nu_{v} = \varepsilon \overline{\nu}_{v}, \qquad 
 \kappa = \varepsilon \overline{\kappa}, \qquad 
 M = \varepsilon \overline{M}.
 \label{eq:damp_eps}
\end{equation}
We keep terms up to first order in $\varepsilon$ 
for the rest of the section.
We further assume that $\alpha_{c;n}$ can be written as
\begin{equation}
 \alpha_{c;n} = \pm i \omega_{c;n} + \varepsilon 
 \overline{\alpha}_{c;n;d},
 \label{eq:damp_alphacn_general}
\end{equation}
where $\omega_{c;n}$ is the angular velocity 
and $\alpha_{c;n;d} = \varepsilon \overline{\alpha}_{c;n;d}$ 
is the damping factor. 

It can be seen that the off-diagonal elements of the matrix $\bm{\mathsf{A}}$ are at least 
one order higher with respect to $\varepsilon$ than 
the diagonal elements, being proportional 
to $\varepsilon \overline{\nu}$.
When computing the 
determinant, the leading order contribution 
comes from the diagonal elements, while any off-diagonal 
contribution comes with an $O(\varepsilon^2)$ penalty, such that
\begin{equation}
 \det{\bm{\mathsf{A}}} = \mathsf{A}_{11} \times \mathsf{A}_{22} \times \mathsf{A}_{33} \times \ldots + O(\varepsilon^2).
\end{equation}
Thus, up to first order in $\varepsilon$, the eigenvalues 
$\alpha_{c;n}$ can be found by requiring that each diagonal element $\mathsf{A}_{nn}$ vanishes.
We further note that there are typically multiple solutions stemming from $\mathsf{A}_{nn} = 0$.
The acoustic modes correspond to complex solutions for $\alpha_{c;n}$, allowing 
the corresponding modes to propagate.
There are also real solutions for $\alpha_{c;n}$, such that the respective modes decay 
exponentially.
In the case of the ideal thermal fluid, there is only one such solution, corresponding 
to the thermal mode. There is also one such mode corresponding to the Cahn-Hilliard equation,
which we will refer to as the Cahn-Hilliard mode. 
For simplicity, when we use the Cahn-Hilliard equation, we assume that 
the fluid is isothermal.

We now discuss the $n = 0$ mode, corresponding to the 
incompressible velocity profile. Since $\lambda_{c;0} = 0$, the case $n = 0$ is degenerate. 
There is only one eigenvalue corresponding to this case, which is given by
\begin{equation}
 \alpha_{c;0} = \frac{2\nu}{R^2 - r^2},\label{eq:damp_alpha0}
\end{equation}
where the relation $\mathsf{M}_{0,0} = a^2 / (1 - a^2) = r^2 / (R^2 - r^2)$ 
was employed. There is no imaginary part to 
$\alpha_{c;0}$, showing that the mode corresponding to 
the incompressible velocity profile does not propagate.
Furthermore, since $\alpha_{c;0} > 0$, the amplitude
of this mode decays exponentially through viscous damping.
On the flat geometry, the incompressible one-dimensional 
flow corresponds to a constant velocity, which cannot 
suffer viscous damping due to the Galilean invariance of 
the theory. In contrast, on the torus, Galilean invariance is no 
longer valid. While the inviscid fluid supports (in the linearised regime)
the incompressible flow profile as an exact, time-independent
solution, this zeroth-order mode with respect to the set 
$\{f_n, g_n\}$ is no longer preserved in the case of the viscous 
fluid, since $f_0 (1 + a \cos\theta)$ 
does not provide an eigenfunction of the viscous 
operator in Eq.~\eqref{eq:shear_modes_eq}.
The damping of the zeroth-order mode, given in Eq.~\eqref{eq:damp_alpha0}, 
depends only on the kinematic viscosity and seems to be 
independent of the type of fluid considered. 
Thus, $\alpha_{c;0}^{-1}$ provides a 
fundamental time scale on which, in the absence
of external forcing, the flow on the 
poloidal direction becomes quiescent.

For $n > 0$, the angular frequency $\omega_{c;n}$ for the 
acoustic mode is given by
\begin{equation}
 \omega_{c;n} = \frac{\lambda_{c;n} c_{s;\kappa;c;n}}{r},\qquad 
 c_{s;\kappa;c;n}^2 = c_{s;0}^2 + 
 \frac{\kappa \lambda_{c;n}^2}{\rho_0r^2} \phi_0.
 \label{eq:damp_omega}
\end{equation}
The acoustic damping coefficient $\alpha_{c; n; a} = 
\varepsilon \overline{\alpha}_{c;n;a}$ 
(as a shorthand, we remove the subscript $d$ and 
add a subscript $a$ to describe the acoustic damping coefficient)
receives 
contributions from the viscous terms, as well as from the energy 
and Cahn-Hilliard terms
\begin{equation}
 \alpha_{c;n; a} = \frac{\nu}{r^2} \mathsf{M}_{n,n} + 
 \frac{\lambda_{c;n}^2}{2r^2} \left[\nu \left(1 + 
 \frac{\gamma P_0 P_{e,0}}{\rho_0^2 c_{s;\kappa;c;n}^2 {\rm Pr}}
 \right) + \nu_{v} - \frac{M \phi_0 P_{\phi,0}}
 {\rho_0 c_{s;\kappa;c;n}^2} \text{A}(1 - 3\phi_0^2)\right].
 \label{eq:damp_alphaa}
\end{equation}
We remind the reader that 
$\alpha_{c;n;a}$ together with the angular
frequency $\omega_{c;n}$ make up the acoustic 
mode, $\alpha_{c;n} \rightarrow \alpha_{c;n;a} \pm i \omega_{c;n}$.
We note that Eqs.~\eqref{eq:damp_omega} and \eqref{eq:damp_alphaa}
are valid for all types of fluids considered in this paper, namely:
the ideal isothermal fluid, the ideal thermal fluid and 
the isothermal fluid coupled with the Cahn-Hilliard equation.

The thermal 
and Cahn-Hilliard modes can be obtained by 
setting, in Eq.~\eqref{eq:damp_matrix_el}, $\alpha_{c;n}$ to $\varepsilon \overline{\alpha}_{c;n;t}$ 
or $\varepsilon \overline{\alpha}_{c;n;\phi}$, respectively,
while setting the angular frequency 
$\omega_{c;n} = 0$.
The values of $\alpha_{c;n}$ satisfying the above ansatz
are found by solving the following equation:
\begin{equation}
 \widetilde{P}_{c;n;0} = 0,
\end{equation}
which is quadratic in $\alpha_{c;n}$. 
In the general case of the thermal flow of a non-ideal (Cahn-Hilliard) fluid, the
solution of this equation is too lengthy to be reproduced here.
In the next section we will specialise the equation to the fluid types 
introduced in Sec.~\ref{sec:inv}, namely an ideal 
isothermal fluid, an ideal fluid with variable temperature and an 
isothermal multicomponent fluid coupled with the Cahn-Hilliard equation, 
allowing for simple expressions to be obtained. 
These solutions are presented in Eqs.~\eqref{eq:damp_ideal_iso},
\eqref{eq:damp_ideal_th} and \eqref{eq:damp_ideal_th}, respectively.

\subsection{Benchmark test}\label{sec:damp:bench}
We now focus on a specific example.
At initial time, $t=0$, we assume that the density, internal energy and 
order parameter fields are unperturbed, while the 
velocity profile is that of the incompressible fluid
\begin{equation}
 \delta \rho_0 = 0, \qquad 
 \delta e_0 = 0, \qquad 
 \delta \phi_0 = 0, \qquad
 u^\htheta_0 = \frac{U_0}{1 + a \cos\theta}.
 \label{eq:damp_init}
\end{equation}
The analysis of the normal modes was performed in 
the limit where the modes become fully decoupled 
(the non-diagonal elements of the matrix $\bm{\mathsf{M}}$ 
were ignored). For the particular case considered here, 
we are also interested in finding the time dependence of 
the amplitudes $U_{c;n}(t)$, defined through Eq.~\eqref{eq:damp_dec}.
To do this, it is sufficient to employ the initial conditions in 
Eq.~\eqref{eq:damp_init} in order to find the full solution.
From Eq.~\eqref{eq:damp_init} and 
\eqref{eq:damp_aux1}, it can be seen that 
\begin{equation}
 U_{c;n}(0) = \frac{U_0 \delta_{n,0}}{(1 -a^2)^{1/4}}, \qquad 
 \dot{U}_{c;n}(0) = -\frac{2\nu U_0}{r^2 (1 - a^2)^{1/4}}
 \mathsf{M}_{n,0}.
 \label{eq:damp_Ucn0}
\end{equation}
The time dependence of the amplitude of the $n= 0$
mode is
\begin{equation}
 U_{c;0}(t) = \frac{U_0}{(1 - a^2)^{1/4}} e^{-2\alpha_\nu t}, \qquad 
 \alpha_\nu \equiv \frac{1}{2} \alpha_{c;0} = \frac{\nu}{R^2 - r^2},
 \label{eq:alphanu}
\end{equation}
where $\alpha_\nu$ is the principal damping coefficient 
which will be fundamental for discussing the dynamics of 
the stripe configurations in Sec.~\ref{sec:CH}.

For the higher-order harmonics, and when the temperature or 
Cahn-Hilliard equation is taken into account, a third equation 
is required to fix the integration constant for the thermal or 
Cahn-Hilliard mode.
This can be obtained by taking the time derivative 
of Eq.~\eqref{eq:damp_aux1}, yielding
\begin{equation}
 \ddot{U}_{c;n} + \frac{P_0}{\rho_0 r} \dot{P}_{c;n} 
 + \frac{\kappa \lambda_{c;n}^2 \phi_0}{r^3} \dot{\Phi}_{c;n}
 + \frac{\nu + \nu_{v}}{r^2} \lambda_{c;n}^2 \dot{U}_{c;n}
 + \frac{2\nu}{r^2} \sum_{m = 0}^\infty 
 \mathsf{M}_{n,m} \dot{U}_{c;m} = 0.
 \label{eq:damp_Ucn_ddot}
\end{equation}
The time derivative $\dot{P}_{c;n}$ can be obtained 
in analogy to Eq.~\eqref{eq:damp_Pcn0_aux}, by differentiating 
Eq.~\eqref{eq:deltaP} with respect to $\theta$ and $t$, multiplying it 
by $f_n(\theta)$ and then integrating it with respect to $\theta$:
\begin{equation}
 \dot{P}_{c;n} = \frac{\rho_0 P_{\rho,0}}{P_0} \dot{R}_{c;n} + 
 \frac{e_0 P_{e,0}}{P_0} \dot{E}_{c;n} + \frac{P_{\phi,0}}{P_0} 
 \dot{\Phi}_{c;n}.\label{eq:damp_dPcn_aux}
\end{equation}
The time derivatives $\dot{R}_{c;n}$, $\dot{E}_{c;n}$ and $\dot{\Phi}_{c;n}$ 
can be obtained by differentiating all three relations in Eq.~\eqref{eq:damp_aux2} 
with respect to $\theta$, multiplying them by $f_n(\theta)$ and integrating 
them with respect to $\theta$. Noting that, at initial time, the perturbations 
$\delta e$, $\delta \rho$ and $\delta \phi$ vanish, the right hand sides of 
the relations in Eq.~\eqref{eq:damp_aux2} cancel, such that the following results are 
obtained:
\begin{equation}
  \begin{pmatrix}
   \dot{R}_{c;n}(0)\\
   \dot{E}_{c;n}(0)\\
   \dot{\Phi}_{c;n}(0)
  \end{pmatrix} = \frac{\lambda_{c;n}^2}{r} 
  \frac{\delta_{n,0} U_0}{(1- a^2)^{1/4}} 
  \begin{pmatrix}
   1 \\ P_0 / \rho_0 e_0 \\ \phi_0
  \end{pmatrix} = 
  \begin{pmatrix}
   0 \\ 0 \\ 0
  \end{pmatrix}.
  \label{eq:damp_IC_aux}
\end{equation}
The latter equality follows after taking into account 
that $\lambda_{c;0}  = 0$. Substituting the above results 
in Eq.~\eqref{eq:damp_dPcn_aux}, it can be seen that $\dot{P}_{c;n}(0) = 0$. 
Since $\dot{\Phi}_{c;n}(0)$ also cancels by virtue of Eq.~\eqref{eq:damp_IC_aux},
the second and third terms in Eq.~\eqref{eq:damp_Ucn_ddot} can be dropped.

The fourth and fifth terms in \eqref{eq:damp_Ucn_ddot} are of second order with respect to the damping coefficients
$\nu$ and $\nu_v$, and thus of order $O(\varepsilon^2)$ in the language of 
Eq.~\eqref{eq:damp_eps}. For consistency, we approximate 
$\ddot{U}_{c;n}(0) = O(\varepsilon^2) \simeq 0$.
Thus, the solution which is accurate to first order in $\varepsilon$ is
\begin{equation}
 U_{c;n}(t) = \frac{2\nu U_0 \lambda_{c;n}^2}{\omega_{c;n} r^2} 
 \frac{I_{c;0;n}}{1 - a^2} \sin(\omega_{c;n}t) e^{-\alpha_{c;n;a} t}, \qquad 
 \forall n > 0.
 \label{eq:damp_Ucn}
\end{equation}
The above solution was obtained under general considerations and 
therefore it applies to all types of fluids studied in this 
paper. The full solution can be constructed via the expansion 
in Eq.~\eqref{eq:damp_dec}:
\begin{align}
 u^\htheta(t,\theta) = \frac{1}{1 + a\cos\theta} \sum_{n = 0}^\infty 
 U_{c;n}(t) f_n(\theta).
\end{align}
Below we give a set of tests for the ideal isothermal fluid,
the ideal fluid with variable temperature and the isothermal 
multicomponent fluid. The initial velocity amplitude is set to $U_0 = 10^{-5}$.

\begin{table}
\begin{center}
\begin{tabular}{lccc}
 & $T$ & $\nu [\times 10^{-3}]$ & $\alpha_{\nu} [\times 10^{-3}]$ \\\hline
Iso & $1$ & $10$ & $2.976$ \\
Th & $0.5$ & $4$ & $1.190$ \\
CH & $0.4112$ & $6.486$ & $1.930$
\end{tabular}
\end{center}
\caption{Values for the background temperature $T$, 
kinematic viscosity $\nu$ and principal damping coefficient 
$\alpha_\nu$ defined in Eq.~\eqref{eq:alphanu},
for the isothermal ideal fluid (Iso), variable temperature ideal fluid (Th) and 
isothermal multicomponent fluid (CH) on the torus with $a = 0.4$. The background density is in 
all cases $\rho = 1$. The heat conductivity and adiabatic index 
for the thermal model are $\text{k} = 0.012$ and $\gamma = 2$, corresponding to
${\rm Pr} = 2/3$. The parameters for the multicomponent fluid are
$M = \nu \simeq 6.486 \times 10^{-3}$, $\text{A} = 1$ and $\kappa = 5 \times 10^{-4}$.
The parameters are chosen such that $c_s^2 = 1$.
\label{tab:damp_params}
}
\end{table}

\begin{table}
\begin{center}
\begin{tabular}{lccccc}
$n$ & $I_{c;0;n}$ & $\mathsf{M}_{n;n} [\times 10^{-2}]$ & 
$U_{c;n;0;a} / U_0 [\times 10^{-3}]$ 
& $\alpha_{c;n;a}(\nu_v = 0)$ & $\alpha_{c;n;a}(\nu_v = 0.02)$ \\\hline
$1$ & $0.2883$ & $6.015$ & $8.158$ & $8.64 \times 10^{-3}$ & $0.02404$\\
$2$ & $-0.01949$ & $8.261$ & $-1.163$ & $3.27 \times 10^{-2}$ & $0.09554$\\
$3$ & $2.156 \times 10^{-3}$ & $8.808$ & $0.1926$ & $0.2128$ & $0.2128$
\end{tabular}
\end{center}
\caption{Values of various parameters required to build the solution 
in Eq.~\eqref{eq:damp_Ucn} when $a = 0.4$. The bulk kinematic viscosity 
$\nu_v$ required to compute the coefficient $\alpha_{c;n;a}$ in the 
last column is set to $\nu_v = 0.02$. The amplitudes are 
computed by dividing the prefactors in Eq.~\eqref{eq:damp_Ucn}
by the initial velocity amplitude $U_0 = 10^{-5}$.
\label{tab:damp_vals}
}
\end{table}

For the isothermal ideal fluid,
Eqs.~\eqref{eq:damp_omega} and \eqref{eq:damp_alphaa} reduce to:
\begin{align}
 c_{s,\kappa;c;n}^2 =& c_{s,0}^2 = \frac{k_B T_0}{m},\nonumber\\
 \alpha_{c;n;a} =& \frac{\nu}{r^2} \mathsf{M}_{n,n} + 
 \frac{\lambda_{c;n}^2(\nu + \nu_v)}{2r^2}.
 \label{eq:damp_ideal_iso}
\end{align}
We set the background density and temperature to $\rho_0 = 1$ 
and $T_0 = 1$, respectively, and take units such that $c_{s,0} = 1$. 
The kinematic viscosity is set to $\nu = 0.01$ and we consider 
two test cases, corresponding to $\nu_{v} = 0$ and $0.02$. 

In the case of the variable temperature ideal fluid,
Eqs.~\eqref{eq:damp_omega} and 
\eqref{eq:damp_alphaa} reduce to:
\begin{align}
 c_{s,\kappa;c;n}^2 =& c_{s,0}^2 = \frac{\gamma K_B T_0}{m},\nonumber\\
 \alpha_{c;n;a} =& \frac{\nu}{r^2} \mathsf{M}_{n,n} + 
 \frac{\lambda_{c;n}^2}{2r^2} \left[\nu\left(1 + 
 \frac{\gamma - 1}{\rm Pr} \right) +\nu_{v}\right],\nonumber\\
 \alpha_{c;n;t} =& \frac{\lambda_{c;n}^2 \nu}
 {r^2 {\rm Pr}}.\label{eq:damp_ideal_th}
\end{align}
We consider the case when $c_v = k_B / m$, such that $\gamma = 2$.
In order to match the sound speed of the isothermal fluid ($c_{s,0} = 1$), 
the background temperature is set to $T_0 = 0.5$. The background 
density is also kept at $\rho_0 = 1$. We further consider the case 
when the Prandtl number is ${\rm Pr} = 2/3$, such that 
$\text{k}_0 = 3\nu$. In order to ensure that $\alpha_{c;n;a}$ matches the 
value corresponding to the isothermal case, the kinematic viscosity 
is set to $\nu = 0.004$, such that $\text{k}_0 = 0.012$. As before, 
we consider two values for the bulk kinematic viscosity, 
namely $\nu_{v} = 0$ and $0.02$.

In the case of the isothermal multicomponent fluid, 
Eqs.~\eqref{eq:damp_omega} and \eqref{eq:damp_alphaa}
reduce to:
\begin{align}
 c_{s,\kappa;c;n}^2 =& c_s^2 + \frac{\kappa\lambda_{c;n}^2}{\rho_0 r^2} \phi_0^2 =
 \frac{k_B T_0}{m} - \frac{\phi_0^2}{\rho_0}\left[\text{A} (1 - 3\phi_0^2) - 
 \frac{\kappa\lambda_{c;n}^2}{r^2}\right],\nonumber\\
 \alpha_{c;n;a} =& \frac{\nu}{r^2} \mathsf{M}_{n,n} +
 \frac{\lambda_{c;n}^2}{2r^2} \left[\nu + \nu_{v} +
 \frac{M \text{A}^2}{\rho_0 c_{s;\kappa;c;n}^2} \phi_0^2(1 - 3\phi_0^2)^2
 \right],\nonumber\\
 \alpha_{c;n;\phi} =& \frac{M P_{\rho,0} \lambda_{c;n}^2}{r^2 c_{s,0}^2}
 \text{A}(3\phi_0^2 - 1) = 
 \frac{M \lambda_{c;n}^2}{r^2} \frac{k_B T_0}{m c_{s,0}^2} 
 \text{A}(3\phi_0^2 - 1).
 \label{eq:damp_iso_CH}
\end{align}
It can be seen that within the spinodal region, where 
$-\frac{1}{\sqrt{3}} < \phi_0 < \frac{1}{\sqrt{3}}$,
$\alpha_{c;n;\phi} < 0$ and spontaneous domain decomposition 
can occur through an exponential growth of fluctuations.
We thus conduct the simulations outside this region, namely 
for the background value $\phi_0 = 0.8$ of the order parameter.
Keeping the density at $\rho_0 = 1$, the interaction strength 
$\text{A} = 1$ and the surface tension parameter $\kappa = 5 \times 10^{-4}$, 
the temperature required to match the isothermal sound speed 
$c_{s;\kappa;c;n} = 1$ is $T_0 \simeq 0.4112$ (this is true only 
for the zeroth-order mode, when $\lambda_{c;0} = 0$). 
We consider the case when the mobility parameter $M$ is equal to the 
kinematic viscosity. In order to obtain the same acoustic damping 
coefficients as in the isothermal case, we set 
$M = \nu \simeq 6.486 \times 10^{-3}$. As before, $\nu_v$ takes the 
values $0$ and $0.02$.

The parameter values discussed above are also summarised in table~\ref{tab:damp_params}.
The other quantities required to compute the solutions $U_{c;n}$ (for $n > 0$), given in Eq. 
\eqref{eq:damp_Ucn}, are summarised in table~\ref{tab:damp_vals}.

\begin{figure}
\begin{center}
\begin{tabular}{cc}
 \includegraphics[width=0.48\linewidth]{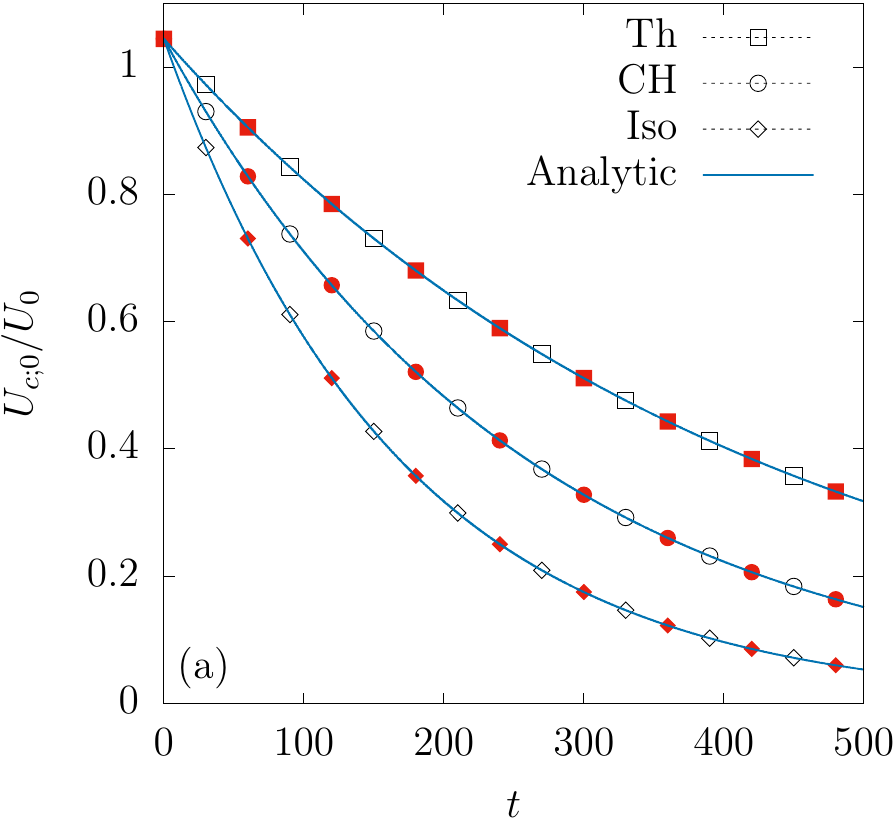} &
 \includegraphics[width=0.48\linewidth]{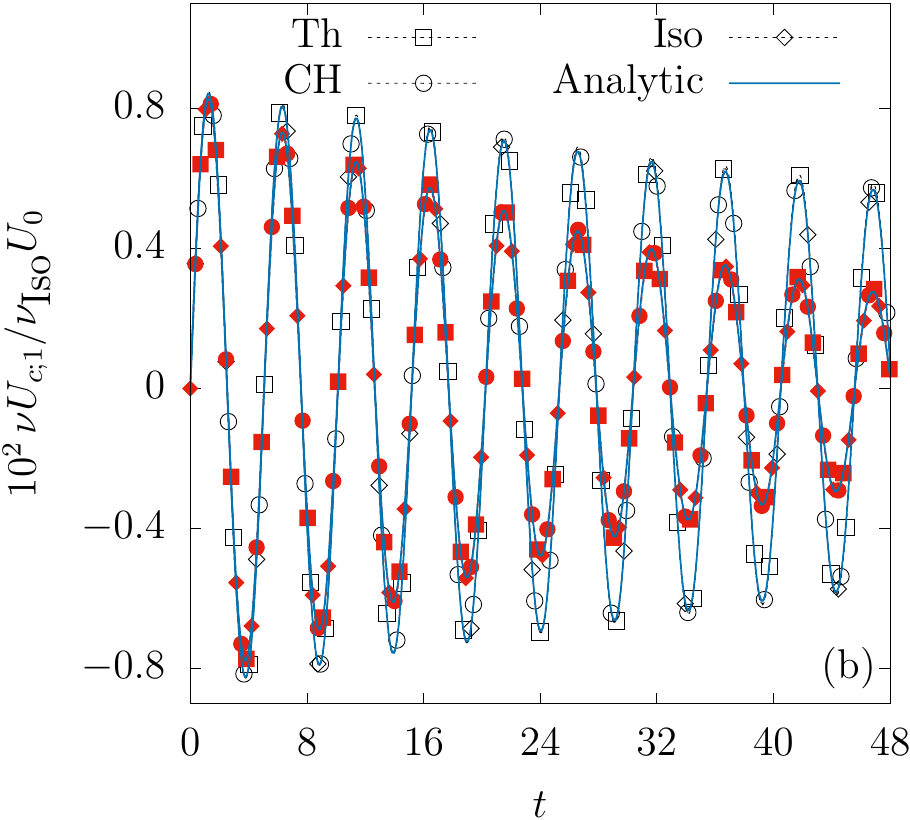} \\
 \includegraphics[width=0.48\linewidth]{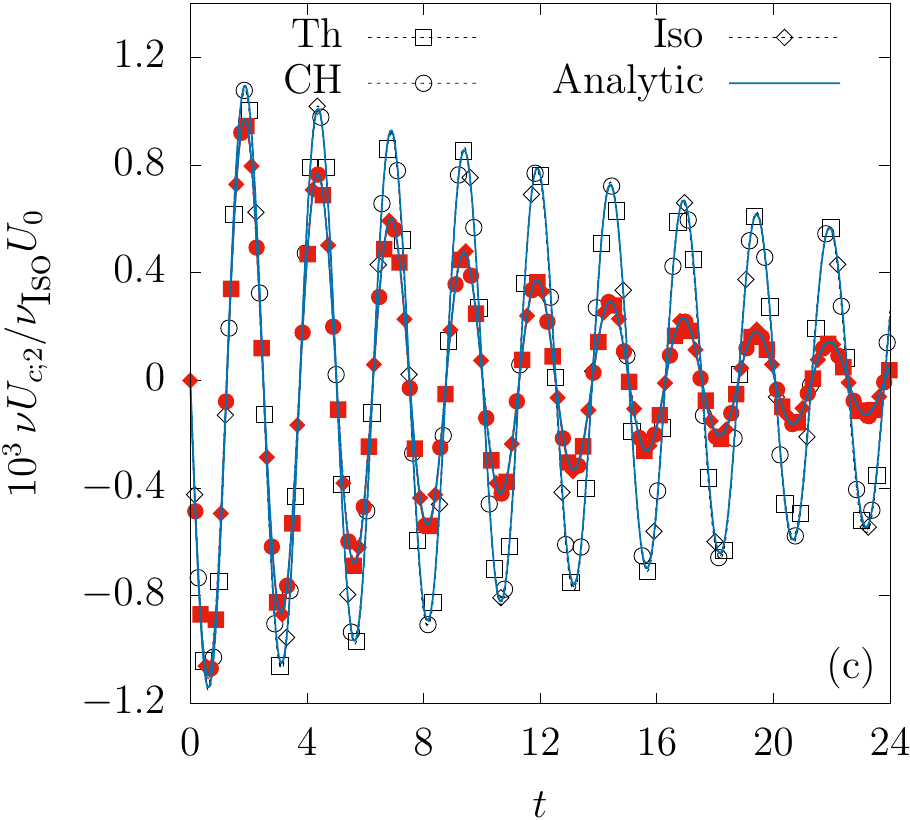} &
 \includegraphics[width=0.48\linewidth]{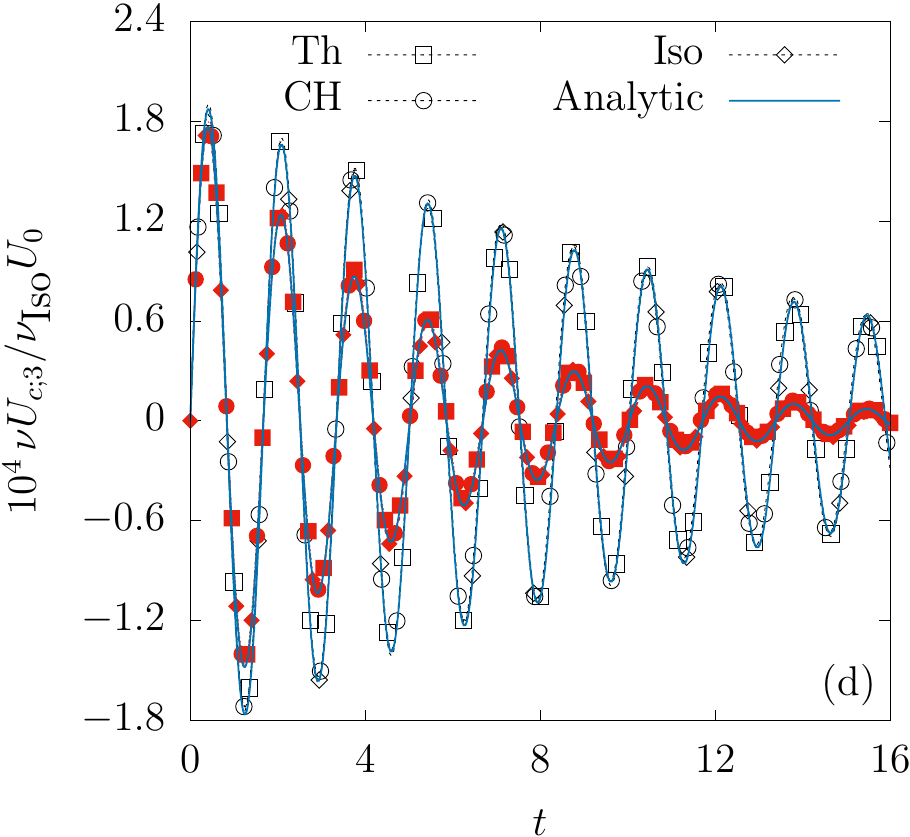}
\end{tabular}
\caption{Time evolution of the ratio $U_{c;n}(t)/U_0$ for the initial 
velocity profile given in Eq.~\eqref{eq:damp_init}, for 
(a) $n = 0$, (b) $n = 1$, (c) $n = 2$ and (d) $n = 3$, 
on the torus with $a = 0.4$. The simulation 
results for $\nu_v = 0$ are shown with dashed black lines 
and empty symbols, while those for $\nu_v = 0.02$ are shown 
with dotted red lines and filled symbols. The analytic predictions 
for $U_{c;0}$ \eqref{eq:alphanu} and $U_{c;n>0}$ \eqref{eq:damp_Ucn}
are shown with solid blue lines. The results corresponding 
to the variable temperature (Th), multicomponent (CH) and 
isothermal (Iso) fluids are shown using squares, circles and 
rhombi, respectively.
\label{fig:damp}
}
\end{center}
\end{figure}

We now discuss the benchmark test results. In figure~\ref{fig:damp}, 
we validate the analytic solution using numerical simulations for the 
$2\times 3$ cases discussed above. The simulations were conducted using $N_\theta = 320$ nodes 
and a time step $\delta t = 5 \times 10^{-4}$ on the torus with $a = 0.4$.
The amplitude of the $n = 0$ 
mode is shown in figure~\ref{fig:damp}(a). As predicted by Eq.~\eqref{eq:alphanu},
the damping coefficient $2\alpha_\nu$ of $U_{c;0}$ depends only on 
the kinematic viscosity. This is natural since the bulk viscosity cannot 
affect the mode corresponding to the incompressible velocity profile.
Thus, the results for $\nu_v = 0$ and $\nu_v = 0.02$ are overlapped and 
only three distinct curves can be seen in figure~\ref{fig:damp}(a), 
corresponding to the differing values of the background kinematic viscosity 
$\nu$ employed in the three fluids discussed above (these values 
are summarised in table~\ref{tab:damp_params}).
The careful choice of parameters discussed above and summarised 
in table~\ref{tab:damp_params} ensures that the acoustic damping 
coefficients $\alpha_{c;n;a}$ corresponding to the higher-order modes 
have the same values. Thus, only two distinct curves can be 
seen in figures~\ref{fig:damp}(b)--\ref{fig:damp}(d), corresponding 
to $\nu_{v} = 0$ (lesser damping, shown with dashed black lines and empty 
symbols) and to $\nu_{v} = 0.02$ (stronger damping, shown with dotted 
red lines and filled symbols). The results for the isothermal (Iso),
variable temperature (Th) and multicomponent fluids (CH), shown 
with squares, circles and rhombi, are overlapped at fixed 
values of $\nu_v$. In all cases, the analytic predictions 
are shown with a continuous blue line and the agreement 
with the numerical results is excellent.

\section{Stripe configurations in equilibrium: Laplace 
pressure test} \label{sec:laplace}

This section starts the series of benchmark problems concerning an isothermal 
multicomponent fluid in axisymmetric ring-type configurations. 
We begin this section by discussing the properties of the equilibrium position in
Subsec.~\ref{sec:laplace:eq}. The stability of these 
equilibria with respect to non-axisymmetric configurations, as well as 
with respect to azimuthal perturbations, is addressed in Subsec.~\ref{sec:laplace:inst}.
The benchmark test proposed in Subsec.~\ref{sec:laplace:bench} concerns a 
generalisation of the Laplace-Young pressure law, giving the 
difference between the pressures measured inside and outside 
of the considered stripe configuration.

\subsection{Equilibrium position}\label{sec:laplace:eq}
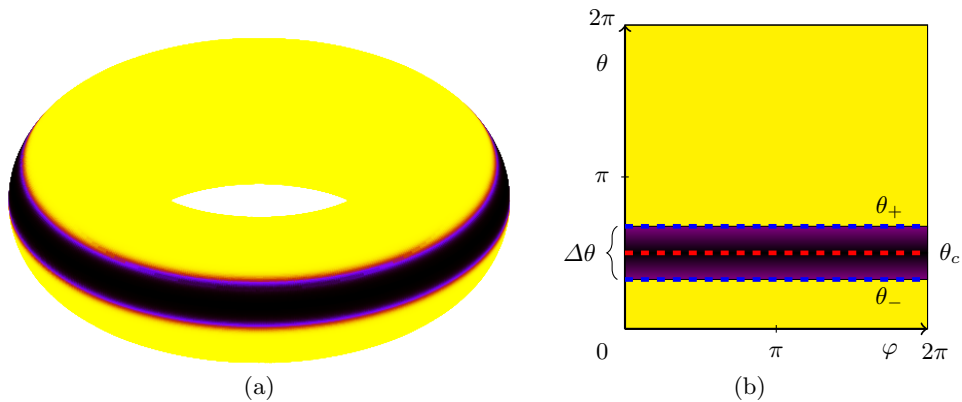
\begin{figure}
 \begin{tabular}{cc}
\includegraphics[width=0.5\linewidth]{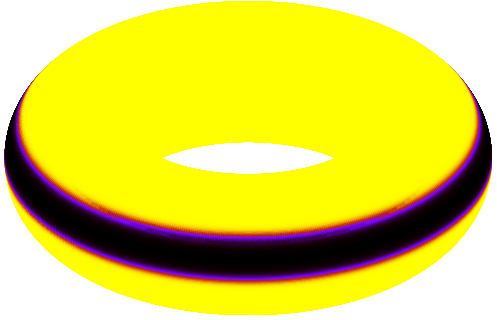}
 &\quad
  \begin{tikzpicture}
 \draw[fill=yellow](-2,-2) rectangle(2,2);
 \draw[thick,->] (-2,-2)--(-2,2);
 \draw[thick,->] (-2,-2)--(2,-2);
\draw[top color=violet, bottom color=black](-2,-1.) rectangle(2,-0.65);
\draw[top color=black, bottom color=violet](-2,-1.35) rectangle(2,-1);
 \draw (-2,-2) rectangle(2,2);
    \draw[dashed,color=red,line width=1.75] (-2,-1) -- (2,-1);
    \draw[dashed,line width=1.75,color=blue] (-2,-1.35) -- (2,-1.35);
    \draw[dashed,line width=1.75,color=blue] (-2,-0.65) -- (2,-0.65);
    \draw (2.3,-1) node {$\theta_c$}; 
 \draw (1.5,-1.6) node {$\theta_-$};
 \draw (1.5,-0.4) node {$\theta_+$};
 \draw (-2.3,1.5) node {$\theta$};
\draw (1.5,-2.3) node {$\varphi$};
 \draw (2.1,-2.3) node {$2\pi$};
 \draw (0,-1.95)--(0,-2.05);
 \draw (0,-2.3) node {$\pi$};
\draw (-2.3,2.1) node {$2\pi$};
 \draw (-2.05,0)--(-1.95,0);
 \draw (-2.3,0) node {$\pi$};
\draw (-2.3,-2.3) node {$0$};
\draw [decorate,decoration={brace,amplitude=4pt},xshift=0pt,yshift=0pt]
(-2.1,-1.35) -- (-2.1,-0.65) node [black,midway,xshift=-0.5cm] {$\Delta \theta$}; 
 \end{tikzpicture}\\
 (a)&(b)
 \end{tabular}
\caption{The axisymmetric ring-type configurations: (a) torus view and (b) unwrapped view, with the color mapping the value of the order parameter.}
\label{fig:stripe_setup}
\end{figure}

Let the stripe interfaces be located at 
\begin{equation}
 \theta_- = \theta_c - \Delta \theta / 2, \qquad
 \theta_+ = \theta_c + \Delta \theta / 2, \label{eq:stripe_thetapm}
\end{equation}
where $\Delta \theta$ is the angular span of the stripe and 
$\theta_c$ is its centre. The remaining part of the fluid domain 
consists of a stripe of width $2\pi - \Delta \theta$, 
centred on $\theta_c + \pi$, which is conjugate to the main stripe. 
For consistency, we only refer to the domain for which $0 < \Delta \theta < \pi$
as `the stripe' in what follows. 
A snapshot of a typical stripe configuration on the torus 
is shown figure~\ref{fig:stripe_setup}(a). The notation introduced above is highlighted in 
a $(\varphi,\theta)$ plot in figure~\ref{fig:stripe_setup}(b).

Since the torus is not geometrically homogeneous with respect to the $\theta$ 
direction, there will be preferred locations where the stripe can be 
in static equilibrium. These locations are found by imposing the 
minimisation of the total interface length subject to fixed stripe 
area $\Delta A$, which is a universal requirement for all fluids 
where interfaces are present. The stripe area can be found by integrating 
over the domain spanned by the stripe
\begin{equation}
 \Delta A = 2\pi r R \int_{\theta_-}^{\theta_+} d\theta (1 + a \cos\theta) 
 = 2\pi r R [\Delta \theta + 2a \sin(\Delta \theta / 2) \cos \theta_c].
 \label{eq:stripe_area}
\end{equation}
On the other hand, the total interface length $\ell_{\rm total}$ 
can be found by adding the circumferences $\ell_+$ and $\ell_-$ corresponding 
to $\theta = \theta_+$ and $\theta = \theta_-$, respectively
\begin{align}
 \ell_{\rm total} =& \ell_+ + \ell_- = 2\pi R (1 + a \cos \theta_+)
 + 2\pi R (1 + a \cos\theta_-) \nonumber \\
  =& 4\pi R \left(1 + a \cos \frac{\Delta \theta}{2} \cos\theta_c \right).
 \label{eq:stripe_ltot}
\end{align}
It can be expected that the minimisation of the interface length 
is required in order for the free energy, Eq.~\eqref{eq:Landau}, to 
reach a minimum. In Sec.~\ref{SM:sec:free} of
the supplementary material, we show that
this is indeed the case to leading order with respect to $\xi_0$. The correction is
due to the fact that the interface shape profile, and hence the line
tension, in principle have a weak dependence on the curvature of the surface.

In order to derive the equilibrium positions, we impose a fixed area 
$\Delta A$. Taking the differential of Eq.~\eqref{eq:stripe_area} gives
\begin{equation}
 d\Delta A = 4\pi r R \left[\left(1 + a \cos \frac{\Delta \theta}{2} \cos\theta_c \right)
 d \frac{\Delta \theta}{2} - a \sin\frac{\Delta \theta}{2} \sin\theta_c d\theta_c \right].
 \label{eq:stripe_dA}
\end{equation}
Setting $d\Delta A = 0$ allows infinitesimal changes $d (\Delta \theta)$ 
in the stripe width 
to be expressed in terms of changes in the position of the stripe centre through
\begin{equation}
 d\frac{\Delta \theta}{2} = 
 \frac{a \sin \frac{\Delta \theta}{2} \sin \theta_c}
 {1 + a \cos\frac{\Delta\theta}{2} \cos\theta_c } d\theta_c.
 \label{eq:stripe_dA0}
\end{equation}
At equilibrium, the interface length $\ell_{\rm total}$ 
[Eq.~\eqref{eq:stripe_ltot}] is minimised. Mathematically, this implies 
\begin{equation}
 d \ell_{\rm total} = -4\pi r \left(\sin \frac{\Delta \theta}{2} 
 \cos\theta_c d \frac{\Delta \theta}{2}
 + \cos\frac{\Delta \theta}{2} \sin \theta_c d\theta_c\right) = 0.
 \label{eq:stripe_dltot}
\end{equation}
Substituting Eq.~\eqref{eq:stripe_dA0} into Eq.~\eqref{eq:stripe_dltot} yields
\begin{equation}
 \left(a \cos\theta_c + \cos\frac{\Delta \theta}{2}\right) \sin\theta_c  = 0,
 \label{eq:minimaring}
\end{equation}
where it is understood that $\Delta \theta$ and $\theta_c$ 
are measured when the stripe is already at its equilibrium
position.

One possibility for Eq.~\eqref{eq:minimaring} to be satisfied is when $\sin\theta_c =0$.
This corresponds to two potential solutions, $\theta_c = 0$ and $\theta_c = \pi$.
From Eq.~\eqref{eq:stripe_ltot}, it can be seen that 
$\theta_c = 0$ corresponds to an unstable equilibrium for 
stripes with $\Delta \theta < \pi$. Conversely, 
$\theta_c = \pi$ is unstable for the conjugate stripes,
having $\Delta \theta > \pi$.
Thus, for stripes with small areas, the minimum energy 
configuration is attained for
\begin{equation}
\theta^{eq}_{c} = \pi. \label{eq:minimapi} 
\end{equation}

We now argue that the above solution is not universally 
valid for all stripe widths. Since the conjugate stripe, having 
width $2\pi - \Delta \theta$, does not equilibrate at $\theta_c^{eq} = \pi$,
it is clear that increasing the stripe area must change the equilibrium 
position from $\theta^{eq}_c = \pi$ towards $\theta^{eq}_c = 0$ (or $2\pi$). 
To illustrate this point, let us consider the case of a maximally wide stripe 
with $\Delta \theta = \pi$. In this case, the conjugate stripe 
also has width $2\pi - \Delta \theta = \pi$, and should thus be 
obtained via a symmetry transformation from the initial stripe. 
The only symmetry of the torus geometry 
is $z \rightarrow -z$. Thus, it is clear that the stripe can 
sit either on the upper half of the torus (centred on $\theta_c^{eq} = \pi/2$),
or on its bottom half (where $\theta_c^{eq} = 3\pi/2$). Both 
configurations are equally stable and it can be seen that 
Eq.~\eqref{eq:minimaring} is satisfied because the 
expression between the parentheses vanishes, while 
the term $\sin \theta_c^{eq} = 1$ is non-vanishing. 

We expect that the equilibrium positions 
at $\theta_c^{eq} = \pi$ for small stripes and 
at $\theta_c^{eq} = \pi \pm \pi/2$ are connected 
smoothly as the area is increased. Thus, $\theta_c^{eq}$ must 
detach from $\pi$ when the equilibrium stripe width exceeds 
a critical value, $\Delta\theta_{\rm crit}$. We can 
deduce that this critical stripe width $\Delta \theta_{\rm crit}$
corresponds to the case where both terms in Eq.~\eqref{eq:minimaring}
vanish simultaneously, leading to
\begin{equation}\label{eq:stripe_Dth_crit}
 \Delta \theta_{\rm crit} = 2\arccos(a).
\end{equation}
Substituting the above value into Eq.~\eqref{eq:stripe_area} yields a critical area,
\begin{equation}
 \Delta A_{\rm crit} = 4\pi r R (\arccos a - a\sqrt{1 - a^2}).
 \label{eq:stripe_dA_crit}
\end{equation}

\begin{figure}
\begin{center}
\hspace{-20pt}\includegraphics[width=0.5\columnwidth]{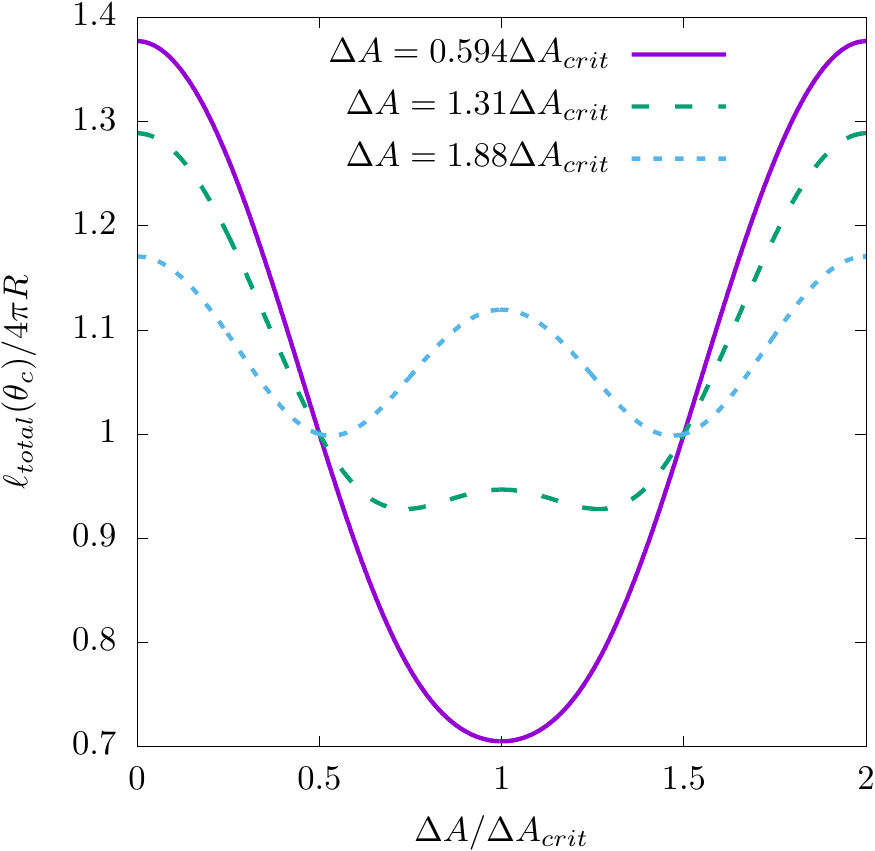}
\end{center}
\caption{
Interface length $\ell_{\rm total}$ on a torus with $a = 0.4$ for various ratios of 
$\Delta A / \Delta A_{\rm crit}$. For $\Delta A / \Delta A_{\rm crit} < 1$
the global minimum is located at $\theta_c = \pi$, while for 
$\Delta A / \Delta A_{\rm crit} > 1$ there are two equivalent minima,
as given by  Eq.~\eqref{eq:stripe_thceq}.
\label{fig:stripe_ltot}}
\end{figure}

When $\Delta A > \Delta A_{\rm crit}$, the point $\theta_c = \pi$ 
corresponds to a local maximum value for $\ell_{\rm total}$. Instead
the global minima correspond to the case where only the parenthesis in 
Eq.~\eqref{eq:minimaring} goes to zero
\begin{equation}
 \theta^{eq}_{c} = \pi \pm \arccos\left[\frac{1}{a} 
 \cos\frac{\Delta \theta_{eq}}{2}\right],
 \label{eq:stripe_thceq}
\end{equation}
where $\Delta \theta_{eq} \equiv \Delta \theta(\theta_c^{eq})$ is the stripe width when 
it is located at the equilibrium position.
We can further show that the total interface length when $\theta_c = \theta^{eq}_{c}$ is
\begin{equation}
 \ell_{\rm total; min} = 4\pi R \sin^2 \frac{\Delta \theta_{eq}}{2},
\end{equation}
with $\Delta \theta_{eq}$ satisfying
\begin{equation}
 \Delta \theta_{eq} - a \sin \Delta \theta_{eq} = \frac{\Delta A}{2\pi r R} =
 \frac{2\Delta A}{\Delta A_{\rm crit}} (\arccos a - a \sqrt{1 - a^2}).
\end{equation}
To better understand the nature of the solutions 
of Eq.~\eqref{eq:minimaring}, figure~\ref{fig:stripe_ltot} shows 
the total interface length $\ell_{\rm total}$ for 
various ratios of $\Delta A / \Delta A_{\rm crit}$. For $\Delta A / \Delta A_{\rm crit} < 1$,
the global minimum configuration is unique and corresponds to $\theta^{eq}_c = \pi$.
Then, as we increase $\Delta A / \Delta A_{\rm crit}$ beyond 1, there is a second-order phase 
transition. The minimum energy configurations become bistable, as given in 
Eq.~\eqref{eq:stripe_thceq}.

\subsection{Stability of stripe configurations}\label{sec:laplace:inst}

In this section, we consider a relaxation of the axial symmetry constraint in order 
to explore the viability of the stripe configurations discussed in the previous 
subsection in the context of $2D$ flows. We first discuss the stability of the 
stripe configurations with respect to small perturbations. 
The main idea is to see the effects of increasing the amplitude of 
azimuthal interface perturbations at the level of orthogonal 
modes. Those modes whose growth causes the interface length to decrease 
lead to instability. Our analysis is limited to the linear growth regime.

Since the upper ($\theta_+$) and lower $(\theta_-)$ interfaces are separated by 
the stripe domain, it is reasonable to neglect the back reaction caused by perturbing 
one interface on the shape of the other. For definiteness, we focus on the lower interface 
$\theta_-$ and assume that it is perturbed according to
\begin{equation}
 \theta_-(\varphi) = \theta_{-;0} + \delta \theta_-(\varphi),
\end{equation}
where $\theta_{-;0}$ is the average value of $\theta_-(\varphi)$, while 
$\delta \theta_{-}(\varphi)$ is a small position-dependent fluctuation, which 
admits the following Fourier decomposition:
\begin{equation}
 \delta \theta_-(\varphi) = \frac{\delta}{\pi} \sum_{n = 1}^\infty A_n \cos (n \varphi + \varphi_{n;0}),
\end{equation}
where $\delta > 0$ is an overall positive infinitesimal factor, while 
the coefficients $A_n = O(1)$ are not necessarily small. 
We assume that $\theta_{-;0}$ changes under the perturbation such that the domain area,
\begin{align}
 \Delta A =& r R \int_0^{2\pi} d\varphi \int_{\theta_-(\varphi)}^{\theta_+} d\theta\, (1 + a \cos\theta) 
 \nonumber\\
 =& 2\pi r R \left[\theta_+ - \theta_{-;0} + 
 a \sin\theta_+ - a \sin\theta_{-;0} \left(1 - \frac{\delta^2}{4\pi^2} \sum_{n = 1}^\infty A_n^2\right) 
 + O(\delta^3)\right],
\end{align}
remains constant. Keeping in mind that the back reaction on $\theta_+$ is negligible,
imposing $d \Delta A / d\delta = 0$ implies that
\begin{equation}
 \frac{d \theta_{-;0}}{d\delta} = \frac{\delta}{2\pi^2} \frac{a \sin\theta_{-;0}}{1 + a \cos\theta_{-;0}}
 \sum_{n = 1}^\infty A_n^2.
 \label{eq:instab_dthdd}
\end{equation}
Let us now compute the length $\ell_-$ of the lower interface
\begin{align}
 \ell_- =& \int_{0}^{2\pi} d\varphi 
 \sqrt{R^2 [1 + a \cos \theta(\varphi)]^2 + r^2 \left(\frac{d \theta_-}{d\varphi}\right)^2}\nonumber\\
 =& 2\pi R(1 + a \cos\theta_{-;0}) + 
 \frac{r \delta^2}{2\pi} \sum_{n=1}^\infty \left( 
 \frac{a n^2}{1 + a \cos\theta_{-;0}} - \cos\theta_{-;0}\right) A_n^2 + O(\delta^3).
\end{align}
Taking the differential of $\ell_-$ with respect to $\delta$ while imposing 
Eq.~\eqref{eq:instab_dthdd} yields
\begin{equation}
 \frac{d\ell_-}{d\delta} = \frac{r \delta}{\pi} \sum_{n = 1}^\infty 
 \frac{a(n^2 - 1) - \cos\theta_{-;0}}{1 + a\cos\theta_{-;0}} A_n^2.
 \label{eq:instab_n}
\end{equation}
The first term in the numerator has a stabilising effect, acting only on
the Fourier modes with $n > 1$. The second term can be related to the Gaussian 
curvature $K$, given by
\begin{equation}
 K \equiv K(\theta) = \frac{\cos\theta}{rR (1 + a \cos\theta)},
 \label{eq:gauss_curv}
\end{equation}
The $n = 1$ mode becomes unstable when $K > 0$ and $\ell_-$ decreases 
when $\delta$ is increased, i.e. in the region of the torus given by 
$-\frac{\pi}{2} < \theta_{-;0} < \frac{\pi}{2}$. The higher-order modes 
become unstable deeper in the region of positive $K$, i.e. when 
$\cos \theta_{-;0}$ exceeds $a(n^2 - 1)$. An equivalent analysis can be performed 
for the upper interface, located at $\theta_+ = \theta_c + \Delta \theta / 2$.
Focussing now only on the onset of instability 
due to the first mode, Eq.~\eqref{eq:instab_n} can be written as
\begin{equation}
 \left(\frac{d\ell_\pm}{d\delta}\right)_{n=1} = 
 - \frac{r^2 R A_1^2 \delta}{\pi} K(\theta_{\pm;0}).
 \label{eq:instab_n1}
\end{equation}
Eq.~\eqref{eq:instab_n1} indicates that the upper and lower interfaces can become unstable simultaneously only 
when the stripe is completely contained in the region where $K > 0$
(i.e. on the outer side of the torus).

We now discuss the stability of stripes with equilibrium position characterised by 
Eq.~\eqref{eq:minimaring}, as derived in the previous subsection. Essentially, instability occurs 
when $\theta_{-}^{eq} = \theta_c^{eq} - \frac{\Delta \theta_{eq}}{2} < \frac{\pi}{2}$ or 
$\theta_{+}^{eq} = \theta_c^{eq} + \frac{\Delta \theta_{eq}}{2} > \frac{3\pi}{2}$.
The subcritical stripes (having $\Delta A < \Delta A_{\rm{crit}}$, 
which stabilise at $\pi$) do not suffer from the instability described by Eq.~\eqref{eq:instab_n1}. For the critical stripe,  as described by Eq.~\eqref{eq:stripe_Dth_crit},
it can be seen that the instability condition on both the upper and 
lower interfaces reduces to ${\rm arccos}\, a > \frac{\pi}{2}$, which is 
marginally satisfied only in the case $a \rightarrow 0$.  Next, supercritical stripes (having $\Delta A > \Delta A_{\rm{crit}}$, which stabilise away from $\pi$) are stable only when
\begin{gather}
\theta_{c,t}^{\rm inst} < \theta_{c}^{eq} < \theta^{\rm inst}_{c,b}, \qquad  \Delta\theta_{eq} < \Delta\theta_{\rm inst},  \nonumber \\ 
\theta_{c,t}^{\rm inst}  = \pi - {\rm arctan}\,a, \qquad \theta_{c,b}^{\rm inst} = \pi + {\rm arctan}\,a, \qquad \Delta\theta_{\rm inst} = 2\, {\rm arctan}\, \frac{1}{a}. \label{eq:inst_thetac} 
\end{gather}
The interface length and area of the stripe, corresponding to the instability condition in Eq.~\eqref{eq:inst_thetac}, are given by
\begin{equation}
 \ell_{\rm inst} = \frac{4\pi R}{1 + a^2}, \qquad
 \Delta A_{\rm inst} = 4\pi r R\left(\arctan\frac{1}{a} - 
 \frac{a}{1 + a^2}\right).
 \label{eq:inst_params}
\end{equation}

\begin{figure*}
\begin{center}
\begin{tabular}{c}
 \includegraphics[width=0.47\linewidth]{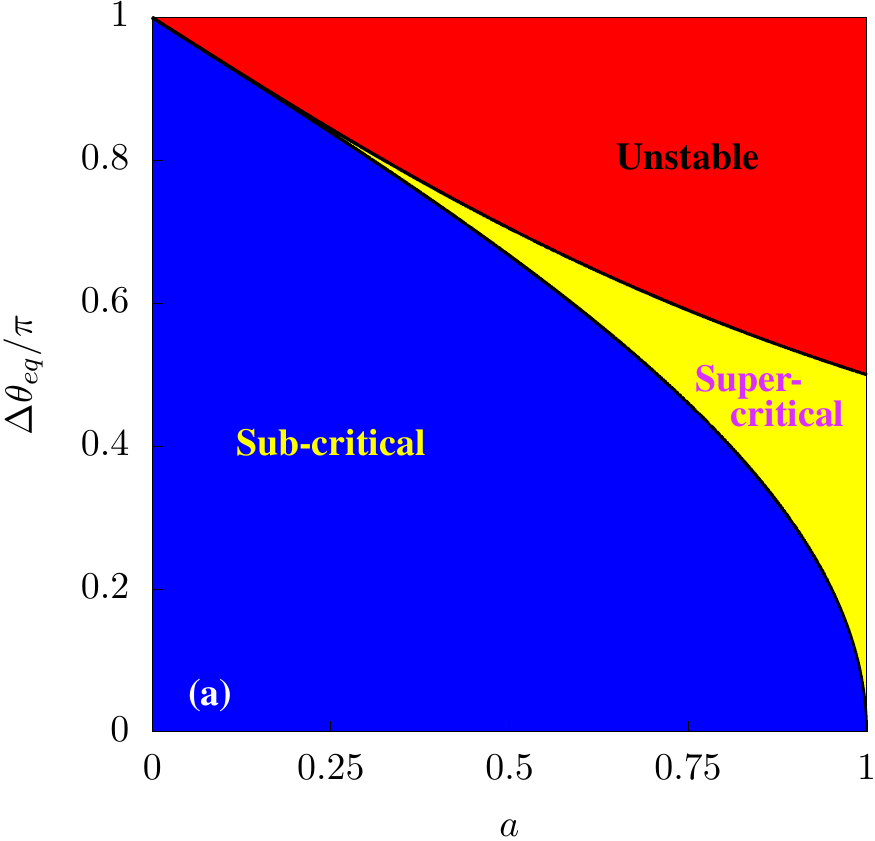} 
 \includegraphics[width=0.49\linewidth]{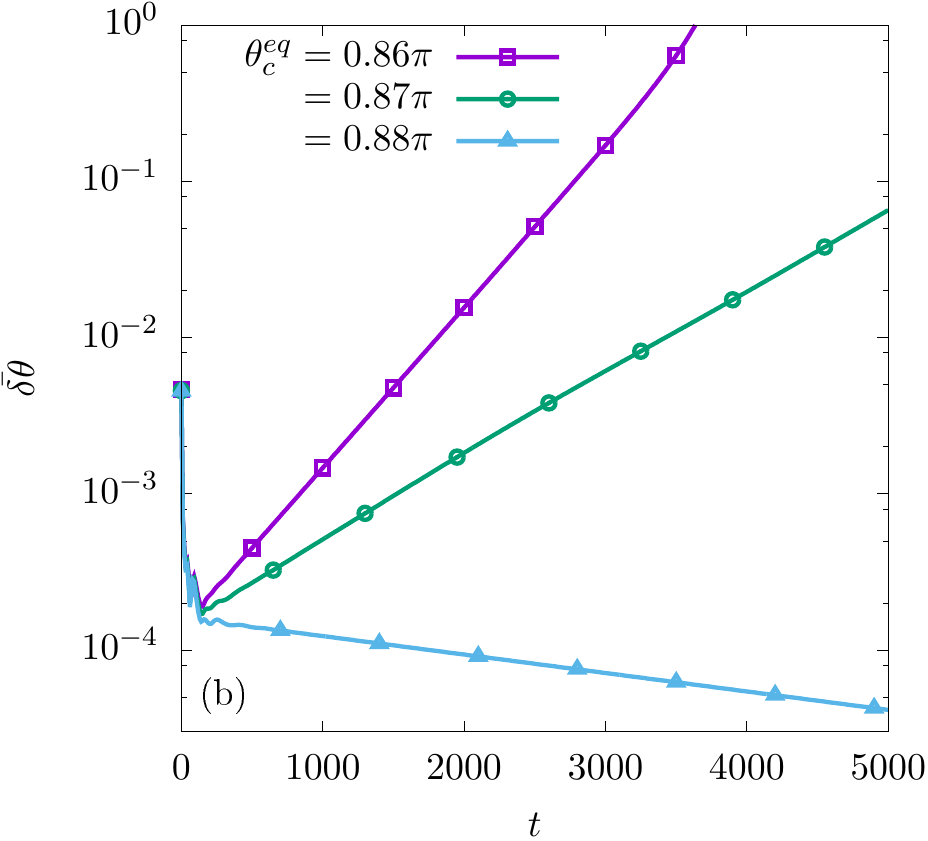} 
\end{tabular}
\end{center}
\caption{(a) Phase diagram showing the regions 
where the stripe is unstable (top right), as given by 
Eq.~\eqref{eq:inst_thetac}, and where it is stable (or at least metastable). The latter region is further divided into two subregions, where the stripes are subcritical ($\Delta A < \Delta A_{\rm crit}$, bottom left) and supercritical ($\Delta A > \Delta A_{\rm crit}$, central right). (b) Time evolution of the root mean square
of the perturbations on the lower interfaces 
($\theta_- = \theta_c - \Delta \theta / 2$), as given by Eq.~\eqref{eq:inst_rms}, for 
stripes centred on $\theta_c^{eq} = 0.86\pi$, $0.87\pi$ and $0.88\pi$, 
on the torus with $a =0.4$.
\label{fig:inst_diagram}}
\end{figure*}

Figure~\ref{fig:inst_diagram}(a) shows a separation of the 
$(a, \Delta \theta_{eq})$ plane into $3$ regions: The 
subcritical region (where the stripes stabilise at $\pi$), shown in blue 
in the bottom left part of the plot; the super-critical region 
(where the stripes stabilise away from $\pi$), shown with yellow; and 
the unstable region (where stripes destabilise under small perturbations),
shown with red in the top right part of the plot. The line separating the red and 
yellow regions is defined by Eq.~\eqref{eq:inst_thetac}, while the line between 
the yellow and blue regions is given by Eq.~\eqref{eq:stripe_Dth_crit}.

To verify the validity of Eq.~\eqref{eq:inst_thetac}, we perform some numerical experiments 
on the torus with $R = 1$ and $r = 0.4$ ($a = 0.4$). The stripes become unstable when 
$\theta_c < \theta_c^{\rm inst} \simeq 0.8788\pi$, therefore we consider three stripes 
initialised at $\theta_c^{eq} = 0.86\pi$, $0.87\pi$ and $0.88\pi$, with their 
corresponding equilibrium widths $\Delta \theta_{eq}=\{0.764,0.761,0.757\}\pi$.
The order parameter is initialised with the hyperbolic tangent profile given in
Eq.~\eqref{eq:stripe_tanh_lap}, but the stripe width 
$\Delta \theta(\varphi) = \Delta \theta_0 + \varepsilon(\varphi)$ is 
allowed to vary with respect to the $\varphi$ coordinate. The perturbation $\varepsilon(\varphi)$
is taken as a random distribution with amplitude $0.001\pi$.
The system is discretised using $N_\theta = 192 $ and $N_\varphi = 288$ equidistant 
values for the $\theta$ and $\varphi$ coordinates.
After generating the values $\varepsilon_q = \varepsilon(\varphi_q)$, where 
$1 \le q \le N_\varphi$, the base width
$\Delta \theta_0$ is computed such that the perturbed stripe has the area $\Delta A$
corresponding to the axisymmetric stripe with the given values for $\theta_c^{eq}$
and $\Delta \theta_{eq}$.
The numerical simulations indicate that the perturbations on the upper interface,
located at $\theta_+ = \theta_c$, are quickly suppressed for all stripes, confirming 
the prediction of the analysis presented above. On the lower interface ($\theta_-$),
we quantify the growth of the perturbation at the level 
of the root-mean-square deviation, computed via
\begin{equation}
 (\bar{\delta\theta^2})^{1/2} = 
 \sqrt{\frac{1}{N_\varphi} \sum^{N_\varphi}_{q=1} |\theta_-^q-\theta_-^{\rm{avg}}|^2},
 \label{eq:inst_rms}
\end{equation}
where $\theta_-^{\rm{avg}}$ is the average position of the lower interface. 
The results are presented in figure~\ref{fig:inst_diagram}(b). It can be seen that, in 
the case of the stripes located at $0.86\pi$ and $0.87\pi$, the perturbations grow exponentially 
with time, while in the case of the stripe centred on $0.88\pi$, the perturbations are suppressed, 
confirming that the onset of the instability is given by Eq.~\eqref{eq:inst_thetac}.

\begin{figure}
\begin{tabular}{cccc}
\multicolumn{4}{c}{(a) $\theta_c^{eq}=0.86\pi$}\\
 & & & \\
\hspace{15pt}(i) $t=0$& (ii) $t=2300$& (iii) $t=2800$& (iv) $t=5000$\\
    \includegraphics[width=0.26825\linewidth]{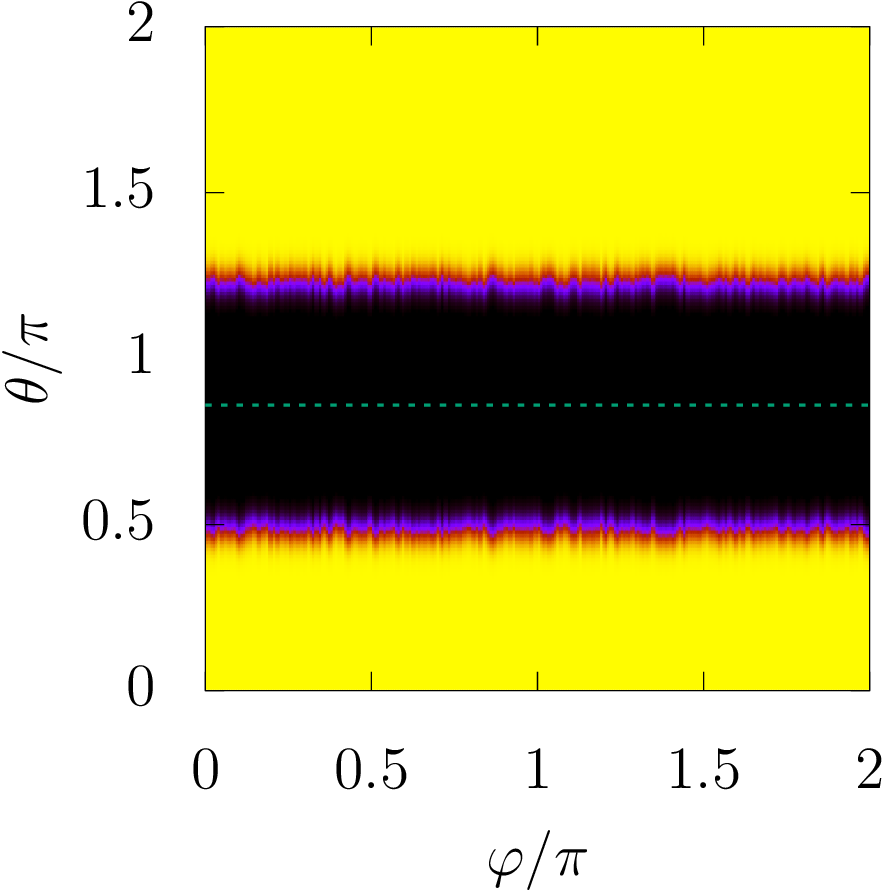}&
    \includegraphics[width=0.21\linewidth]{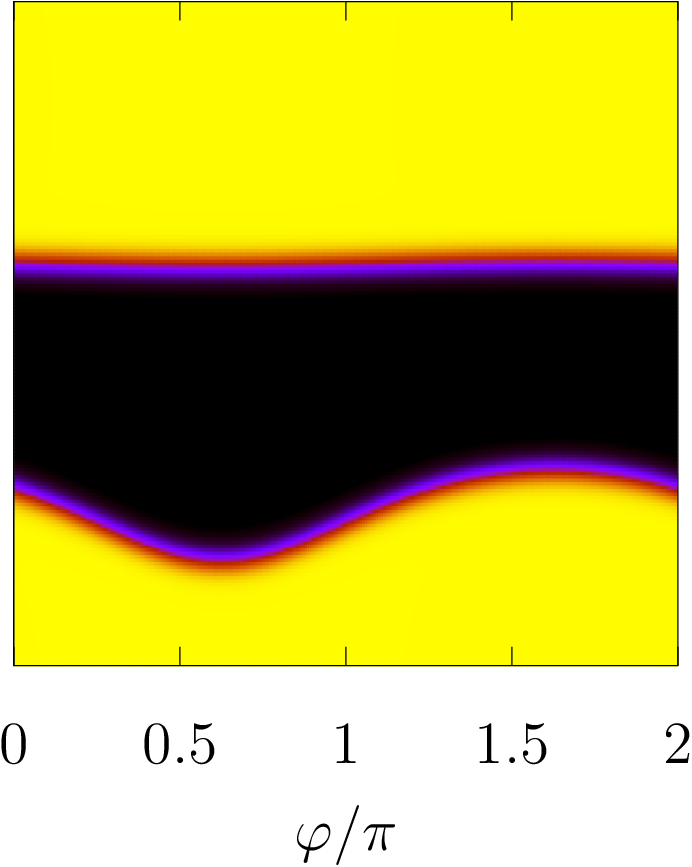}&
    \includegraphics[width=0.21\linewidth]{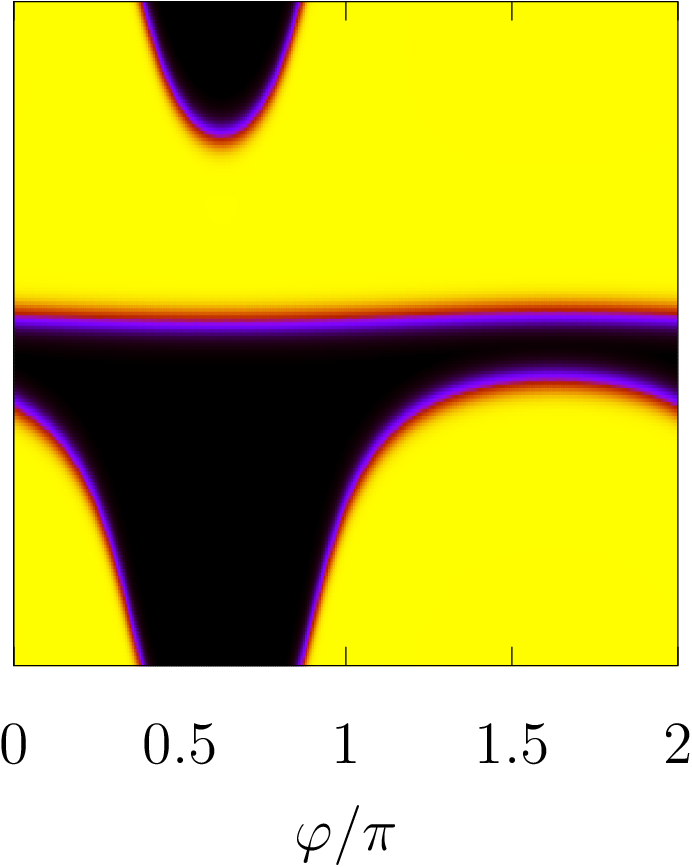}&
    \includegraphics[width=0.21\linewidth]{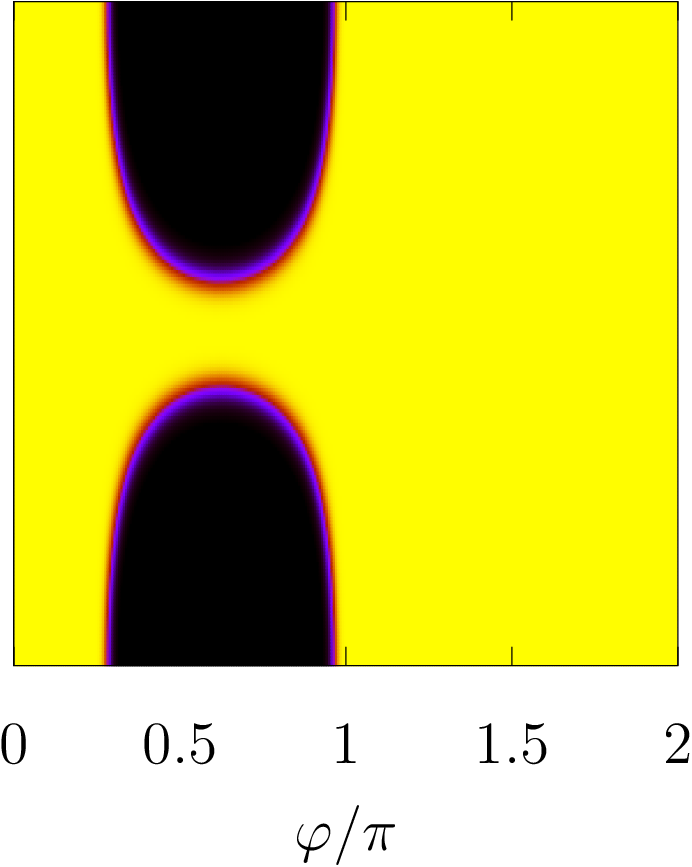}\\
 & & & \\
 \hspace{15pt}    \includegraphics[width=0.22\linewidth]{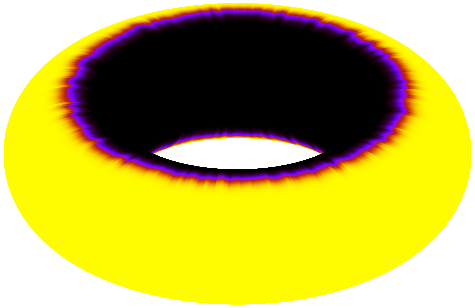}&
    \includegraphics[width=0.22\linewidth]{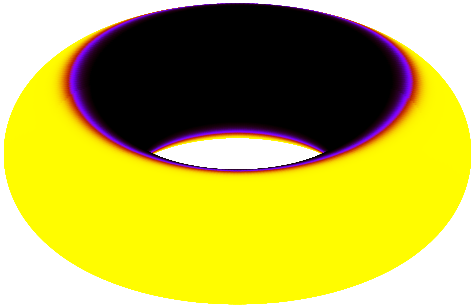}&
    \includegraphics[width=0.22\linewidth]{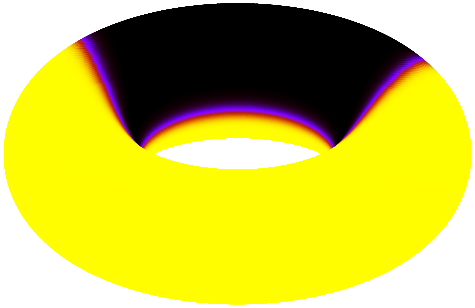}&
    \includegraphics[width=0.22\linewidth]{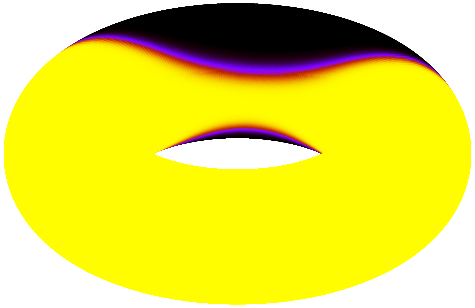}\\
\multicolumn{4}{c}{------------------------------------------------------------------------------------------------------------------------}\\
     \multicolumn{4}{c}{(b) $\theta_c^{eq}=0.65\pi$}\\
 & & & \\
 \hspace{15pt} (i) $t=0$ & (ii) $t=800$ & (iii) $t=1100$ & (iv) $t=5000$\\
     \includegraphics[width=0.26825\linewidth]{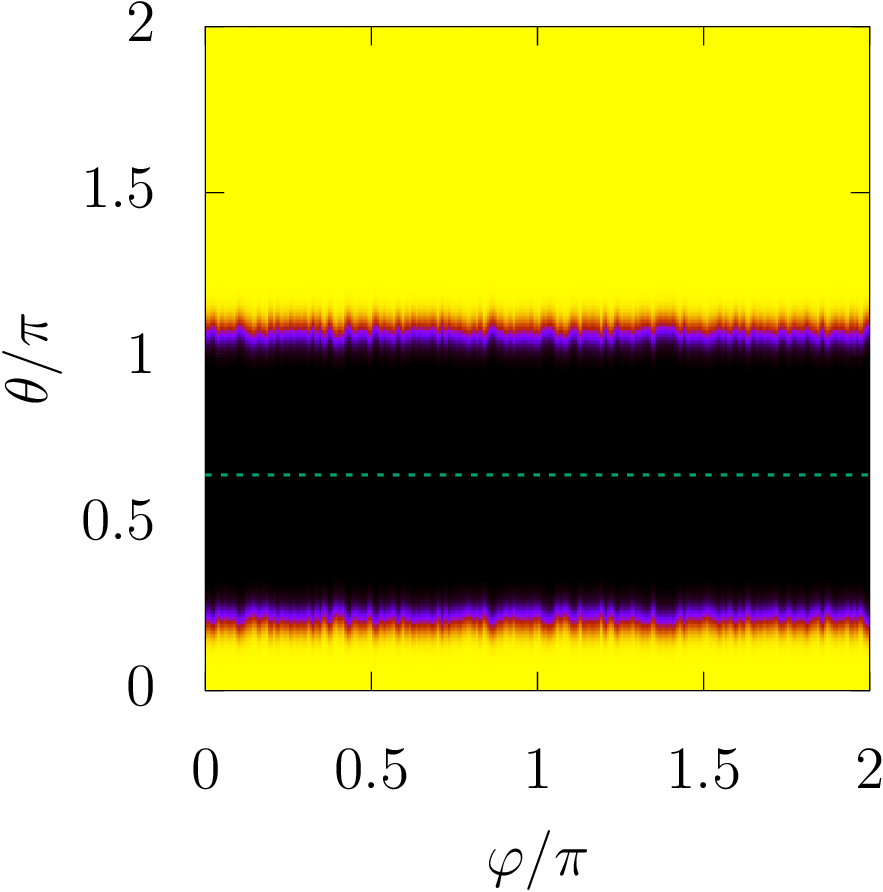}&
    \includegraphics[width=0.21\linewidth]{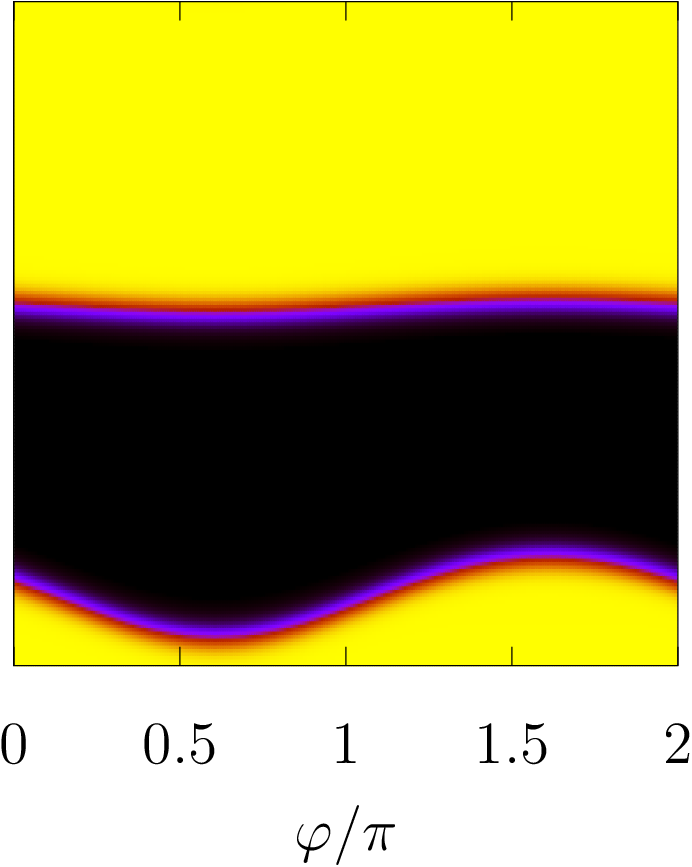}&
    \includegraphics[width=0.21\linewidth]{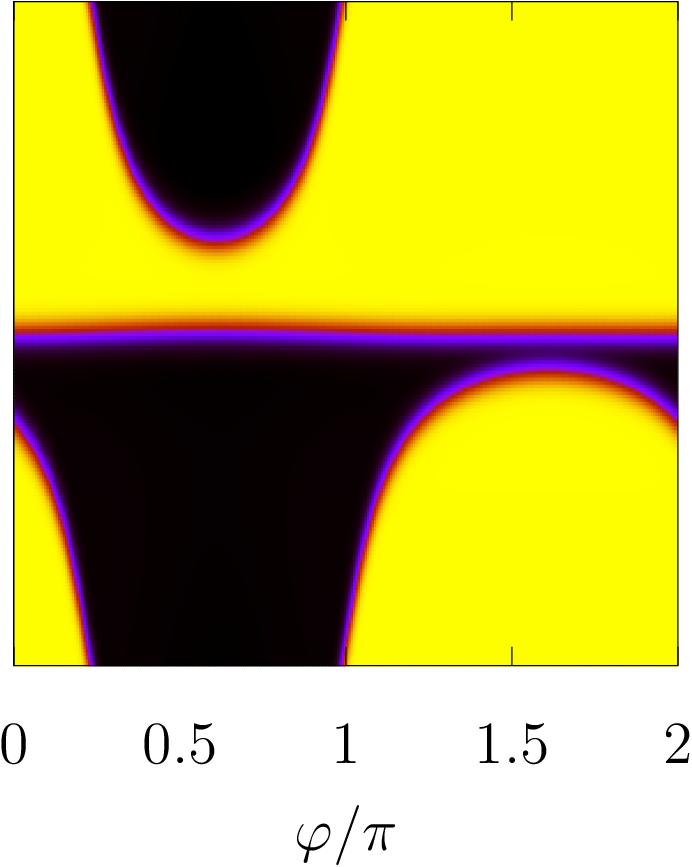}&
    \includegraphics[width=0.21\linewidth]{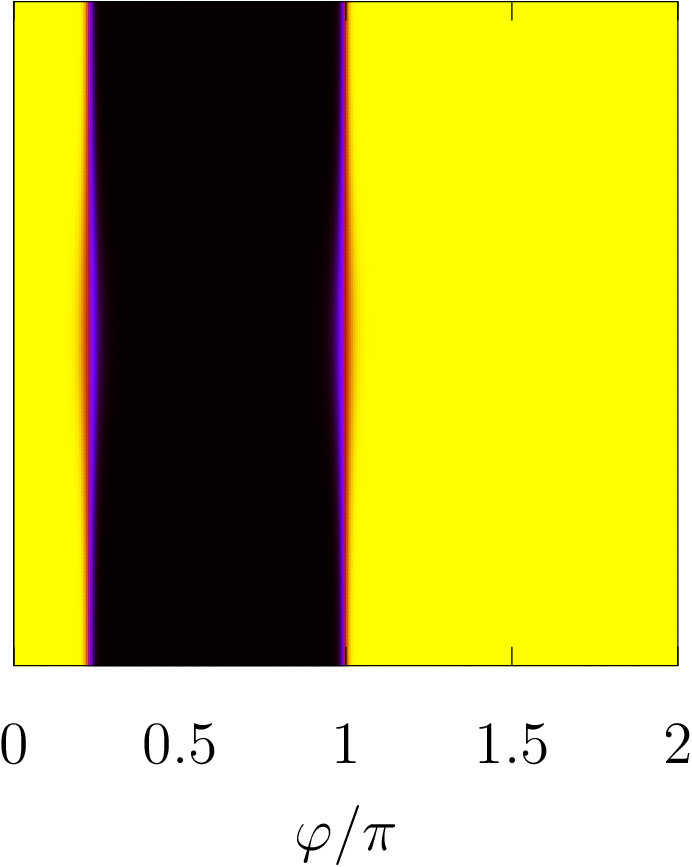}\\
 & & & \\
 \hspace{15pt}    \includegraphics[width=0.22\linewidth]{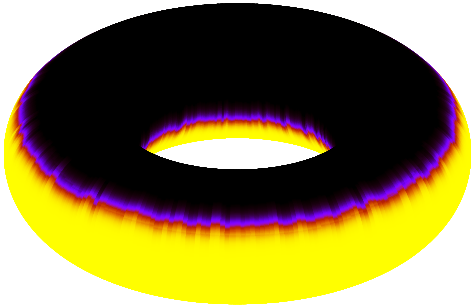}&
    \includegraphics[width=0.22\linewidth]{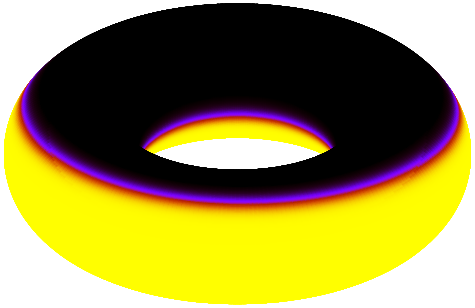}&
    \includegraphics[width=0.22\linewidth]{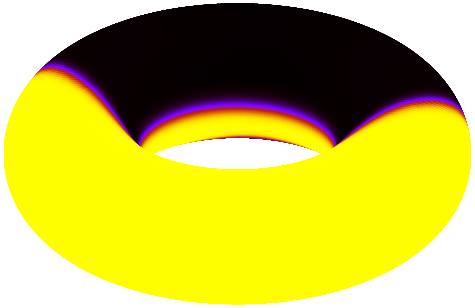}&
    \includegraphics[width=0.22\linewidth]{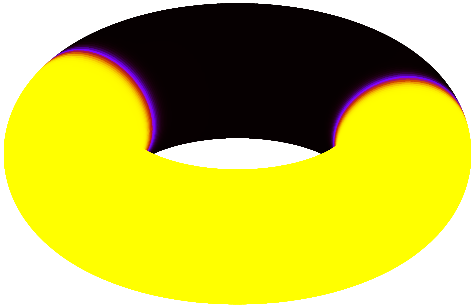}
   \end{tabular}
\caption{Time evolution of two unstable stripes on the torus with $a = 0.4$. 
(a) The stripe is initialised at $\theta_c^{eq}=0.86\pi$ and leads after breaking 
to a droplet configuration equilibrated on the outer side of the torus (animation 
available as Movie~1 on the publisher's website).
(b) The stripe is initialised at $\theta_c^{eq}=0.65\pi$ and merges 
on the poloidal direction after breaking 
to form a band configuration (animation 
available as Movie~2 on the publisher's website). 
In (a-b), column (i) shows the initial conditions,
with perturbations along the $\varphi$ direction
on both interfaces. 
Columns (ii) and (iii) contain intermediate snapshots of the configurations. Column (iv) shows the final equilibrium configurations.
\label{fig:inst_break}}
\end{figure}

The instability invariably causes the stripe to break. The final configuration 
must correspond to a smaller value of the total free energy. Figure~\ref{fig:inst_break} presents snapshots of the evolution of two unstable fluid stripes, initialised at 
(a) $\theta_c = 0.86\pi$ and (b) $\theta_c = 0.65\pi$, on the torus with $a = 0.4$.
For convenience, the order 
parameter $\phi$ is shown using a colour map in a two-dimensional representation (top rows) and in the three-dimensional representation, on the torus (bottom rows).
The stripe widths are set to the equilibrium values, $\Delta \theta = 0.764236\pi$
and $0.883748\pi$, respectively, while the interfaces are perturbed as described 
in the previous paragraphs, with initial perturbation amplitude $\varepsilon = 0.02\pi$.
The initial states are shown in panels (ai,bi). Panels (aii,bii) and(aiii,biii) show intermediate stages in the development of the perturbations. From figure~\ref{fig:inst_break} (aii,bii), it can be seen that the perturbations are dominated by 
the first Fourier mode, corresponding to $\cos(\varphi + \varphi_{1;0})$,
thus confirming that the higher-order modes are suppressed compared to the 
first order one.
Panels (aiii,biii) depict the configurations just before the stripes break.
Finally, column (iv) shows the equilibrium configurations, which are a drop 
for the smaller stripe and a band, wrapping around the torus along the 
$\theta$ coordinate, for the larger one. 
Animations of the time development of the instability for the 2 cases 
shown in figure~\ref{fig:inst_break} are available as Movie~1 and Movie~2 
on the publisher's website \citep{SM:JFM}.

The fact that the stripe configurations lead to droplets or bands
indicates that these latter configurations correspond to lower values 
of the free energy. Under the assumption that the free energy is related to 
the interface length\footnote{We note that, as revealed in Sec.~\ref{SM:sec:free},
the free energy for the stripe configurations is just $\Psi = \sigma \ell_{\rm total} + O(\xi^2)$.
This simple relation may not hold for more general domain shapes.},
we note that, according to Eq.~\eqref{eq:stripe_ltot}, the interface length for the stripe configuration can vary as the stripe area grows between 
$\ell_{\rm stripe}^{\rm min} = 4\pi (R - r)$ 
for infinitesimally small stripes (the two interfaces are at $\theta = \pi$) and 
$\ell_{\rm stripe}^{\rm max} = 4\pi R$ for the largest stripe, covering
half of the torus and having the interfaces at $\theta = 0$ and $\pi$.

For sufficiently small domain areas, the interface length of a droplet configuration
grows with the domain area roughly as 
$\ell_{\rm drop} \sim \sqrt{\Delta A}$, vanishing as $\Delta A \rightarrow 0$. 
Thus, at sufficiently small domain areas, the droplet is energetically preferred.

The band configuration has a domain area-independent interface length,
given by the two boundary circles located at constant $\varphi$, 
$\ell_{\rm band} = 4\pi r$. For sufficiently large domain areas, $\ell_{\rm band}$ 
will be smaller than $\ell_{\rm stripe}$, since 
$\ell_{\rm stripe}^{\rm max} = 4\pi R  > 4\pi r$. In fact, the band configuration can be 
energetically preferable to the stripe configurations for any domain size when 
$\ell_{\rm band} < \ell_{\rm stripe}^{\rm min}$, which is always satisfied when 
$a < \frac{1}{2}$.

A more comprehensive analysis of the energy landscape, indicating 
which configurations correspond to the minimum of the free energy, would require a 
detailed study of the droplet and band configurations, which is beyond the 
scope of this work. However, based on the discussion in the previous paragraph, 
it is safe to conclude that there are domains of the subcritical and supercritical 
regions shown in figure~\ref{fig:inst_diagram}(a) where the stripe configurations are 
actually only metastable. 

\subsection{Laplace pressure}\label{sec:laplace:bench}

We now seek for an expression for the pressure difference $\Delta P$ between the 
two fluid components. For a small increase $\delta \Delta A$ of the stripe 
area, let $\delta \ell_{\rm total}$ be the increase in the interface length.
These two quantities can be related through the equation
\begin{equation}
 \Delta P \delta \Delta A = \sigma \delta \ell_{\rm total}.
\end{equation}
The variations $\delta \Delta A$ and $\delta \ell_{\rm total}$ can 
be computed using Eqs.~\eqref{eq:stripe_area} and \eqref{eq:stripe_ltot}:
\begin{equation}
 \delta \Delta A = 4 \pi r R \left(1 + a \cos\theta_c 
 \cos \frac{\Delta \theta}{2}\right) \delta \frac{\Delta \theta}{2}, \qquad 
 \delta \ell_{\rm total} = -4\pi r \cos \theta_c \sin\frac{\Delta \theta}{2} 
 \delta \frac{\Delta \theta}{2}.
\end{equation}
Thus, the pressure difference $\Delta P$ can be written as
\begin{equation}
 \Delta P = -\frac{\sigma}{R} \frac{\cos\theta_c \sin(\Delta \theta / 2)}
 {1 + a \cos\theta_c \cos(\Delta \theta / 2)}.\label{eq:stripe_laplace_gen}
\end{equation}
The above expression is valid regardless of where the stripe is 
positioned. 

Assuming that the stripe is already in its equilibrium position, 
Eq.~\eqref{eq:stripe_laplace_gen} reduces to
\begin{equation}
 \Delta P = 
 \begin{cases}
  {\displaystyle \frac{\sigma}{R} \frac{\sin(\Delta \theta_{eq} / 2)}
  {1 - a \cos(\Delta \theta_{eq} / 2)}}, & \Delta A < \Delta A_{\rm crit} \text{ and }
  \theta_c^{eq}= \pi,\\
  {\displaystyle \frac{\sigma}{r} \cot \frac{\Delta \theta_{eq}}{2}}, 
  & \Delta A > \Delta A_{\rm crit} \text{ and } 
  a \cos \theta_c^{eq} + \cos\tfrac{\Delta \theta_{eq}}{2} = 0.
 \end{cases}\label{eq:stripe_laplace_eq}
\end{equation}
Equation~\eqref{eq:stripe_laplace_eq} loses relevance 
in the domain of stripe instability discussed in Subsec.~\ref{sec:laplace:inst}, 
unless strict axisymmetry is enforced. On the instability line, where 
Eqs.~\eqref{eq:inst_thetac} and \eqref{eq:inst_params} hold, we find
\begin{equation}
 \Delta P_{\rm inst} = \frac{\sigma}{R},
 \label{eq:stripe_laplace_DP_inst}
\end{equation}
which, remarkably, is independent of $a$. 

\begin{figure}
 \begin{center}
 \begin{tabular}{c}
  \includegraphics[width=0.48\columnwidth]{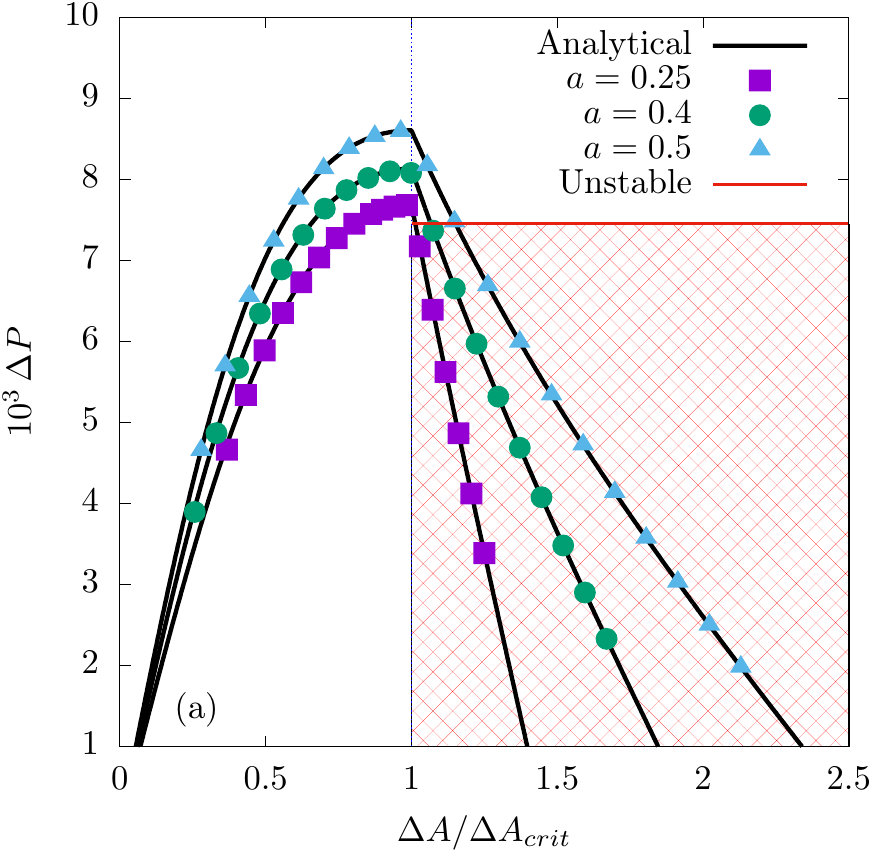}
\end{tabular}
  \end{center}
\caption{
Comparison of numerical results obtained for 
$a = 0.25$ (solid squares), $a=0.4$ (solid circles)
and $a=0.5$ (solid triangles) against 
the analytic formula \eqref{eq:stripe_laplace_eq}.
The system parameters are $\text{A} = 0.5$, $\kappa = 5 \times 10^{-4}$ and $M = \tau = 2.5 \times 10^{-3}$.
The simulations are performed using $N_\theta=\{200,320,400\}$ nodes along the 
$\theta$ direction for $r=\{0.5,0.8,1.0\}$, while the time step was set to 
$\delta t = 2 \times 10^{-3}$. 
The shaded region corresponds to the stripes which are unstable when the axisymmetric assumption is lifted.
\label{fig:stripe_laplace}}
\end{figure}

We now propose the benchmark test concerning stripe configurations, 
consisting of the generalisation of the Laplace-Young pressure test. 
An alternative derivation of Eq.~\eqref{eq:stripe_laplace_gen} in the 
context of the Cahn-Hilliard model considered in this paper is provided 
in Sec.~\ref{SM:sec:phi0} of the supplementary material. 
It is interesting to note that the 
Laplace pressure is related to a non-vanishing value of the chemical 
potential when the stripe is in equilibrium. This in turn induces an 
offset in the order parameter, denoted using $\phi_0$ and 
computed in Eq.~\eqref{SM:eq:stripe_phi0} of the supplementary material.

We perform a series of numerical simulations in the absence of hydrodynamics at three values of $a$, 
namely $a=0.25$, $0.4$ and $0.5$, by fixing the outer radius to $R=2$ 
and setting the inner radius to $r=\{0.5,0.8,1.0\}$, while keeping 
$M = 2.5 \times 10^{-3}$, $\kappa = 5 \times 10^{-4}$ and $\text{A} = 0.5$ 
unchanged. We consider stripes of various areas $\Delta A$. For each 
value of $\Delta A$, the equilibrium position $\theta_c^{eq}$ 
is computed and the stripe is initialised using 
a hyperbolic tangent profile, 
\begin{equation}
 \phi = \phi_0 + \tanh \zeta, \qquad 
 \zeta = \frac{r}{\xi_0 \sqrt{2}}\left(
 |\widetilde{\theta - \theta_c}| - \frac{\Delta \theta}{2}\right), 
 \label{eq:stripe_tanh_lap}
\end{equation}
and centred on $\theta_c = \theta_c^{eq} - \delta \theta$,
with $\delta \theta = 0.05 \pi$.
The notation $\widetilde{\theta - \theta_c}$ indicates that the angular difference $\theta - \theta_c$ takes values between $-\pi$ and $\pi$. The initial width 
$\Delta \theta$ is obtained by numerically solving 
Eq.~\eqref{eq:stripe_area} for fixed $\Delta A$ and
$\theta_c$. The value of $\phi_0$ corresponding to the 
initial stripe centre $\theta_c$ and initial width $\Delta \theta$ 
is derived in Sec.~\eqref{SM:sec:phi0} of the supplementary material.
It is given by
\begin{equation}
 \phi_0 = \frac{\xi_0}{3R \sqrt{2}} 
 \frac{\cos\theta_c \sin(\Delta \theta / 2)}
 {1 + a \cos\theta_c \cos (\Delta \theta/2)}.
 \label{eq:stripe_phi0}
\end{equation}
After initialisation, the stripes slowly migrate towards 
the equilibrium positions, as discussed in the Sec.~\ref{SM:sec:noh} 
of the supplementary material. 
In order to reach the stationary state, we performed 
$4 \times 10^9$ iterations at $\delta t = 0.002$. 
After the stationary state was reached, we measured the pressure 
$ P_{\rm binary} = \text{A}(-\frac{1}{2} \phi^2 + 
\frac{3}{4} \phi^4)$ in the interior and exterior of the stripe and 
computed the difference $\Delta P$ between these two values. 
The results are shown using dotted lines and symbols in figure~\ref{fig:stripe_laplace}.
The shaded region indicates the region where the stripes become unstable once the 
axisymmetric assumption is removed in the model. It is bounded from above by the 
pressure difference value $\Delta P_{\rm inst}$ on the instability line, given in 
Eq.~\eqref{eq:stripe_laplace_DP_inst}.
We observe an excellent agreement with the analytic result, Eq.~\eqref{eq:stripe_laplace_eq}, 
which is shown using solid lines. 

It is worth noting that the second-order phase 
transition observed in the stripe equilibrium positions when 
$\Delta A = \Delta A_{\rm crit}$ is also visible in the dependence of 
$\Delta P$ on $\Delta A$ in figure~\ref{fig:stripe_laplace}.
Its non-monotonic behaviour can be understood as follows.
For infinitesimally small stripes, the torus curvature is negligible and 
no pressure difference can be seen across the interface, as is also 
the case for the Cartesian (flat space) geometry. 
As the stripe width $\Delta \theta$ increases, $\Delta P$ also increases.
In general, a turning point in the Laplace pressure can be expected. This is because 
the pressure difference vanishes for infinitesimal stripes ($\Delta \theta \rightarrow 0$), 
as well as in the opposite case, when the stripe occupies the top or bottom halves of the 
torus ($\Delta \theta \rightarrow \pi$). In the latter case, 
the conjugate domain can be obtained from the stripe by employing a symmetry transformation, $z \leftrightarrow -z$, which also changes the torus into itself. Thus, the configurations corresponding to the stripe and its conjugate are perfectly equivalent and one can expect there to be no pressure difference across the interface. 
When the equilibrium position of the stripe is always centred on 
$\theta_c^{eq} = \pi$, a smooth dependence of $\Delta P$ on $\Delta \theta$ can be 
expected. However, the phase transition at 
$\Delta A = \Delta A_{\rm crit}$ which causes the stripe to detach form $\theta_c = \pi$ 
leads to the sharp change observed in figure~\ref{fig:stripe_laplace}.

\section{Evolution of fluid stripes in a Cahn-Hilliard multicomponent fluid}
\label{sec:CH}

In this section we consider the dynamics of the axisymmetric 
fluid stripes discussed in Sec. \ref{sec:laplace}.
Here, we focus on the case where the Cahn-Hilliard equation is fully coupled with hydrodynamics,
when the stripes undergo underdamped oscillatory motion towards their equilibrium positions. 
The relaxation dynamics in the absence of hydrodynamics is discussed in detail in 
Sec.~\ref{SM:sec:noh} of the supplementary material, where we are able to obtain a semi-analytical description of how the stripes relax exponentially to their equilibrium positions. From the perspective of benchmarking Navier-Stokes 
solver on non-uniform curved surfaces, this section is a culmination of the various ingredients 
developed in Sec. \ref{sec:inv}, \ref{sec:damp} and \ref{sec:laplace}. 
In particular, we find that the dynamics is governed to leading order by the 
zeroth-order mode of the velocity derived in Eq.~\eqref{eq:tor_inv_vinc}, which 
corresponds to incompressible flow.
This section is structured as follows. The general solution for the underdamped oscillatory motion is presented in Subsec. \ref{sec:CH:general}. 
A benchmark test is proposed in Subsec. \ref{sec:CH:benchmark}.

\subsection{General solution} \label{sec:CH:general}

To derive the stripe dynamics, our starting point is the Cauchy equation in 
the linearised regime
\begin{multline}
 \frac{\partial u^\htheta}{\partial t} = 
 -\frac{k_B T}{m r} \frac{\partial \delta \rho}{\partial \theta} 
 + \frac{\nu}{r^2 (1 + a\cos\theta)^2} \frac{\partial}{\partial\theta} 
 \left[(1 + a \cos\theta)^3 \frac{\partial}{\partial \theta} 
 \left(\frac{u^\htheta}{1 + a\cos\theta}\right)\right] \\
 +\frac{\nu_{v}}{r^2} \frac{\partial}{\partial \theta}
 \left\{\frac{\partial_\theta[u^\htheta(1 + a \cos\theta)]}
 {1 + a \cos\theta}\right\} - 
 \frac{\phi}{\rho_0 r} \frac{\partial \mu}{\partial \theta}.
 \label{eq:stripe_hydro_cauchy}
\end{multline}
As in Sec.~\ref{sec:inv} and \ref{sec:damp}, we will employ the 
decomposition written in Eq.~\eqref{eq:inv_utheta_series} for 
$u^\htheta$. Moreover, we will also take advantage of the fact that 
the higher-order terms $U_{c,n}$ and $U_{s,n}$ ($n > 0$) are damped at a
significantly higher rate than the fundamental term $U_0$. 
Then, in order to track the evolution of $U_0$, we multiply
Eq.~\eqref{eq:stripe_hydro_cauchy} with $f_0 / 2\pi = (1-a^2)^{1/4} / 2\pi$, 
and integrate it over $\theta$ between $0$ and $2\pi$, to obtain
\begin{equation}
 \dot{U}_0 + \frac{2\nu a^2}{r^2(1 -a^2)} U_0 +
 \frac{(1-a^2)^{1/4}}{2\pi \rho_0 r} I_\mu \simeq 0, \qquad 
 I_\mu = \int_0^{2\pi} d\theta\, \phi 
 \frac{\partial \mu}{\partial \theta},
 \label{eq:stripe_hydro_cauchyU}
\end{equation}
where the $\simeq$ sign indicates that the nonlinear terms, as 
well as the components of $u^\htheta$ with $n > 0$, have been neglected.
Employing integration by parts, $I_\mu$ can be written as
\begin{equation}
 I_\mu = \int_0^{2\pi} d\theta \left[\text{A} \frac{\partial}{\partial \theta} 
 \left(\frac{\phi^2}{2} - \frac{\phi^4}{4}\right) + 
 \frac{\kappa}{r^2}\frac{\partial \phi}{\partial \theta} 
 \frac{\partial^2 \phi}{\partial \theta^2} - 
 \frac{\kappa}{r^2}
 \frac{a \sin\theta}{1 + a\cos\theta} \left(\frac{\partial \phi}{\partial \theta}\right)^2
 \right].
\end{equation}
The first and second terms above do not contribute to the integral.
To evaluate the integral of the third term, we assume that $\phi$ 
is approximately given by the hyperbolic tangent profile 
in Eq.~\eqref{eq:stripe_tanh_lap} and employ the procedure 
introduced in Sec.~\ref{SM:sec:phi0} of the supplementary material, which we briefly review here.
First, the integration variable is changed to $\vartheta = \theta - \theta_c$
and the integration domain is shifted to $-\pi < \vartheta < \pi$. 
Then, the flip $\vartheta \rightarrow -\vartheta$ is performed on the 
negative ($\vartheta < 0$) branch, yielding

\begin{equation}
 I_\mu = -\pi a \text{A} \int_0^\pi \frac{d\vartheta}{2\pi}\left[
 \frac{\sin (\theta_c + \vartheta)}{1 + a \cos(\theta_c + \vartheta)} + 
 \frac{\sin (\theta_c - \vartheta)}{1 + a \cos(\theta_c - \vartheta)}\right]
 \frac{1}{\cosh^4 \zeta}.
\end{equation}
Next, the integration variable is changed to $\zeta = r \varsigma / \xi_0 \sqrt{2}$, where 
$\varsigma = \vartheta - \Delta \theta / 2$, such that the integration domain is 
$- r \Delta \theta / \xi_0 \sqrt{8} < \zeta < r(2\pi - \Delta \theta) / \xi_0 \sqrt{8}$.
Noting that $\xi_0 \ll r \Delta \theta$, the integration domain can be extended 
to $(-\infty, \infty)$ and $I_\mu$ becomes
\begin{equation}
 I_\mu = -\frac{3\sigma}{4R} \int_{-\infty}^\infty 
 \frac{d\zeta}{\cosh^4 \zeta} \left[ 
 \frac{\sin\left(\theta_+ + \frac{\xi_0 \zeta\sqrt{2}}{r}\right)}
 {1 + a \cos \left(\theta_+ + \frac{\xi_0 \zeta\sqrt{2}}{r}\right)} + 
 \frac{\sin\left(\theta_- - \frac{\xi_0 \zeta\sqrt{2}}{r}\right)}
 {1 + a \cos \left(\theta_- - \frac{\xi_0 \zeta\sqrt{2}}{r}\right)}\right],
\end{equation}
where $\sigma = \sqrt{8 \kappa \text{A} / 9}$ is the line tension and 
$\theta_\pm = \theta_c \pm \Delta \theta / 2$.
We now consider an expansion of the integrand with respect to $\xi_0 \zeta / r$.
The dominant contribution comes from the zeroth-order term.
Since the integration domain is even with respect to $\zeta$, the first-order term of the expansion does not contribute. Considering that 
$\xi_0 / r \ll 1$, the higher-order terms can be discarded and 
$I_\mu$ can be approximated through
\begin{equation}
 I_\mu \simeq -\frac{2\sigma}{R} 
 \frac{\sin\theta_c [a \cos\theta_c + \cos(\Delta \theta/2)]}
 {(1 + a \cos\theta_-)(1 + a \cos\theta_+)}. \label{eq:Imuapprox}
\end{equation}
Substituting Eq.~\eqref{eq:Imuapprox} into Eq.~\eqref{eq:stripe_hydro_cauchyU}, we obtain
\begin{equation}
 \dot{U}_0 + 2\alpha_{\nu} U_0 -
 \frac{\sigma (1-a^2)^{1/4}}{\pi r R \rho_0} 
 \frac{\sin\theta_c [\cos(\Delta \theta / 2) + a \cos\theta_c]}
 {(1 + a \cos\theta_-) (1 + a \cos\theta_+)} = 0,
 \label{eq:stripe_hydro_cauchyU0}
\end{equation}
where the viscous damping coefficient $\alpha_\nu = \nu / (R^2 - r^2)$ is 
introduced in Eq.~\eqref{eq:alphanu}.

The relation between $U_0(t)$ and $\theta_c(t)$ can be established 
by evaluating the Cahn-Hilliard equation on the 
top and bottom interfaces $\theta = \theta_\pm$:
\begin{equation}
 \frac{r}{\xi_0\sqrt{2}} \left[-\dot{\theta}_+ + 
 \frac{u^\htheta_+}{r} \right] = M (\Delta \mu)_+, \qquad 
 \frac{r}{\xi_0\sqrt{2}} \left[\dot{\theta}_- -
 \frac{u^\htheta_-}{r} \right] = M (\Delta \mu)_- ,\label{eq:Uthetarel}
\end{equation}
where we have kept the leading-order term of the time derivative of $\phi$, assuming
it takes the hyperbolic tangent profile in Eq.~\eqref{eq:stripe_tanh_lap} and evaluating it on the
two interfaces
\begin{equation}
 \left.\frac{\partial \phi}{\partial t}\right\rfloor_{\theta_\pm}\simeq 
 \mp \frac{r \dot{\theta}_\pm}{\xi_0 \sqrt{2}}.
 \label{eq:stripe_noh_dphidt_thpm}
\end{equation}
Subtracting the two equations in Eq.~\eqref{eq:Uthetarel}, we obtain
\begin{equation}
 \frac{u_+^\htheta + u_-^\htheta}{2r} = 
 \dot{\theta}_c + \frac{\xi_0 M}{r \sqrt{2}} 
 \left[(\Delta \mu)_+ - (\Delta \mu)_-\right].
 \label{eq:stripe_hydro_CH}
\end{equation}
On the left hand side, the velocity profile can be approximated through 
its zeroth-order term, corresponding to the velocity profile of an 
incompressible flow
\begin{equation}
 u_+^\htheta \simeq \frac{U_0(t) f_0(\theta)}{1 + a \cos\theta_+}, \qquad 
 u_-^\htheta \simeq \frac{U_0(t) f_0(\theta)}{1 + a\cos\theta_-},
\end{equation}
as discussed in Eq.~\eqref{eq:tor_inv_vinc}. The function
$f_0(\theta) = (1-a^2)^{1/4}$ is introduced in Eq.~\eqref{eq:tor_f0}.
Thus, the left hand side of Eq.~\eqref{eq:stripe_hydro_CH} can be written as
\begin{equation}
 \frac{u_+^\htheta + u_-^\htheta}{2r} \simeq 
 \frac{U_0(t) (1 -a^2)^{1/4} [1 + a \cos\theta_c \cos (\Delta \theta / 2)]}
 {r(1 + a \cos\theta_+)(1 + a \cos\theta_-)}.
\end{equation}

The right hand side of Eq.~\eqref{eq:stripe_hydro_CH} is identical to 
the equation for the stripe relaxation dynamics in the absence of hydrodynamics,
as discussed in Sec.~\ref{SM:sec:noh} of the supplementary material. 
When hydrodynamics is present,
which is the case in this section, the term on the left hand side dominates over 
the second term on the right hand side of Eq.~\eqref{eq:stripe_hydro_CH}.
We will also now consider the linearised limit when $\delta \theta = \theta_c - \theta_c^{eq}$ is 
a small quantity. In this case, Eq.~\eqref{eq:stripe_hydro_CH} yields
\begin{equation}
 U_0(t) = \frac{r(1 + a \cos\theta_+^{eq})(1 + a \cos\theta_-^{eq})}
 {(1 -a^2)^{1/4} [1 + a \cos\theta_c^{eq} \cos (\Delta \theta_{eq} / 2)]} \dot{\delta \theta}.
 \label{eq:stripe_hydro_U0}
\end{equation}
Taking the derivative of Eq.~\eqref{eq:stripe_hydro_U0} allows 
$\dot{U}_0$ to be expressed in the linearised limit as
\begin{equation}
 \dot{U}_0(t) = \frac{r(1 + a \cos\theta_+^{eq})(1 + a \cos\theta_-^{eq})}
 {(1 -a^2)^{1/4} [1 + a \cos\theta_c^{eq} \cos (\Delta \theta_{eq} / 2)]} \ddot{\delta \theta}.
 \label{eq:stripe_hydro_dU0}
\end{equation}
Eqs.~\eqref{eq:stripe_hydro_U0} and \eqref{eq:stripe_hydro_dU0} 
can be inserted into Eq.~\eqref{eq:stripe_hydro_cauchyU0} to obtain an equation 
governing the evolution of $\delta \theta$. The last term 
in Eq.~\eqref{eq:stripe_hydro_cauchyU0} can be linearised 
using Eqs.~\eqref{eq:stripe_minima_lin} and \eqref{eq:stripe_noh_lin2}, as 
follows:
\begin{align}
 \sin \theta_c \left(a \cos\theta_c + \cos\frac{\Delta \theta}{2}\right) \simeq&
 -\delta \theta \times 
 \begin{cases}
  \cos(\Delta \theta_{eq}/2) - a, & \Delta A < \Delta A_{\rm crit},\\
  2a \sin^2\theta_c^{eq}, & 
  \Delta A > \Delta A_{\rm crit}.
 \end{cases},\label{eq:stripe_minima_lin}\\
 (1 + a \cos\theta_+)(1 + a \cos\theta_-) \simeq&
 \begin{cases}
  [1 - a \cos(\Delta \theta_{eq}/2)]^2,& \Delta A < \Delta A_{\rm crit},\\
  (1 - a^2) \sin^2 (\Delta \theta_{eq} / 2),& \Delta A > \Delta A_{\rm crit}.
 \end{cases}
 \label{eq:stripe_noh_lin2}
\end{align}
After some rearrangements, the following equation is obtained for $\delta \theta$:
\begin{equation}
 \ddot{\delta \theta} + 2 \alpha_\nu \dot{\delta \theta} + \omega_0^2 \delta \theta = 0. \label{eq:dampedharmonic}
\end{equation}
When $\Delta A < \Delta A_{\rm crit}$, $\theta_c^{eq} = \pi$ and $\omega_0^2$ is given by
\begin{equation}
 \omega_0^2 = \frac{\sigma \sqrt{1 - a^2}}{\pi r^2 R \rho_0} 
 \frac{\cos(\Delta \theta_{eq} / 2) - a}{[1 - a \cos(\Delta \theta_{eq} / 2)]^3}.
 \label{eq:stripe_hydro_omega0_pi}
\end{equation}
For $\Delta A > \Delta A_{\rm crit}$, the equilibrium position is at $\cos(\Delta \theta_{eq}/2) + a \cos\theta_c^{eq} = 0$ and
$\omega_0^2$ is given by:
\begin{equation}
 \omega_0^2 = \frac{2\sigma}{\pi r^3 \rho_0 (1 - a^2)^{3/2}} 
 \frac{a^2 - \cos^2 (\Delta \theta_{eq} / 2)}{\sin^2(\Delta \theta_{eq} / 2)}.
 \label{eq:stripe_hydro_omega0_npi}
\end{equation}
On the instability line, characterised by Eqs.~\eqref{eq:inst_thetac} and 
\eqref{eq:inst_params}, we find
\begin{equation}
 \omega_0^2 = \frac{2\sigma a}{\pi R^3 \rho_0 (1 - a^2)^{3/2}}.
 \label{eq:stripe_hydro_omega0_inst}
\end{equation}
The general solution of Eq.~\eqref {eq:dampedharmonic} is 
\begin{equation}
 \delta \theta = \delta \theta_0 \cos(\omega_0 t + \vartheta) e^{-\alpha_\nu t},
 \label{eq:stripe_hydro_sol}
\end{equation}
where $\delta \theta_0$ and $\vartheta$ are integration constants. 
It is understood that, in the unstable region given by 
$\theta_c < \pi - \arctan\,a$ or $\theta_c > \pi + \arctan\,a$, the above 
solution is valid only for strictly axisymmetric flows.
In principle,
there is a correction to the exponential decay term due to the second term on the
right hand side of Eq.~\eqref{eq:stripe_hydro_CH}. However, we find that this correction
is approximately one or two orders of magnitude smaller than $\alpha_\nu$,
(a more detailed analysis of the dynamics of stripes in the 
absence of hydrodynamics can be found in Subsec.~\ref{SM:sec:noh} of the supplementary 
material).

\begin{figure}
 \begin{center}
 \begin{tabular}{cc}
  \includegraphics[width=0.48\columnwidth]{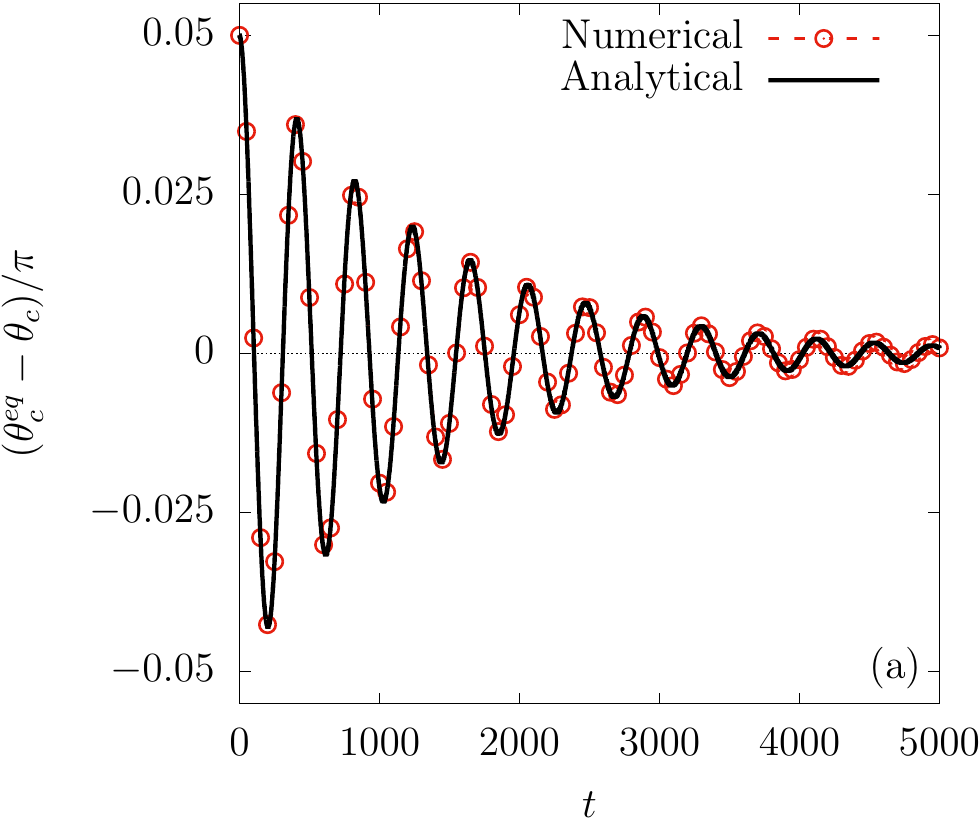} &
  \includegraphics[width=0.48\columnwidth]{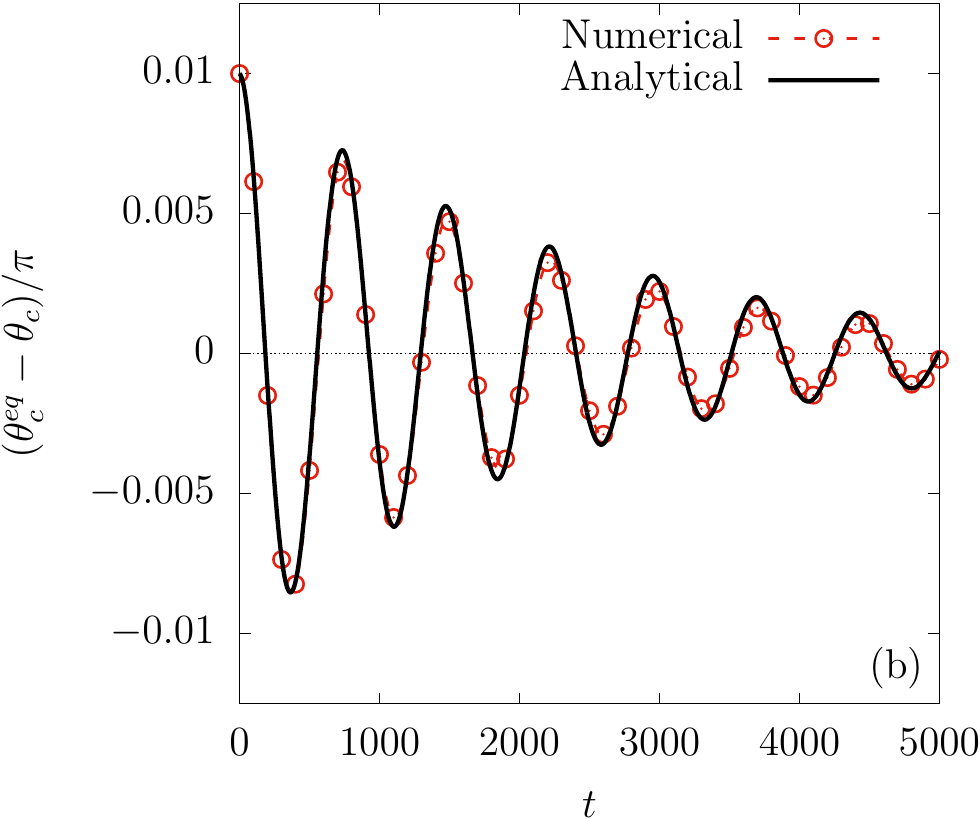}
 \end{tabular}
 \end{center}
\caption{Time evolution of the stripe center $\theta_c$ for stripes
initialised at (a) $\theta_0 = 0.95\pi$ with $\Delta \theta_0 = 0.280\pi$
(equilibrating at $\theta_c^{eq} = \pi$, on the torus with $a = 0.4$);
and (b) $\theta_0 = 0.79\pi$ with $\Delta\theta_0= 0.552\pi$ (equilibrating at $\theta_c^{eq} = 4\pi/5$, on the torus with a = 0.8).
The numerical results are shown using dotted lines and symbols, 
while the analytic solution Eq.~\eqref{eq:stripe_hydro_sol} is shown 
with solid lines.
\label{fig:stripe_hydro_dth}}
\end{figure}

\subsection{Benchmark test} \label{sec:CH:benchmark}

The solution derived in Eq.~\eqref{eq:stripe_hydro_sol} can serve as a benchmark 
for solvers involving interface dynamics. This benchmark test is particularly 
difficult since the dynamics of the interface can be significantly altered by 
numerical artefacts, such as the spurious velocity at the interface,
which are known to plague numerical solutions \citep{sofonea04,shan06}.

In the numerical tests discussed below, the velocity field is initialised with $u^{\hat{\varphi}} = 0$ 
and $u^{\hat{\theta}} = U_0 f_0(\theta) / (1 + a \cos\theta)$, where 
$f_0(\theta) = (1 - a^2)^{1/4}$ is the zeroth-order harmonic derived 
in Eq.~\eqref{eq:tor_f0} and 
$U_0$ is computed based on Eq.~\eqref{eq:stripe_hydro_U0} 
using the solution in Eq.~\eqref{eq:stripe_hydro_sol} with
$\vartheta = 0$ and $\dot{\delta \theta} = -\alpha_\nu \delta \theta_0$.
The order parameter is initialised with the hyperbolic tangent 
profile in Eq.~\eqref{eq:stripe_tanh_lap}.

In the first test, we consider a stripe equilibrating at $\theta_c^{eq} = \pi$, 
on the torus with $R = 2$ and $r = 0.8$ ($a = 0.4$). The stability region for this torus
is $0.8789\pi < \theta_c^{eq} < 1.1211 \pi$. 
We choose an initial amplitude of $\delta \theta_0 = -0.05 \pi$
(the initial position is $\theta_0 = 0.95 \pi$). The initial stripe width is set 
to $\Delta \theta_0 = 0.280406 \pi$ (at equilibrium, 
$\Delta \theta_{eq} \simeq 0.282296 \pi \simeq 
0.38 \Delta \theta_{\rm crit}$). The simulation parameters are $\kappa = 2.5 \times 10^{-4}$, 
$\text{A} = 0.5$, $\nu = M = 2.5 \times 10^{-3}$,
$\nu_v = 0$ and $\rho_0 = 20$, resulting in 
$\omega_0 \simeq 0.0152$ and $\alpha_\nu = 7.44 \times 10^{-4}$. 
The number of nodes and time step are 
$N_\theta = 480$ and $\delta t = 5\times10^{-4}$.
The numerical results, shown with red dashed lines and empty circles, are shown 
alongside the analytical curve corresponding to Eq.~\eqref{eq:stripe_hydro_sol}
with $\vartheta = 0$ and angular velocity $\omega_0$ computed
using Eq.~\eqref{eq:stripe_hydro_omega0_pi} in figure~\ref{fig:stripe_hydro_dth}(a).
Without resorting to any fitting routines, it can be seen that 
the analytic expression provides an excellent match to the 
simulation results.

For the second test, we choose a stripe equilibrating away from $\pi$. 
In order for this test to be meaningful also when axisymmetry is not 
strictly imposed, we seek to ensure that the stripe evolution occurs 
exclusively in the region of stability. For this reason, we increase $a$ 
to $0.8$ ($R = 2$ remains the same as before and $r$ is increased to $1.6$),
such that the stability region is now $0.7853 \pi \le \theta_c^{eq} \le 1.2147 \pi$.
Taking $\theta_c^{eq} = 0.8\pi$ (corresponding to the 
equilibrium width $\Delta \theta_{eq} \simeq 0.551868\pi \simeq 1.98 
\Delta \theta_{\rm crit}$),
we choose an initial amplitude of $\delta \theta_0 = -0.01\pi$,
such that $\theta_0 = 0.79\pi$ and $\Delta \theta_0 = 0.539376\pi$.
At larger initial amplitudes, the evolution of the stripe becomes visibly 
asymmetric, due to the inequivalence between the left and right sides of 
the equilibrium position. The fluid parameters are set to
$\kappa = 1.25 \times 10^{-4}$, 
$\text{A} = 0.25$, $\nu = M = 6.25 \times 10^{-4}$,
$\nu_v = 0$ and $\rho_0 = 20$, resulting in 
$\omega_0 \simeq 8.49 \times 10^{-3}$ and 
$\alpha_\nu \simeq 4.34 \times 10^{-4}$.  The number of nodes and time step are set to
$N_\theta = 960$ and $\delta t = 5\times10^{-3}$.
The simulation results, shown using a red dashed line with empty circles,
are shown alongside the analytic result, given by Eq.~\eqref{eq:stripe_hydro_sol} 
with $\omega_0$ computed using Eq.~\eqref{eq:stripe_hydro_omega0_npi}, are in 
good agreement, as can be seen from figure~\ref{fig:stripe_hydro_dth}(b).

We now discuss some of the properties of the oscillation frequency, $\omega_0$.
As can be seen from Eqs.~\eqref{eq:stripe_hydro_omega0_pi} and \eqref{eq:stripe_hydro_omega0_npi},
$\omega_0^2$ is proportional to the line tension, $\sigma$, and inversely proportional to 
the fluid density, $\rho_0$. No explicit dependence can be seen on the viscosities $\nu$ and 
$\nu_v$. This is to be expected, since the line tension is responsible for the driving force,
while the local mass density is a measure of the fluid inertia. 

\begin{figure}
 \begin{center}
 \begin{tabular}{cc}
  \includegraphics[width=0.45\columnwidth]{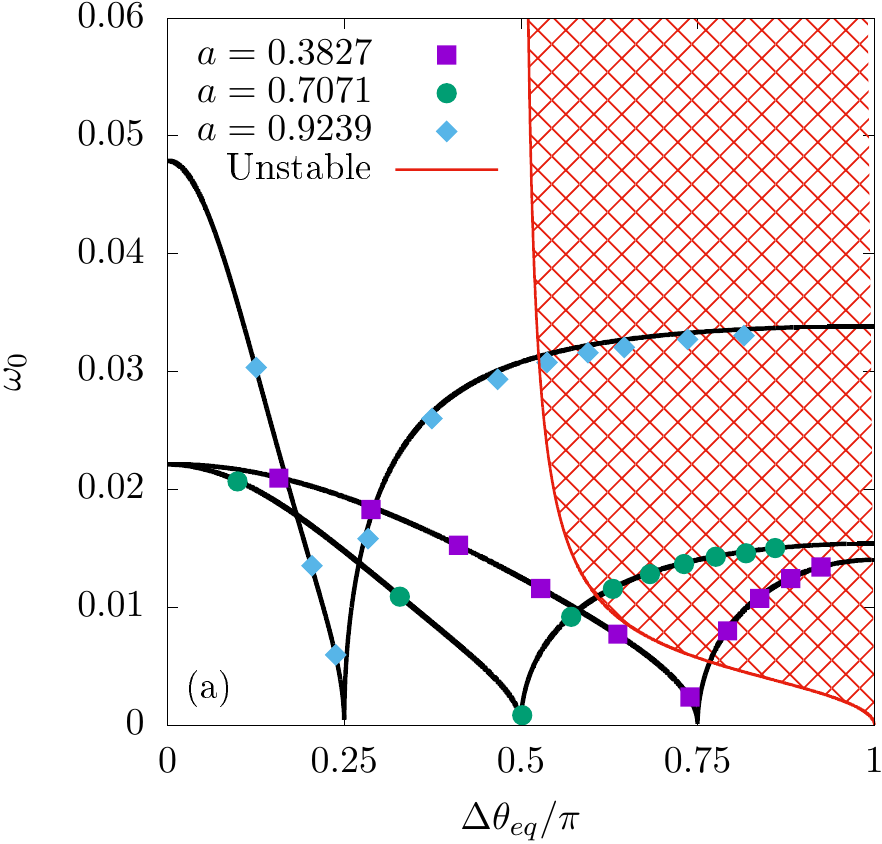} &
  \includegraphics[width=0.51\columnwidth]{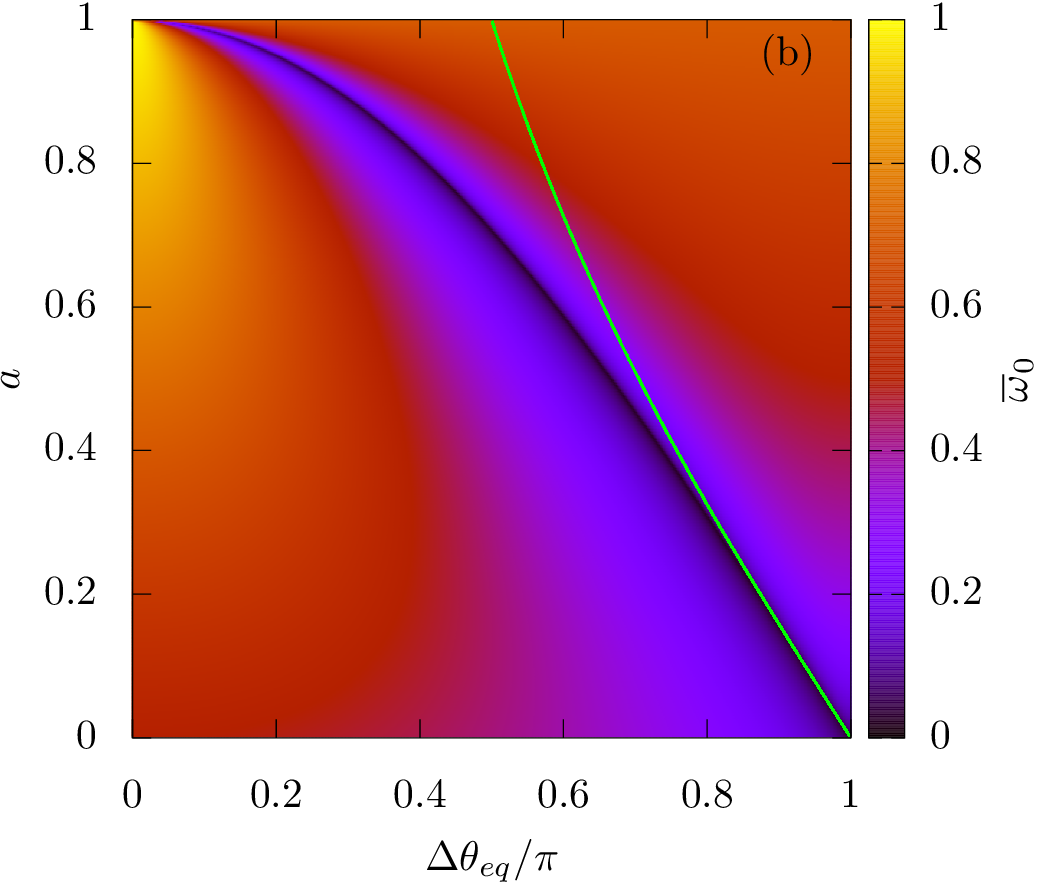}
 \end{tabular}
 \end{center}
\caption{(a) Comparison between the values of $\omega_0$ obtained by fitting 
Eq.~\eqref{eq:stripe_hydro_sol} to the numerical results, shown 
with points, and the analytic expressions, Eq.~\eqref{eq:stripe_hydro_omega0_pi} 
for $\Delta \theta< \Delta \theta_{\rm crit}$ (on the descending branch)
and Eq.~\eqref{eq:stripe_hydro_omega0_npi} for $\Delta \theta > \Delta \theta_{\rm crit}$
(on the ascending branch). 
The radii ratios where chosen such that $ \Delta \theta_{\rm crit}=\{0.25\pi,0.5\pi,0.75\pi\}$. 
The shaded area indicates the region where the stripe configurations 
are unstable.
(b) Colour plot representation of the regularised angular velocity, 
$\overline{\omega}_0$, defined in Eq.~\eqref{eq:stripe_w0_reg}, with respect to 
$\Delta \theta_{eq} / \pi$ (horizontal axis) and $a = r / R$ (vertical axis).
The green dashed line separates the stability (bottom left) from the 
instability (top right) regions of the parameter space.
\label{fig:stripe_w0_map}}
\end{figure}

Keeping $\rho_0$ and $\sigma$ fixed and considering fixed values of the torus radii, 
$r$ and $R$, $\omega_0$ exhibits a non-monotonic dependence 
on the stripe width at equilibrium, $\Delta \theta_{eq}$. 
Considering that the stripes of negligible width are always subcritical, 
we have $\lim_{\Delta \theta_{eq} \rightarrow 0} \omega_0 = \sigma \sqrt{1-a^2} / \pi r^2 R \rho_0 (1 - a)^2$.
At the other end of the spectrum, stripes with 
$\Delta \theta_{eq} = \pi$ have 
$\lim_{\Delta \theta_{eq} \rightarrow \pi} \omega_0 = 2\sigma / \pi r R^2 \rho_0 (1 - a^2)^{3/2}$.
In between, it can be seen that $\omega_0$ vanishes for critical stripes on both the subcritical 
[Eq.~\eqref{eq:stripe_hydro_omega0_pi}] and supercritical [Eq.~\eqref{eq:stripe_hydro_omega0_npi}] 
branches. This is highlighted in figure~\ref{fig:stripe_w0_map}(a), where $\omega_0$ is represented 
as a function of $\Delta \theta_{eq}$ for three values of $a = r / R$, namely 
$0.3827$ (purple squares), $0.7071$ (green circles) and $0.9239$ (blue rhombi).
These values are chosen such that the critical stripe width is 
$\Delta \theta_{eq} = 3\pi / 4$, $\pi /2$ and $\pi / 4$, respectively.
The shaded region marks the instability region, being bounded 
from below by Eq.~\eqref{eq:stripe_hydro_omega0_inst}.
The numerical values of $\omega_0$ are obtained by performing a two-parameter fit 
of Eq.~\eqref{eq:stripe_hydro_sol} with respect to $\alpha_\nu$ and $\omega_0$
(the offset is set to $\varsigma = 0$) on the numerical data. The other fluid 
parameters are $\rho_0 = 20$, $\nu = 2.5\times 10^{-3}$, $\nu_v = 0$, 
$\kappa = 5 \times 10^{-4}$ and $\text{A} = 0.5$, while 
$R = 2$ is kept fixed. The corresponding analytic results 
are shown with solid black lines. An excellent agreement can be seen, even for 
the nearly critical stripe, for which $\omega_0$ is greatly decreased. 

In order to further explore the properties of $\omega_0$, we focus on its dependence 
on the stripe width at equilibrium, $\Delta \theta_{eq}$, and on the torus aspect 
ratio $a = r / R$. From Eq.~\eqref{eq:stripe_hydro_omega0_pi}, it is clear that 
$\omega_0$ diverges as $r^{-1} = (a R)^{-1}$ when $a \rightarrow 0$.
This is to be expected, since $\omega_0$ is proportional to the number of oscillations 
per unit time, which increases as $r$ is decreased. 
Furthermore, Eq.~\eqref{eq:stripe_hydro_omega0_npi} shows that when 
$a \rightarrow 1$, $\omega_0$ diverges as $(1 - a^2)^{-3/4}$. 
From the above discussion, it is 
instructive to introduce the dimensionless, regularised oscillation frequency, 
$\overline{\omega}_0$, through
\begin{equation}
 \overline{\omega}_0^2 \equiv \frac{\pi \rho_0}{4\sigma} 
 r^2 R (1 - a^2)^{3/2} \omega_0^2 =
 \begin{cases}
  {\displaystyle 
  \frac{[\cos(\Delta \theta_{eq}/2) - a] (1 - a^2)^2}
  {4[1 - a \cos(\Delta \theta_{eq}/2)]^3}},& 
  \Delta \theta_{eq} < \Delta \theta_{\rm crit},\medskip\\
  {\displaystyle \frac{[a^2 - \cos^2(\Delta \theta_{eq}/2)]}
  {2a \sin^2(\Delta \theta_{eq}/2)}},& 
  \Delta \theta_{eq} > \Delta \theta_{\rm crit},
 \end{cases}
 \label{eq:stripe_w0_reg}
\end{equation}
where the factor $\pi / 4$ was introduced for normalisation 
purposes. It can be seen that $\overline{\omega}_0$ attains the 
maximum value with respect to $\Delta \theta_{eq}$ when 
$\Delta \theta_{eq} \rightarrow 0$. This value is
\begin{equation}
 \lim_{\Delta \theta_{eq} \rightarrow 0} \overline{\omega}_0 = \frac{1+a}{2}.
\end{equation}
The regularised angular velocity $\overline{\omega}_0$ is represented 
in figure~\ref{fig:stripe_w0_map}(b) 
as a function of the stripe width $\Delta \theta_{eq}/\pi$ (on the horizontal axis) 
and the radii ratio $a = r/R$ (on the vertical axis). Due to the chosen normalisation,
the colour map spans $[0,1]$. The dark line joining the 
bottom right and top left corners corresponds to the parameters of the critical stripe.
The green dashed line delimits the regions of stability (bottom left) 
and instability (top right).

\section{Conclusions}\label{sec:conc}

In this work, we focussed on a series of axisymmetric flows on the torus geometry
which are solvable analytically. The analytical results are also directly and systematically compared  
against numerical results obtained using a finite-difference Navier-Stokes solver.

Starting with perfect fluids, we first investigated the propagation of 
sound waves, identifying the discrete set of frequencies allowed on the torus geometry. 
In contrast to the planar geometry, the even and odd modes are no longer degenerate.
Moreover, since the ratios of the eigenfrequencies are not integers, the periodicity in the
fluid flows is lost. We also showed that the sound speed can be altered when changing the equation of 
state by considering isothermal and thermal ideal fluids, as well as 
multicomponent flows described via the Cahn-Hilliard equation.

We next looked at the equivalent of the popular shear wave damping 
problem in Cartesian coordinates. Here, we considered a fluid 
flowing along the azimuthal direction, with vanishing poloidal velocity. 
Under the assumption of axial symmetry, we showed that the velocity 
can be expanded with respect to a discrete set of basis functions which 
are the eigenfunctions of a second-order differential operator 
with respect to the poloidal coordinate $\theta$. The eigenvalues 
corresponding to these eigenfunctions control the damping rate of 
the associated velocity components. In particular, we highlighted the 
relaxation of an initially constant velocity profile towards the 
zeroth order eigenfunction, corresponding to a vanishing eigenvalue, 
which corresponds to a non-dissipative flow.

The third problem concerns the damping of sound waves. Here, 
we discussed the effect of the various dissipative terms appearing in the 
Navier-Stokes, energy and Cahn-Hilliard equations.
Generally, the fluid flow can be decomposed into acoustic modes, which propagate,
and thermal/Cahn-Hilliard modes, which simply decay exponentially.
The extension of the methodology to other types of fluids is straightforward.

The fourth and fifth phenomena we have studied concern multicomponent flows governed 
by the Cahn-Hilliard equation. The typical multicomponent axisymmetric 
configuration that we considered is the stripe, centred on poloidal 
coordinate $\theta_c$ and having angular span $\Delta \theta$. 

We showed that, for a general class of multiphase and multicomponent models,
the requirement of minimisation of interface length while preserving the 
stripe area determines the equilibrium position of the stripe. For stripes 
having a total area less than a critical area $\Delta A_{\rm crit}$, 
the equilibrium position is on the inside of the torus ($\theta_c^{eq} = \pi$). 
As the stripe area is increased above $\Delta A_{\rm crit}$, two equilibrium 
positions become possible, highlighting a second-order phase transition in 
this class of systems. We also generalise the Laplace pressure law. 
Our analysis gives an exact expression for the difference between the 
pressure inside of the (minority phase) stripe and the pressure outside 
of the stripe (i.e. in the majority phase), for both subcritical ($\Delta A < \Delta A_{\rm crit}$) 
and supercritical ($\Delta A > \Delta A_{\rm crit}$) stripes.

We have also shown that the stripe configurations are not always stable, or even metastable, when axisymmetry is not strictly
enforced. For example, the droplet 
configuration is energetically favoured at small domain areas,
while the band configurations, which wrap around 
the torus along the $\theta$ direction, are favoured at large 
domain areas. Moreover, we highlighted that the stripe configurations 
become unstable to small perturbations when either one of their interfaces 
crosses the boundary from the region of negative 
Gaussian curvature ($\frac{\pi}{2} < \theta < \frac{3\pi}{2}$) towards the region of 
positive Gaussian curvature ($-\frac{\pi}{2} < \theta < \frac{\pi}{2}$). 

Finally, we considered the dynamics of stripes in the presence of
hydrodynamics, when the approach to equilibrium of the stripes 
is achieved through underdamped harmonic oscillations. 
Using analytical techniques, we find expressions for both the 
angular velocity and damping coefficient. This is in contrast to the case in the absence of
hydrodynamics (detailed in Sec.~\ref{SM:sec:noh} of the supplementary material), 
where the approach to equilibrium is an exponential relaxation. 

We believe that the results presented here provide non-trivial problems
for developing computational methods for flows on curved surfaces (including the torus), 
and for benchmarking their 
accuracy and performance. For instance, the first three flow phenomena in this paper
can be used for convergence testing of numerical codes implementing hydrodynamics on curved surfaces.
To this end, we present a recipe for performing such tests in Appendix~\ref{app:conv},
where we perform a convergence analysis for the numerical scheme employed in this
paper. The multicomponent flow phenomena also provide a good example
for cases where the Navier-Stokes equation is coupled to other equations capturing more complex physics. 
For instance, this approach can be adapted to study complex flows on lipid membranes, or to investigate 
passive and active liquid crystal flows on curved surfaces. 
Here, the analytical results are limited to the torus geometry and primarily for axisymmetric flows. 
In the future, it would be interesting to apply and extend the methodology employed here to non-symmetric flow configurations, as well as to other manifolds.

{\bf Declaration of Interests:} The authors report no conflict of interest.

{\bf Acknowledgements:} HK acknowledges funding from EPSRC (EP/J017566/1 and EP/P007139/1).
HK and VEA also thank the EU COST action MP1305 Flowing Matter (VEA and HK; Short Term Scientific Mission 38607).
VEA expresses gratitude towards Professor L.-S. Luo (Old Dominion University, Norfolk, VA, USA)
for useful discussions and hospitality during the partial completion of this work, as well 
as towards the Romanian-U.S. Fulbright Commission for generous 
support through The Fulbright Senior Postdoctoral Program for Visiting Scholars,
Grant number 678/2018. SB acknowledges funding from EPSRC, grant number EP/R007438/1.
VEA and SB thank Professor Victor Sofonea (Romanian Academy, Timi\cb{s}oara Branch) for 
encouragement, as well as for sharing with us the computational infrastructure 
available at the Timi\cb{s}oara Branch of the Romanian Academy.
This research was supported by the Research Computing clusters at Old Dominion University.
The authors thank Professor A. J. Wagner (North Dakota State University, Fargo, ND, USA) for useful discussions. 
We thank an anonymous referee for suggesting the stability analysis 
for the stripe configurations.
\appendix

\section{Convergence test} \label{app:conv}

This section of the Appendix illustrates a procedure 
for using the benchmark problems introduced in Sections~\ref{sec:inv},
\ref{sec:shear} and \ref{sec:damp} for convergence tests of numerical 
codes designed for hydrodynamics on curved surfaces. 
The validation is done against the analytic solutions derived in the 
aforementioned sections, which are constructed using expansions of the mode 
functions $\{f_\ell, g_\ell\}$ (for longitudinal waves) and $\{F_\ell, G_\ell\}$ (for the 
shear waves) including terms up to order $n$ ($3 \le n \le 8$) with respect to the torus 
radii ratio, $a = r / R$. For definiteness, we restrict our convergence study to the 
amplitudes of the first even harmonic, $U_{c;1}(t)$ and $V_{c;1}(t)$.

In the first part of this section, we present 
the validation of our numerical scheme with respect to the spatial resolution.
We consider the three benchmark tests described in Sections~\ref{sec:inv:b1},
\ref{sec:shear:b1} and \ref{sec:damp:bench}. Unless otherwise stated, 
the fluid parameters and initial state 
are identical to those described in these sections. The numerical values of the 
amplitudes $U_{c;1}(t)$ and $V_{c;1}(t)$ are obtained as follows. 
The total simulation time, $t_{\rm max}$, is divided 
into $S$ intervals $\Delta t = t_{\rm max} / S$, numbered using $0 \le s \le S$. 
At each time $t_s = s \Delta t$, the numerical solution for the profile 
of $u^\htheta$ or $u^\hvarphi$ (for the longitudinal or shear wave benchmarks) 
are projected onto the basis functions $f_1(\theta)$ and $F_1(\theta)$ using 
rectangle integration
\begin{equation}
 U^{\rm num}_{c;1}(t_s) = \frac{1}{N_\theta} \sum_{i = 1}^{N_\theta} 
 \frac{u^\htheta_{\rm num}(t_s, \theta_i) f_1(\theta_i)}{1 + a \cos\theta_i}, \qquad 
 V^{\rm num}_{c;1}(t_s) = \frac{1}{N_\theta} \sum_{i = 1}^{N_\theta}
 \frac{u^\hvarphi_{\rm num}(t_s, \theta_i) F_1(\theta_i)}{(1 + a\cos\theta_i)^{-2}},
 \label{eq:conv_ampl}
\end{equation}
where `${\rm num}$' indicates that the amplitudes are determined numerically.
The mode functions $f_1(\theta)$ and $F_1(\theta)$ are
computed via the eighth-order expansions with respect to $a$ given in 
Eqs.~\eqref{SM:eq:inv_modes_1} and \eqref{SM:eq:shear_modes_1}.

\begin{figure}
\begin{tabular}{ccc}
\includegraphics[width=0.32\columnwidth]{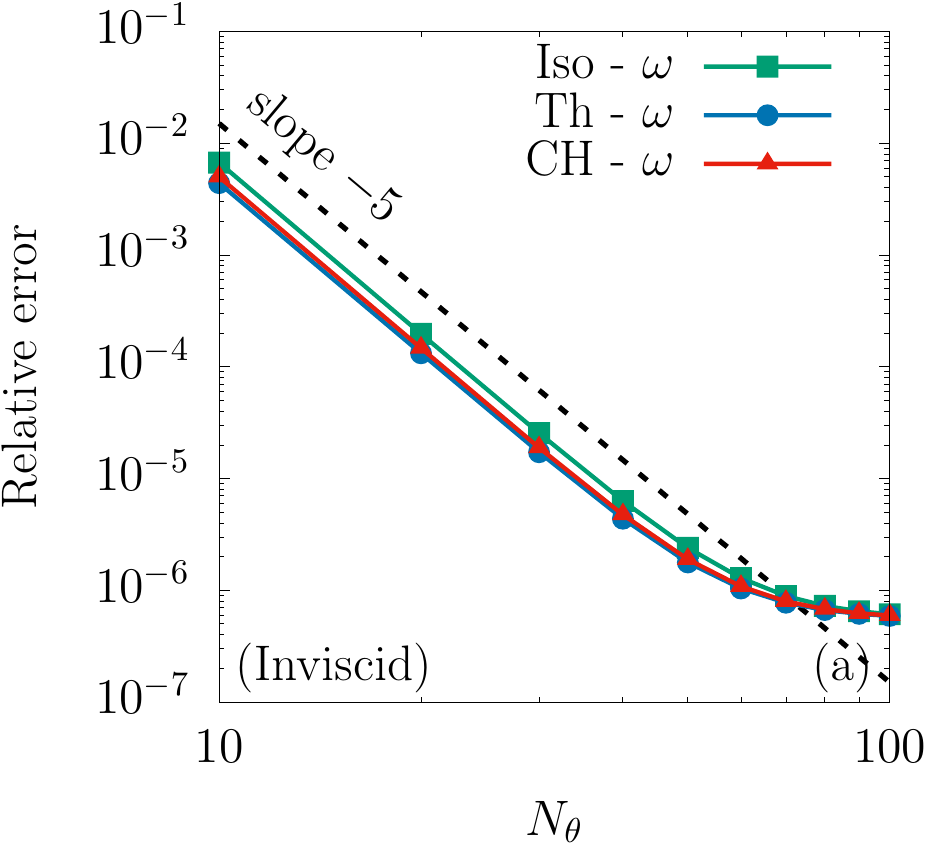}& 
\includegraphics[width=0.32\columnwidth]{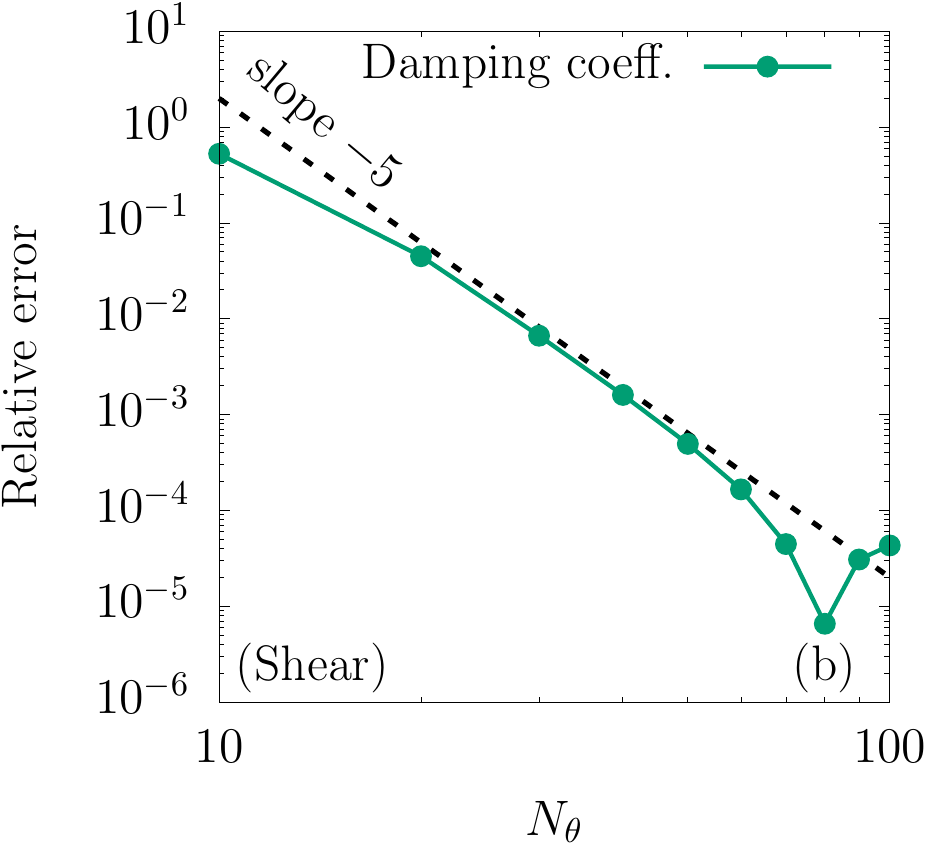}& 
\includegraphics[width=0.32\columnwidth]{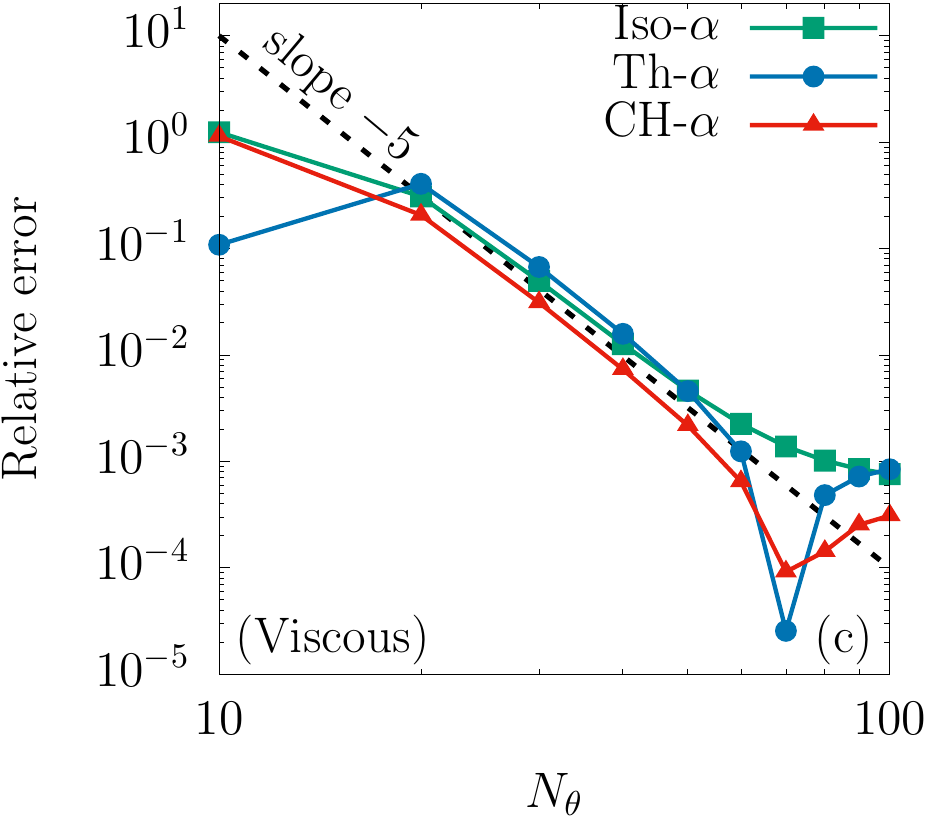} 
\end{tabular}
\caption{
(a) The relative error of $\omega_{c;1}$ 
in the context of the inviscid propagation of longitudinal waves, 
for an isothermal ideal fluid (Iso), an ideal fluid with variable temperature (Th) 
and an isothermal Cahn-Hilliard multicomponent fluid (CH);
(b) The relative error of the damping coefficient 
$\nu \chi_{c;1}^2 / r^2$ in the context of shear waves damping;
(c) The relative error of $\alpha_{c;1;a}$ 
in the context of the viscous damping of longitudinal waves, 
for the same 3 fluid described in (a).
Each panel contains a dashed line indicating fifth order convergence.
\label{fig:conv}}
\end{figure}

In the context of the propagation of longitudinal waves along the poloidal ($\theta$) 
direction through a perfect fluid, figure~\ref{fig:conv}(a) shows 
the relative error of the angular frequency $|\omega_{c;1}^{\rm num}/\omega_{c;1}^{\rm an}-1|$,
where $\omega_{c;1}^{\rm an} = c_s \lambda_{c;1} / r$ is computed using the 
eighth-order expansion of $\lambda_{c;1}$ in Eq.~\eqref{SM:eq:inv_modes_1Nl}, while the numerical 
value $\omega_{c;1}^{\rm num}$ is obtained using a two-parameter fit of the 
numerical amplitudes $U^{\rm num}_{c;1}(t_s)$ to the analytic prediction 
in Eq.~\eqref{eq:inv_sol}, i.e.:
\begin{equation}
 U_{c;1}(t) = \mathcal{A} \cos(\omega_{c;1}^{\rm num} t),
\end{equation}
where $\mathcal{A}$ and $\omega_{c;1}^{\rm num}$ are free parameters. 
The time interval and total simulation time are taken as $\Delta t = 0.05$
(corresponding to $100$ simulation steps at $\delta t = 5\times 10^{-4}$)  and
$t_{\rm max} = 18$, such that the total number of intervals is $S = 360$.
For completeness, we present the results for the isothermal and thermal 
ideal fluid cases, as well as for the isothermal Cahn-Hilliard 
multicomponent fluid with background order parameter $\phi_0 = 0.8$. 
The values of the parameters are identical to those considered in Sec.~\ref{sec:inv:b1}.
It can be seen that all curves are parallel to the slope $-5$ dashed line, 
indicating that our numerical scheme has fifth order accuracy.

We now consider the benchmark problem presented in Sec.~\ref{sec:shear:b1} concerning the 
damping of shear waves. figure~\ref{fig:conv}(b) shows the decrease in the relative 
error of the damping coefficient $\nu (\chi^{\rm num}_{c;1})^2 / r^2$ for the amplitude 
of the first mode, $V_{c;1}(t)$, as a function of the number of grid points.
The numerical values for the damping coefficient are obtained by fitting the numerical data 
using the analytic formula obtained by combining Eqs.~\eqref{eq:shear_intndef_aux} and 
\eqref{eq:shear_sol}, i.e.:
\begin{equation}
 V_{c;1}(t) = \mathcal{A} e^{-\alpha t},
\end{equation}
where $\mathcal{A}$ and $\alpha$ are free parameters. 
The time interval and total simulation time are taken as $\Delta t = 5$
(corresponding to $1000$ simulation steps at $\delta t = 5 \times 10^{-3}$)  and
$t_{\rm max} = 1800$, such that the total number of intervals is $S = 360$.
The analytic prediction for 
$\alpha$ is $\nu \chi_{c;1}^2 / r^2$. In the $\log$-$\log$ plot of figure~\ref{fig:conv}(b), 
the relative error of the damping coefficient follows the slope $-5$ dashed line,
also indicating the scheme is fifth-order accurate.
The simulation parameters are identical to those considered in Sec.~\ref{sec:shear:b1}.

Finally, we consider the sound waves damping benchmark problem
introduced in Sec.~\ref{sec:damp:bench}. 
figure~\ref{fig:conv}(c) presents the relative error for 
the acoustic damping coefficient $|1 - \alpha_{c;1;a}^{\rm num}/\alpha_{c;1;a}^{\rm an}|$, 
where $\alpha_{c;1;a}^{\rm an}$ is listed in Sec.~\ref{sec:damp:bench} for the various fluid types considered. 
Considering the three types of fluids discussed in the first paragraph,
these relative errors are plotted with respect to $N_\theta$. 
The values $\alpha^{\rm num}_{c;1;a}$ are obtained by fitting the numerical data 
using the analytic formula, given in Eqs.~\eqref{eq:damp_Ucn}:
\begin{equation}
 U_{c;1}(t) = \mathcal{A} e^{-\alpha^{\rm num}_{c;1;a} t} \sin (\omega_{c;1}^{\rm num} t),
\end{equation}
where $\mathcal{A}$, $\alpha^{\rm num}_{c;1;a}$ and $\omega_{c;1}^{\rm num}$
are free parameters. The parameters used in this benchmark test are identical 
to those in Sec. \ref{sec:damp:bench} and for definiteness, we focus only 
on the case when the volumetric kinematic viscosity $\nu_v = 0.02$ 
(the other transport coefficients change from one type of fluid to the other, as 
described in Sec.~\ref{sec:damp:bench}).
The time interval and total simulation time are taken as $\Delta t = 0.05$
(corresponding to $100$ simulation steps at $\delta t = 5 \times 10^{-4}$)  and
$t_{\rm max} = 48$, such that the total number of intervals is $S = 960$.
It can be seen that the relative error $|1 - \alpha^{\rm num}_{c;1;a} / \alpha^{\rm an}_{c;1;a}|$
in the acoustic damping coefficient generally follows the slope $-5$ dashed line.

In the second part of this section, we consider the effect of varying the expansion 
order $n$ of the eigenfunctions, eigenfrequencies and all derived quantities.
This study is performed at the level of the $L_2$ norms of the errors
$[U_{c;1}^{\rm an}(t) - U_{c;1}^{\rm num}(t)]/U_0$ and 
$1 - V_{c;1}^{\rm num}(t)/V_{c;1}^{\rm an}(t)$ between the numerical values 
and analytic predictions for the amplitudes of the first even mode.
These norms are computed by integrating over the simulation time 
using the trapezoidal rule
\begin{align}
 L_2^{\rm long} =& \left\{
 \frac{1}{S} \sum_{s = 0}^S \mathfrak{f}_s 
 \left[\frac{U_{c;1}^{\rm num}(t_s) - U_{c;1}^{\rm an}(t_s)}{U_0}\right]^2
 \right\}^{1/2}, \nonumber\\
 L_2^{\rm shear} =& \left\{
 \frac{1}{S} \sum_{s = 0}^S \mathfrak{f}_s 
 \left[\frac{V_{c;1}^{\rm num}(t_s)}{V_{c;1}^{\rm an}(t_s)} - 1\right]^2
 \right\}^{1/2},\label{eq:conv_L2}
\end{align}
where $\mathfrak{f}_s = 1/2$ when $s = 0$ or $s = S$ and $1$ otherwise.
The reason why $L_2^{\rm long}$ is computed using absolute [$U_{c;1}^{\rm num}(t_s) - U_{c;1}^{\rm an}(t_s)$] rather than relative [$U_{c;1}^{\rm num}(t_s)/ U_{c;1}^{\rm an}(t_s) - 1$] differences 
is that due to the oscillatory nature of $U_{c;1}^{\rm an}(t_s)$, there are in principle values 
of $t_s$ where $U_{c;1}^{\rm an}(t_s)$ is arbitrarily close to $0$. For such values of $t_s$, 
the relative error could be disproportionally large, producing meaningless results.
Instead, the relative difference is preferred for $V_{c;1}(t_s)$ since $V_{c;1}^{\rm an}(t_s)$ 
exhibits an exponential decay with respect to $t_s$. Thus, the absolute differences 
$V_{c;1}^{\rm num}(t_s) - V_{c;1}^{\rm an}(t_s)$ would contribute with an exponentially decreasing 
amplitude at large times and the result of an $L_2$ norm based on the absolute differences 
would therefore be biased towards the early time properties of $V_{c;1}(t)$.
The analytical predictions $U^{\rm an}_{c;1}(t)$ and $V^{\rm an}_{c;1}$ 
can be obtained from Eqs.~\eqref{eq:inv_sol} and \eqref{eq:shear_sol}. 
The numerical amplitudes $U^{\rm num}_{c;1}(t_s)$ and $V^{\rm num}_{c;1}(t_s)$ 
are obtained by projecting $u^\htheta_{\rm num}(t_s, \theta)$ and $u^\hvarphi_{\rm num}(t_s, \theta)$ 
onto the basis functions $f_1(\theta)$ and $F_1(\theta)$, as described in 
Eq.~\eqref{eq:conv_ampl}.
Both the basis functions and the analytic solutions are obtained using the 
expansions in Eqs.~\eqref{SM:eq:inv_modes_1} and \eqref{SM:eq:shear_modes_1},
truncated at power $n$ of the radii ratio $a$.

\begin{figure}
\begin{tabular}{ccc}
\includegraphics[width=0.32\columnwidth]{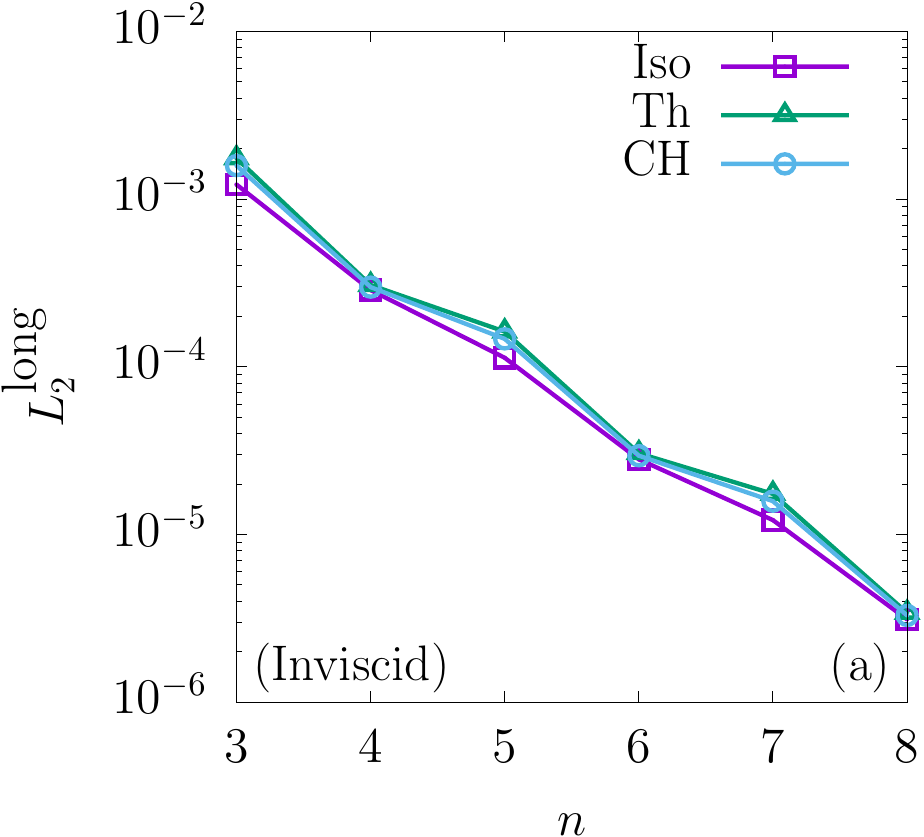}&
\includegraphics[width=0.32\columnwidth]{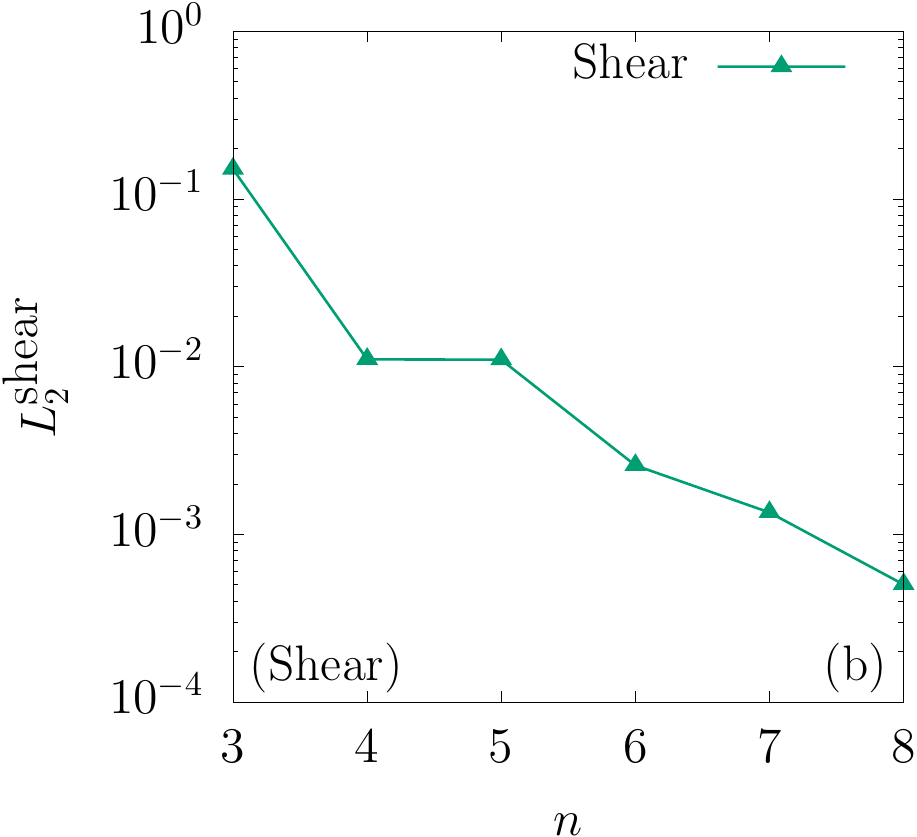}&  
\includegraphics[width=0.32\columnwidth]{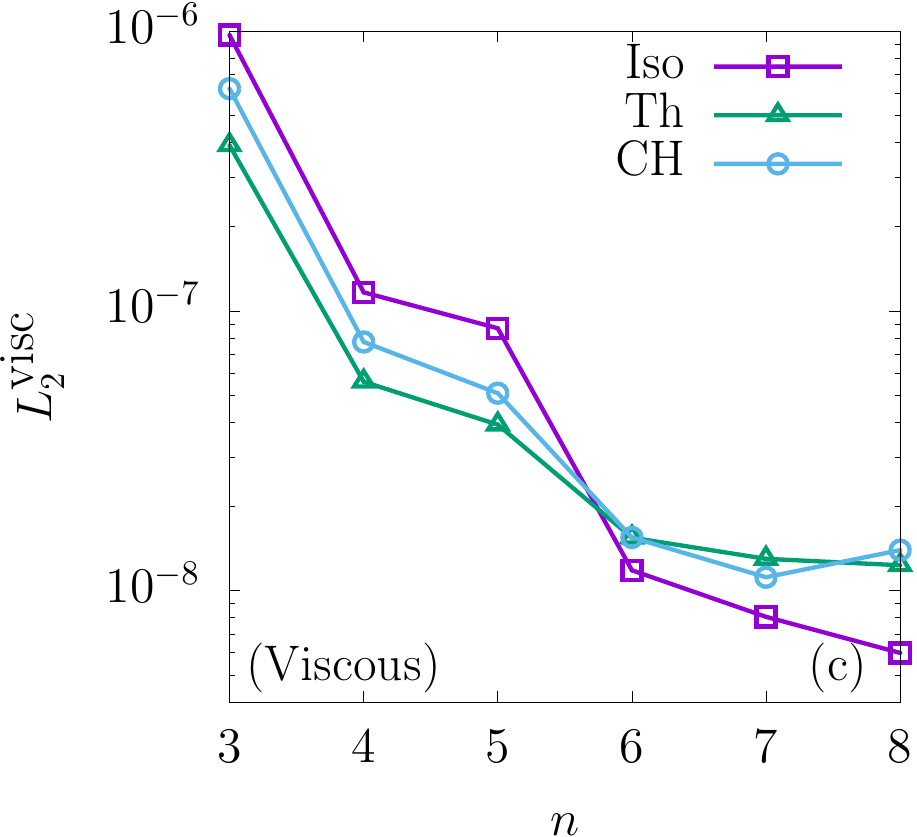} 
\end{tabular}
\caption{
The $L_2$ norm computed using Eq.~\eqref{eq:conv_L2} in the context of 
(a) propagation of inviscid longitudinal waves;
(b) damping of shear wave and (c) damping of longitudinal waves.
In (a) and (c), we consider the cases of the isothermal ideal fluid (Iso), 
ideal fluid with variable temperature (Th) and 
isothermal Cahn-Hilliard multicomponent fluid (CH).
In (b), only the isothermal fluid is considered.
\label{fig:conv-vs-a}}
\end{figure}

We begin with the benchmark problem introduced in Sec.~\ref{sec:inv:b1}, concerning 
the propagation of longitudinal waves through a perfect fluid.
figure~\ref{fig:conv-vs-a}(a) shows the variation of $L_2^{\rm long}$ with respect to the truncation order of the expansion, which is varied between $3 \le n \le 8$,
for the isothermal and thermal ideal fluid cases, as well as for the isothermal 
Cahn-Hilliard multicomponent fluid. 
The simulation parameters are identical to those presented in Sec.~\ref{sec:inv:b1},
as well as earlier in this section.
In general, an exponential decay of $L_2^{\rm long}$ with respect 
to $n$ can be observed for all fluid types considered. 
A sharper decrease in the $L_2$ error norm can be observed when 
$n$ is increased from an odd value to an even one. 

In the context of the shear wave damping benchmark introduced in Sec.~\ref{sec:shear:b1}, the analytical expression 
$V_{c;1}^{\rm an}(t)$ is obtained 
by combining Eqs.~\eqref{eq:shear_intndef_aux} and \eqref{eq:shear_sol}.
As before, $V_{c;1}^{\rm num}(t)$ is obtained by projecting the velocity profile 
onto the basis functions $F_1$, given in  Eq.~\eqref{SM:eq:shear_modes_1}, truncated at order $n$. 
The same order $n$ is used to evaluated the analytic prediction $V_{c;1}^{\rm an}(t)$. 
The results are presented in figure~\ref{fig:conv-vs-a}(b), up to order $n=8$.
The $L_2^{\rm shear}$ decays exponentially and again sharper drops are seen when
the expansion order is increased from an odd to an even value. 
The simulation parameters are identical to those employed in Sec.~\ref{sec:shear:b1}. 
A total of $S=360$ time intervals of length $\Delta t = 1000\delta t = 5$
were saved ($t_{\rm max} = 1800$).

Lastly, we investigate the convergence of the first harmonic in 
the context of viscous damping of longitudinal waves.
The $L_2^{\rm visc}$ norm is computed using Eq.~\eqref{eq:conv_L2}, 
where the analytical prediction for $U_{c;1}^{\rm an}(t)$
is given in Eqs.~\eqref{eq:damp_Ucn}. This prediction is evaluated 
using the values for $\omega_1^{\rm an}$ 
and the integral $I_{c;0;1}$ truncated at $n$th order.
The velocity profile is projected using Eq.~\eqref{eq:conv_ampl} 
onto the basis function $f_1$, computed using a truncation of 
Eq.~\eqref{SM:eq:inv_modes_1} at the same order $n$, obtaining $U_{c;1}^{\rm num}$. 
The results for the isothermal and thermal ideal fluid cases, as well as for the 
isothermal Cahn-Hilliard 
multicomponent fluid are summarised in figure~\ref{fig:conv-vs-a}(c). 
Since the linearised theory introduces errors of order $O(U_0,\varepsilon^2)$, 
in order to reveal the error induced by the expansion order, we decrease the kinematic viscosities employed in § 5.2 for each type of fluid by two orders of magnitude, namely $\nu_{\rm Iso}=10^{-4}$, $\nu_{\rm Th}=4\times 10^{-5}$ 
and $\nu_{\rm CH}=M\approx 6.486\times 10^{-5}$, 
while the volumetric kinematic viscosity is 
set to $\nu_v=2\times 10^{-4}$ for all fluid types.
The rest of the simulation parameters are: 
$N_\theta=320$, $R=2$ and $r=0.8$ ($a=0.4$), $U_0=10^{-5}$ and $\delta t=5\times10^{-4}$.
A total of $S=500$ time intervals of length $\Delta t = 1$ were saved ($t_{\rm max} = 500$). 
The exponential decay of the $L_2^{\rm visc}$ can be clearly seen, 
and again, a larger decrease can be seen when $n$ is increased from an odd to an even value.

\section{Eigenfunctions on the torus} \label{app:modes}

This section of the Appendix presents a perturbative 
procedure for constructing solutions of Eqs.~\eqref{eq:inv_modes_eq} 
and \eqref{eq:shear_modes_eq} in powers of $a = r / R$, where $r$ and 
$R$ are the inner and outer radii of the torus. 
Multiplying Eqs.~\eqref{eq:inv_modes_eq} and \eqref{eq:shear_modes_eq} 
by $1 + a\cos\theta$ yields:
\begin{equation}
 (1 + a\cos\theta)\left(\frac{\partial^2 \Psi_n}{\partial \theta^2} 
 + \lambda_n^2 \Psi_n\right) 
 + \alpha a\sin\theta \frac{\partial \Psi_n}{\partial \theta} 
 = 0,\label{eq:modes_pert}
\end{equation}
where $\alpha = 1$ and $-3$ for Eqs.~\eqref{eq:inv_modes_eq} and 
\eqref{eq:shear_modes_eq}, respectively.
We seek solutions of the form
\begin{gather}
 \Psi_n = N_n (1 + a \cos\theta)^\alpha \psi_n, \qquad 
 \psi_n = \psi_n^{(0)} + a \psi_n^{(1)} + a^2 \psi_n^{(2)} + \dots, \nonumber\\
 \lambda^2_{n} = \lambda^2_{n;0} + a \lambda^2_{n;1} + a^2 \lambda^2_{n;2} + \dots,
 \label{eq:modes_pert_Psi}
\end{gather}
where the normalisation constant $N_n$ ensures that 
$\Psi_n$ retains unit norm.
The prefactor $(1 + a \cos\theta)^\alpha$ ensures that all solutions 
$\Psi_n$ with $n > 0$ are exactly orthogonal to the zeroth-order 
solution as long as they do not contain any free terms.
Taking into account this prefactor, Eq.~\eqref{eq:modes_pert} becomes
\begin{equation}
 (1 + a \cos\theta)^2(\psi_n'' + \lambda_n^2 \psi_n) 
 - \alpha a \sin\theta(1 + a \cos\theta) \psi_n' - 
 \alpha a (a + \cos\theta) \psi_n = 0.
\end{equation}
Demanding that the coefficient of each power of $a$ vanishes,
at zeroth order the harmonic equation is recovered
\begin{align}
 \psi_{n;0}'' + \lambda_{n;0}^2 \psi_{n;0} = 0.\label{eq:modes_a0}
\end{align}
Furthermore, demanding that the solution at each level of the perturbative analysis 
be periodic with respect to $\theta$, the general solution 
of Eq.~\eqref{eq:modes_a0} can be written as
\begin{equation}
 \psi_{n;0} = e^{i n \theta}, \qquad 
 \lambda^2_{n;0} = n^2.
\end{equation}
where the real and imaginary parts correspond to the even and odd 
solutions, respectively.

Taking into account Eq.~\eqref{eq:modes_a0}, the first-order 
contribution to Eq.~\eqref{eq:modes_pert} is
\begin{equation}
 \psi_{n;1}'' + n^2 \psi_{n;1} + \lambda_{n;1}^2 e^{i n \theta} 
 - \frac{\alpha}{2} \left[(n+1) e^{i(n+1)\theta} - 
 (n-1)e^{i(n - 1)\theta}\right] = 0.
 \label{eq:modes_a1_eq}
\end{equation}
Since the solution of the homogeneous version of the above equation is proportional to 
$\psi_{n;0}$, it can be seen that $\lambda_{n;1}^2 = 0$, while 
$\psi_{n;1}$ can be found as
\begin{equation}
 \psi_{n;1} = -\frac{\alpha}{2} \left[
 \frac{n+1}{2n+1} e^{i(n+1)\theta} + 
 \frac{n-1}{2n-1} e^{i(n-1)\theta}\right].
 \label{eq:modes_a1_sol}
\end{equation}
At second order, the following equation is obtained:
\begin{multline}
 \psi_{n;2}'' + n^2 \psi_{n;2} + 
 \left[\lambda_{n;2}^2 - \frac{\alpha^2 n^2}{2(4n^2 - 1)}\right] e^{in \theta} + 
 \frac{\alpha (n+2)}{4} \left[1 + \frac{\alpha(n + 1)}{2n+1}\right] 
 e^{i(n+2)\theta} \\
 -\frac{\alpha (n-2)}{4} \left[1 + \frac{\alpha(n - 1)}{2n-1}\right] 
 e^{i(n-2)\theta} = 0.
 \label{eq:modes_a2_eq}
\end{multline}
As before, the coefficient of $e^{in\theta}$ must vanish.
At this point, we note that in the case when $ n = 1$,
$e^{i(n - 2)\theta} = e^{-i\theta}$ and is thus not independent 
of $\psi_{1;0} = e^{i \theta}$. Moreover, there is no value for $\lambda_{1;2}^2$ 
which ensures that the coefficients of $\cos\theta$ and $\sin\theta$ 
vanish simultaneously. Thus, at $n = 1$, the solution is
\begin{equation}
 \psi_{1;2} = \frac{\alpha(3+2\alpha)}{32} e^{3i \theta}, \qquad 
 \lambda_{1;c/s;2}^2 = \frac{\alpha^2}{6} \mp \frac{\alpha}{4},
\end{equation}
where the upper and lower signs refer to the even and odd solutions, 
respectively. For $n > 1$, the solution is
\begin{align}
 \psi_{n;2} =& \frac{\alpha}{16} \left[(n-2)\left(
 \frac{1}{n-1} + \frac{\alpha}{2n-1}\right) e^{i(n-2)\theta}
 + (n+2)\left(
 \frac{1}{n+1} + \frac{\alpha}{2n+1}\right) e^{i(n+2)\theta}\right], \nonumber\\
 \lambda_{n;2}^2 =& \frac{\alpha^2 n^2}{2(4n^2 - 1)}.
 \label{eq:modes_a2}
\end{align}

Keeping into account that at order $O(a^{n+1})$, the corrections to the eigenvectors
of orders up to $n$ must be computed as outlined above for $n = 1$, 
the above procedure can be continued to higher orders. 
Explicit expressions for the mode functions 
for $\alpha = 1$ ($f_n$ and $g_n$) and for $\alpha = -3$ 
($F_n$ and $G_n$) are given in Sections \ref{SM:sec:modes:inv} and \ref{SM:sec:modes:visc} 
of the supplementary material \citep{SM:JFM}.

\nocite{olver10}

\bibliographystyle{jfm}
\bibliography{refs}

\begin{thebibliography}{42}
\expandafter\ifx\csname natexlab\endcsname\relax\def\natexlab#1{#1}\fi
\def\au#1{#1} \def\ed#1{#1} \def\yr#1{#1}\def\at#1{#1}\def\jt#1{\textit{#1}}
  \def\bt#1{#1}\def\bvol#1{\textbf{#1}} \def\vol#1{#1} \def\pg#1{#1}
  \def\publ#1{#1}\def\arxiv#1{#1}\def\org#1{#1}\def\st#1{\textit{#1}}

\bibitem[Al-Izzi {\em et~al.\/}(2018)Al-Izzi, Sens \& Turner]{Izzi18}
{\sc \au{Al-Izzi, S.~C.}, \au{Sens, P.} \& \au{Turner, M.~S.}} \yr{2018}
  \at{Shear-driven instabilities of membrane tubes and dynamin-induced
  scission}.  \jt{arXiv}  \pg{p. 1810.05862}.

\bibitem[Ambru\cb{s} {\em et~al.\/}(2019)Ambru\cb{s}, Busuioc, Wagner,
  Paillusson \& Kusumaatmaja]{Ambrus19}
{\sc \au{Ambru\cb{s}, V.~E.}, \au{Busuioc, S.}, \au{Wagner, A.~J.},
  \au{Paillusson, F.} \& \au{Kusumaatmaja, H.}} \yr{2019}  \at{Multicomponent
  flow on curved surfaces: A vielbein lattice {B}oltzmann approach}.  \jt{Phys.
  Rev. E}  \bvol{100},  \pg{063306}.

\bibitem[Arroyo \& Desimone(2009)]{arroyo09}
{\sc \au{Arroyo, M.} \& \au{Desimone, A.}} \yr{2009}  \at{Relaxation dynamics
  of fluid membranes}.  \jt{Phys. Rev. E}  \bvol{79},  \pg{039906}.

\bibitem[Bertalm\'{i}o {\em et~al.\/}(2001)Bertalm\'{i}o, Cheng, Osher \&
  Sapiro]{Bertalmio01}
{\sc \au{Bertalm\'{i}o, M.}, \au{Cheng, L.-T.}, \au{Osher, S.} \& \au{Sapiro,
  G.}} \yr{2001}  \at{Variational problems and partial differential equations
  on implicit surfaces}.  \jt{J. Comput. Phys.}  \bvol{174}~(2),  \pg{759 --
  780}.

\bibitem[Boozer(2005)]{Boozer05}
{\sc \au{Boozer, A.~H.}} \yr{2005}  \at{Physics of magnetically confined
  plasmas}.  \jt{Rev. Mod. Phys.}  \bvol{76},  \pg{1071--1141}.

\bibitem[Briant \& Yeomans(2004)]{Briant04}
{\sc \au{Briant, A.~J.} \& \au{Yeomans, J.~M.}} \yr{2004}  \at{Lattice
  {B}oltzmann simulations of contact line motion. {II.} {B}inary fluids}.
  \jt{Phys. Rev. E}  \bvol{69},  \pg{031603}.

\bibitem[Busuioc \& Ambru\cb{s}(2019)]{busuioc19}
{\sc \au{Busuioc, S.} \& \au{Ambru\cb{s}, V.~E.}} \yr{2019}  \at{Lattice
  {B}oltzmann models based on the vielbein formalism for the simulation of
  flows in curvilinear geometries}.  \jt{Phys. Rev. E}  \bvol{99},
  \pg{033304}.

\bibitem[Busuioc {\em et~al.\/}(2020{\natexlab{{\em a\/}}})Busuioc,
  Ambru\cb{s}, Biciu\cb{s}c\u{a} \& Sofonea]{busuioc20camwa}
{\sc \au{Busuioc, S.}, \au{Ambru\cb{s}, V.~E.}, \au{Biciu\cb{s}c\u{a}, T.} \&
  \au{Sofonea, V.}} \yr{2020{\natexlab{{\em a\/}}}}  \at{Two-dimensional
  off-lattice {B}oltzmann model for van der {W}aals fluids with variable
  temperature}.  \jt{Comput. Math. Appl.}  \bvol{79},  \pg{111--140}.

\bibitem[Busuioc {\em et~al.\/}(2020{\natexlab{{\em b\/}}})Busuioc,
  Kusumaatmaja \& Ambru\cb{s}]{SM:JFM}
{\sc \au{Busuioc, S.}, \au{Kusumaatmaja, H.} \& \au{Ambru\cb{s}, V.~E.}}
  \yr{2020{\natexlab{{\em b\/}}}} Supplementary material. URL to be made
  available by the publisher.

\bibitem[Cox {\em et~al.\/}(2004)Cox, Weaire \& Glazier]{Cox04}
{\sc \au{Cox, S.}, \au{Weaire, D.} \& \au{Glazier, J.~A.}} \yr{2004}  \at{The
  rheology of two-dimensional foams}.  \jt{Rheologica Acta}  \bvol{43},
  \pg{442--448}.

\bibitem[Dziuk \& Elliott(2007)]{Dziuk07}
{\sc \au{Dziuk, G.} \& \au{Elliott, C.~M.}} \yr{2007}  \at{Surface finite
  elements for parabolic equations}.  \jt{J. Comput. Math.}  \bvol{25},
  \pg{385--407}.

\bibitem[Dziuk \& Elliott(2013)]{Dziuk13}
{\sc \au{Dziuk, G.} \& \au{Elliott, C.~M.}} \yr{2013}  \at{Finite element
  methods for surface {PDE}s}.  \jt{Acta Numerica}  \bvol{22},  \pg{289–396}.

\bibitem[Fonda {\em et~al.\/}(2018)Fonda, Rinaldin, Kraft \& Giomi]{Fonda2018}
{\sc \au{Fonda, P.}, \au{Rinaldin, M.}, \au{Kraft, D.~J.} \& \au{Giomi, L.}}
  \yr{2018}  \at{Interface geometry of binary mixtures on curved substrates}.
  \jt{Phys. Rev. E}  \bvol{98},  \pg{032801}.

\bibitem[Giordanelli {\em et~al.\/}(2018)Giordanelli, Mendoza \&
  Herrmann]{Giordanelli18}
{\sc \au{Giordanelli, I.}, \au{Mendoza, M.} \& \au{Herrmann, H.~J.}} \yr{2018}
  \at{Modelling electron-phonon interactions in graphene with curved space
  hydrodynamics}.  \jt{Sci. Rep.}  \bvol{8},  \pg{12545}.

\bibitem[Gross \& Atzberger(2018)]{gross18}
{\sc \au{Gross, B.~J.} \& \au{Atzberger, P.~J.}} \yr{2018}  \at{Hydrodynamic
  flows on curved surfaces: Spectral numerical methods for radial manifold
  shapes}.  \jt{J. Comput. Phys.}  \bvol{371},  \pg{663--689}.

\bibitem[Henkes {\em et~al.\/}(2018)Henkes, Marchetti \& Sknepnek]{Henkes18}
{\sc \au{Henkes, S.}, \au{Marchetti, M.~C.} \& \au{Sknepnek, R.}} \yr{2018}
  \at{Dynamical patterns in nematic active matter on a sphere}.  \jt{Phys. Rev.
  E}  \bvol{97},  \pg{042605}.

\bibitem[Henle \& Levine(2010)]{Henle10}
{\sc \au{Henle, M.~L.} \& \au{Levine, A.~J.}} \yr{2010}  \at{Hydrodynamics in
  curved membranes: the effect of geometry on particulate mobility}.  \jt{Phys.
  Rev. E}  \bvol{81},  \pg{011905}.

\bibitem[Janssen {\em et~al.\/}(2017)Janssen, Kaiser \& L\"{o}wen]{Janssen17}
{\sc \au{Janssen, L. M.~C.}, \au{Kaiser, A.} \& \au{L\"{o}wen, H.}} \yr{2017}
  \at{Aging and rejuvenation of active matter under topological constraints}.
  \jt{Sci. Rep.}  \bvol{7},  \pg{5667}.

\bibitem[Keber {\em et~al.\/}(2014)Keber, Loiseau, Sanchez, DeCamp, Giomi,
  Bowick, Marchetti, Dogic \& Bausch]{Keber14}
{\sc \au{Keber, F.~C.}, \au{Loiseau, E.}, \au{Sanchez, T.}, \au{DeCamp, S.~J.},
  \au{Giomi, L.}, \au{Bowick, M.~J.}, \au{Marchetti, M.~C.}, \au{Dogic, Z.} \&
  \au{Bausch, A.~R.}} \yr{2014}  \at{Topology and dynamics of active nematic
  vesicles}.  \jt{Science}  \bvol{345},  \pg{1135--1138}.

\bibitem[Koba(2018)]{koba18}
{\sc \au{Koba, H.}} \yr{2018}  \at{On derivation of compressible fluid systems
  on an evolving surface}.  \jt{Quart. Appl. Math.}  \bvol{76},  \pg{303--359}.

\bibitem[Koba {\em et~al.\/}(2017)Koba, Liu \& Giga]{koba17}
{\sc \au{Koba, H.}, \au{Liu, C.} \& \au{Giga, Y.}} \yr{2017}  \at{Energetic
  variational approaches for incompressible fluid systems on an evolving
  surface}.  \jt{Quart. Appl. Math.}  \bvol{75},  \pg{359--389}.

\bibitem[Kr\"{u}ger {\em et~al.\/}(2017)Kr\"{u}ger, Kusumaatmaja, Kuzmin,
  Shardt, Silva \& Viggen]{KrugerBook}
{\sc \au{Kr\"{u}ger, T.}, \au{Kusumaatmaja, H.}, \au{Kuzmin, A.}, \au{Shardt,
  O.}, \au{Silva, G.} \& \au{Viggen, E.~M.}} \yr{2017} {\em Lattice {B}oltzmann
  Method: {P}rinciples and Practice\/}.  \publ{Springer}.

\bibitem[Macdonald \& Ruuth(2010)]{Macdonald10}
{\sc \au{Macdonald, C.} \& \au{Ruuth, S.}} \yr{2010}  \at{The implicit closest
  point method for the numerical solution of partial differential equations on
  surfaces}.  \jt{SIAM Journal on Scientific Computing}  \bvol{31}~(6),
  \pg{4330--4350}.

\bibitem[Marsden \& Hughes(1994)]{marsden94}
{\sc \au{Marsden, J.~E.} \& \au{Hughes, J.~R.}} \yr{1994} {\em Mathematical
  foundations of elasticity\/}.  \publ{Dover publications}.

\bibitem[Nitschke {\em et~al.\/}(2017)Nitschke, Reuther \& Voigt]{nitschke17}
{\sc \au{Nitschke, I.}, \au{Reuther, S.} \& \au{Voigt, A.}} \yr{2017}
  \at{Discrete exterior calculus ({DEC}) for the surface {N}avier-{S}tokes
  equation}.  \bt{In {\em Transport Processes at Fluidic Interfaces\/} (ed.
  \ed{D.~Bothe \& A.~Reusken})},  \pg{pp. 125--263}.  \publ{Birkh\"auser}.

\bibitem[Nitschke {\em et~al.\/}(2019)Nitschke, Reuther \& Voigt]{Nitschke19}
{\sc \au{Nitschke, I.}, \au{Reuther, S.} \& \au{Voigt, A.}} \yr{2019}
  \at{Hydrodynamic interactions in polar liquid crystals on evolving surfaces}.
   \jt{Phys. Rev. Fluids}  \bvol{4},  \pg{044002}.

\bibitem[Nitschke {\em et~al.\/}(2012)Nitschke, Voigt \& Wensch]{Nitschke12}
{\sc \au{Nitschke, I.}, \au{Voigt, A.} \& \au{Wensch, J.}} \yr{2012}  \at{A
  finite element approach to incompressible two-phase flow on manifolds}.
  \jt{J. Fluid Mech.}  \bvol{708},  \pg{418–438}.

\bibitem[Olver {\em et~al.\/}(2010)Olver, Lozier, Boisvert \& Clark]{olver10}
{\sc \au{Olver, F. W.~J.}, \au{Lozier, D.~W.}, \au{Boisvert, R.~F.} \&
  \au{Clark, C.~W.}} \yr{2010} {\em NIST handbook of mathematical functions\/}.
   \publ{New York, NY: Cambridge University Press}.

\bibitem[Pearce {\em et~al.\/}(2019)Pearce, Ellis, Fernandez-Nieves \&
  Giomi]{Pearce19}
{\sc \au{Pearce, D. J.~G.}, \au{Ellis, Perry~W.}, \au{Fernandez-Nieves,
  Alberto} \& \au{Giomi, L.}} \yr{2019}  \at{Geometrical control of active
  turbulence in curved topographies}.  \jt{Phys. Rev. Lett.}  \bvol{122},
  \pg{168002}.

\bibitem[R\"{a}tz \& Voigt(2006)]{Ratz06}
{\sc \au{R\"{a}tz, A.} \& \au{Voigt, A.}} \yr{2006}  \at{{PDE}'s on surfaces-a
  diffuse interface approach}.  \jt{Commun. Math. Sci.}  \bvol{4},  \pg{575 --
  590}.

\bibitem[Rembiasz {\em et~al.\/}(2017)Rembiasz, Obergaulinger, Cerd\'a-Dur\'an,
  Aloy \& M\"uller]{rembiasz17}
{\sc \au{Rembiasz, T.}, \au{Obergaulinger, M.}, \au{Cerd\'a-Dur\'an, Pablo},
  \au{Aloy, M.-\'A.} \& \au{M\"uller, E.}} \yr{2017}  \at{On the measurements
  of numerical viscosity and resistivity in {E}ulerian {MHD} codes}.
  \jt{Astrophys. J. Suppl. S.}  \bvol{230},  \pg{18}.

\bibitem[Reuther \& Voigt(2018)]{reuther18}
{\sc \au{Reuther, S.} \& \au{Voigt, A.}} \yr{2018}  \at{Solving the
  incompressible surface {N}avier-{S}tokes equation by surface finite
  elements}.  \jt{Physics of Fluids}  \bvol{30}~(1),  \pg{012107}.

\bibitem[Rieutord(2015)]{rieutord15}
{\sc \au{Rieutord, M.}} \yr{2015} {\em Fluid Dynamics: An Introduction\/}.
  \publ{Springer}.

\bibitem[Sasaki {\em et~al.\/}(2015)Sasaki, Takehiro \& Yamada]{sasaki15}
{\sc \au{Sasaki, E.}, \au{Takehiro, S.} \& \au{Yamada, M.}} \yr{2015}
  \at{Bifurcation structure of two-dimensional viscous zonal flows on a
  rotating sphere}.  \jt{Journal of Fluid Mechanics}  \bvol{774},
  \pg{224--244}.

\bibitem[Serrin(1959)]{serrin59}
{\sc \au{Serrin, J.}} \yr{1959}  \at{Mathematical principles of classical fluid
  mechanics}.  \bt{In {\em Encyclopedia of physics, Vol. VIII/1 (Fluid dynamics
  I)\/} (ed. \ed{S.~Fl\"ugge \& C.~Truesdell})},  \pg{pp. 125--263}.
  \publ{Springer-Verlag}.

\bibitem[Shan(2006)]{shan06}
{\sc \au{Shan, X.}} \yr{2006}  \at{Analysis and reduction of the spurious
  current in a class of multiphase lattice {B}oltzmann models}.  \jt{Phys. Rev.
  E}  \bvol{73},  \pg{047701}.

\bibitem[Sofonea {\em et~al.\/}(2018)Sofonea, Biciu\cb{s}c\u{a}, Busuioc,
  Ambru\cb{s}, Gonnella \& Lamura]{sofonea18pre}
{\sc \au{Sofonea, V.}, \au{Biciu\cb{s}c\u{a}, T.}, \au{Busuioc, S.},
  \au{Ambru\cb{s}, V.~E.}, \au{Gonnella, G.} \& \au{Lamura, A.}} \yr{2018}
  \at{Corner-transport-upwind lattice {B}oltzmann model for bubble cavitation}.
   \jt{Phys. Rev. E}  \bvol{97},  \pg{023309}.

\bibitem[Sofonea {\em et~al.\/}(2004)Sofonea, Lamura, Gonnella \&
  Cristea]{sofonea04}
{\sc \au{Sofonea, V.}, \au{Lamura, A.}, \au{Gonnella, G.} \& \au{Cristea, A.}}
  \yr{2004}  \at{Finite-difference lattice {B}oltzmann model with flux limiters
  for liquid-vapor systems}.  \jt{Phys. Rev. E}  \bvol{70},  \pg{046702}.

\bibitem[Sofonea \& Sekerka(2003)]{sofonea03}
{\sc \au{Sofonea, V.} \& \au{Sekerka, R.~F.}} \yr{2003}  \at{Viscosity of
  finite difference lattice {B}oltzmann models}.  \jt{J. Comput. Phys.}
  \bvol{184},  \pg{422--434}.

\bibitem[Taylor(2011)]{taylor11}
{\sc \au{Taylor, M.~E.}} \yr{2011} {\em Partial Differential Equations III:
  Nonlinear Equations\/}, 2nd edn.  \publ{Springer-Verlag}.

\bibitem[Torres-S\'anchez {\em et~al.\/}(2019)Torres-S\'anchez, Mill\'an \&
  Arroyo]{torres-sanchez19}
{\sc \au{Torres-S\'anchez, A.}, \au{Mill\'an, D.} \& \au{Arroyo, M.}} \yr{2019}
   \at{Modelling fluid deformable surfaces with an emphasis on biological
  interfaces}.  \jt{J. Fluid Mech.}  \bvol{872},  \pg{218--271}.

\bibitem[Voigt(2019)]{voigt19}
{\sc \au{Voigt, A.}} \yr{2019}  \at{Fluid deformable surfaces}.  \jt{J. Fluid
  Mech.}  \bvol{878},  \pg{1--4}.

\end{thebibliography}

\end{document}